\documentclass[twoside,11pt]{article}

\usepackage{blindtext}

\usepackage{jmlr2e}

 \usepackage{amsmath}
 \usepackage{graphicx}
 \usepackage{bbm}
\usepackage{enumerate}
\usepackage{natbib}
\usepackage{hyperref}
\usepackage{url} 
 \usepackage{algorithm}
\usepackage{algpseudocode}
\usepackage{graphicx}
\usepackage{subcaption}
\usepackage{enumitem,enumerate}
\usepackage{xcolor}
\usepackage{bm}
\newcommand{\xhdr}[1]{\vspace{0.1mm}\noindent{{\bf #1.}}}

\newcommand{\indc}[1]{{\mathbf{1}_{\left\{{#1}\right\}}}}

\newcommand{\wh}{\widehat}
\newcommand{\bA}{\mathbf A}
\newcommand{\bU}{\mathbf U}
\newcommand{\bV}{\mathbf V}
\newcommand{\bS}{\mathbf \Sigma}
\newcommand{\bX}{\mathbf X}
\newcommand{\bL}{\mathbf L}
\newcommand{\bI}{\mathbf I}
\newcommand{\bB}{\mathbf B}
\newcommand{\bO}{\mathbf O}
\newcommand{\bTheta}{\mathbf \Theta}
\newcommand{\bGamma}{\mathbf \Gamma}
\newcommand{\bLambda}{{\mathbf \Lambda}}
\newcommand{\bY}{\mathbf Y}
\renewcommand{\P}{\mathbb{P}}
\newcommand{\wt}{\widetilde}

\newcommand{\fnorm}[1]{\left\|#1\right\|_{F}}
\newcommand{\opnorm}[1]{\left\|#1\right\|_{op}}
\newcommand{\nnorm}[1]{\big\|#1\big\|_{\rm *}}

\newcommand{\rank}{\mathop{\sf rank}}
\newcommand{\Tr}{\mathop{\sf Tr}}

\newcommand{\supp}{{\rm{supp}}}
\newcommand{\iprod}[2]{ \langle #1, #2 \rangle}

\newenvironment{psmallmatrix}
  {\left(\begin{smallmatrix}}
  {\end{smallmatrix}\right)}

\newcommand{\R}{\mathbb{R}}

\newcommand{\minimize}{\operatornamewithlimits{minimize}}
\newcommand{\maximize}{\operatornamewithlimits{maximize}}
\makeatletter
\newcommand{\doublewidetilde}[1]{{%
  \mathpalette\double@widetilde{#1}%
}}
\newcommand{\double@widetilde}[2]{%
  \sbox\z@{$\m@th#1\widetilde{#2}$}%
  \ht\z@=.9\ht\z@
  \widetilde{\box\z@}%
}

\usepackage{lastpage}
\jmlrheading{27}{2026}{1-\pageref{LastPage}}{1/25; Revised
7/26}{8/26}{25-0196}{Claire Donnat and Elena Tuzhilina}

\ShortHeadings{Canonical Correlation Analysis as Reduced Rank Regression in High Dimensions}{Donnat and Tuzhilina}
\firstpageno{1}

\begin{document}

\title{Canonical Correlation Analysis as Reduced Rank Regression in High Dimensions}

\author{\name Claire Donnat \email cdonnat@uchicago.edu\\
       \addr Department of Statistics\\
      University of Chicago\\ Chicago, IL, 60637, USA
       \AND
       \name Elena Tuzhilina\email elena.tuzhilina@utoronto.ca \\
       \addr Department of Statistical Sciences\\ University of Toronto\\ Toronto, ON, M5S1A1, Canada}

\editor{Pradeep Ravikumar}

\maketitle

\begin{abstract}
Canonical correlation analysis is a widespread technique for discovering linear relationships between two sets of variables. In high dimensions, however, standard estimates of the canonical directions cease to be consistent without assuming further structure. In this setting, a possible solution consists in leveraging the presumed sparsity of the solution: only a subset of the covariates span the canonical directions. While the last decade has seen a proliferation of sparse canonical correlation analysis methods, practical challenges regarding the scalability and adaptability of these methods still persist. To circumvent these issues, this paper suggests an alternative strategy that uses reduced rank regression to estimate the canonical directions when one of the datasets is high-dimensional while the other remains low-dimensional.  By casting the problem of estimating the canonical direction as a regression problem, our estimator is able to leverage the rich statistics literature on high-dimensional regression and is easily adaptable to accommodate a wider range of structural priors. Our proposed solution maintains computational efficiency and accuracy, even in the presence of very high-dimensional data. We demonstrate the advantages of our approach through a series of simulated experiments, achieving a substantial increase in accuracy compared to existing methods. Additionally, we highlight its practical utility by applying it to three real-world datasets, where we show that our method is able to uncover scientifically meaningful associations between variables.
\end{abstract}

\begin{keywords}
 canonical correlation analysis; structured sparsity; high-dimensional data; reduced rank regression;
\end{keywords}

\section{Introduction}
\label{sec:intro}

Consider two random vectors \(  x \) in \( \mathbb{R}^{p} \) and \(  y \) in \( \mathbb{R}^{q} \) with respective covariance matrices $\bS_{X}$ and $\bS_{Y}$, and cross-covariance matrix $\bS_{XY}$. The objective of {\it canonical correlation analysis} (CCA) is to identify a set of $r\leq \min(p,q)$ linear combinations of \(  x \) and  \(  y \) that are maximally correlated. Mathematically, for \(i = 1, \ldots, r \), the $i^{\text{th}}$ solution pair $( u_i,  v_i)$ of CCA can be expressed as the maximizer of the following constrained optimization problem:
\begin{equation}
\label{eq:cca}
\begin{split}
\maximize_{{u} \in \R^{p},~{v}\in \R^q} \quad&{u}^\top\bS_{XY}{v}~ \\
\mbox{subject to}\quad & u^\top\bS_{X} u =  v^\top\bS_{Y} v = 1, \quad u^\top\bS_{X} u_j=  v^\top\bS_{Y} v_j = 0~ \mbox{for } j<i.
\end{split}
\end{equation}
The pairs $(x  u_i, y v_i)$ are called the \textit{canonical variates}. The individual vectors $ u_i$ and $ v_i$ are called the \textit{canonical directions}, and the corresponding correlation values $\lambda_i = u_i^\top \bS_{XY}  v_i$ are called the \textit{canonical correlations}.  

Since its introduction by \cite{hotelling1936relations}, canonical correlation analysis has become an indispensable tool for analyzing relationships between sets of measurements. Given its versatility and interpretability, it is unsurprising that CCA has found applications in a wide spectrum of domains, ranging from psychology and social sciences \citep{thorndike2000canonical, fan2018canonical} to genomics \citep{witten2009extensions, parkhomenko2009sparse, lin2013group, gossmann2018fdr} or neuroscience \citep{zhuang2020technical}. 

Under the canonical pair model \citep{chen2013sparse, gao2017sparse}, it can be shown that the CCA model corresponds to a reparameterization of the cross covariance $\bS_{XY}$ as
\begin{equation}\label{eq:cca}
    \bS_{XY} = \bS_{X} \bU \bLambda \bV^\top \bS_{Y}, \mbox{\quad where\quad} \bU^\top \bS_{X} \bU = \bV^\top \bS_{Y} \bV = \bI_r.
\end{equation}  
Here $\bU = [ u_1|  u_2| \ldots|  u_r] \in \R^{p \times r}$ and $\bV = [ v_1|  v_2| \ldots|  v_r] \in \R^{q \times r}$ represent the $r$ canonical directions, and $\bLambda\in \R^{r \times r}$ is a diagonal matrix with canonical correlations $\lambda_i$ on the diagonal ranked by decreasing order of magnitude. It is a classical exercise to show that the canonical directions $(u_i, v_i)$ can be obtained through a transformation of the right and left singular vectors of $\bS_{X}^{-1/2}\bS_{XY}\bS_{Y}^{-1/2}$. Specifically, given the singular value decomposition (SVD)
$    \bS_{X}^{-1/2}\bS_{XY}\bS_{Y}^{-1/2} = {\bU}_0 \bLambda  {\bV}_0^\top,$
the canonical directions $u_i$ and $v_i$ are simply the $i^{\text{th}}$ columns of the matrices $\bU = \bS_{X}^{-1/2} {\bU}_0$ and  $\bV = \bS_{Y}^{-1/2}  {\bV}_0$ respectively. In practice, the canonical directions are typically obtained by replacing the population covariance matrices by their sample estimators, i.e. $\widehat{\bS}_{X}, \widehat{\bS}_{Y}$  and $\widehat{\bS}_{XY}$,  respectively.

Recent studies have become increasingly interested in applying CCA to the analysis of high-dimensional datasets where at least one of the dimensions is significantly larger than the number of observations $n$. It is a well-established fact that the conventional CCA framework breaks down in this regime, both from a theoretical and computational perspective \citep{gao2017sparse}. From a computational standpoint, if, for instance, \( p > n \), the sample covariance matrix \( \widehat{\bS}_{X} \) is no longer invertible, thereby preventing the straightforward calculation of the CCA directions using a singular value decomposition of the transformed cross-covariance matrix $  \wh\bS_{X}^{-1/2}\wh\bS_{XY}\wh\bS_{Y}^{-1/2} $. From a theoretical point of view, the eigenvector estimates obtained from the sample covariance matrices \( \widehat\bS_{X} \) and \( \widehat\bS_{XY} \) are no longer guaranteed to be consistent \citep{gao2015minimax, gao2017sparse}, and replacing the sample estimator $\widehat{\bS}_{X}^{-1/2}$ by its pseudoinverse may not lead to an accurate estimate of the canonical directions.

To overcome the challenge posed by the high-dimensionality of data, several adaptations of the original CCA problem have recently been developed. To keep our discussion concise, a comprehensive discussion of these methods is deferred to Section~\ref{sec:related_works}, which we summarize here. All of these variants incorporate some form of regularization or constraint on the estimators of $\bU$ and $\bV$, and can be classified into one of two categories depending on the type of regularization they employ: {\it ridge-regularized CCA}, which mitigates ill-conditioning by adding \(\ell_2\) penalties, and {\it sparse CCA}, which incorporates \(\ell_1\) penalties to enforce sparsity, focusing on selecting only the most informative variables. 
While sparse CCA methods have proliferated \citep{witten2009extensions, wilms2016, parkhomenko2009sparse,waaijenborg2009}, many rely on heuristics, lack guarantees on the accuracy of the estimated directions, or make simplifying assumptions (for instance, that the covariances $\mathbf{\Sigma}_X$ and $\mathbf{\Sigma}_Y$ are diagonal to enhance computational speed, e.g. \cite{witten2009extensions}). In contrast, theoretical efforts \citep{gao2015minimax, gao2017sparse, gao2021sparse} establish algorithms with provable guarantees and high-probability bounds on the error of the associated estimators of $\bU$  and $\bV$, along with the corresponding minimax rates. However, these methods rely on a prohibitively computationally intensive initialization step (hereafter referred to as the ``Fantope initialization'', detailed in Appendix~\ref{algo:comp}) that first estimates the joint product $\bU \bV^\top$. Because it scales cubically with the dataset dimensions \citep{gao2021sparse}, this initialization step effectively makes this entire class of methods unusable on most real-world datasets.

This highlights an important gap in current CCA methodologies. Existing heuristic methods are fast but tend to yield poor estimates of the canonical directions. Conversely, while theoretical approaches guarantee a more accurate estimation, their computational demands are so substantial that they become impractical even on moderately sized datasets.  Moreover, the methods developed by \cite{gao2015minimax, gao2017sparse} and \cite{gao2021sparse} all assume symmetry in the roles played by $\bX$ and $\bY$, thus imposing sparsity in both.  {
However, many contemporary applications of CCA exhibit substantial asymmetry between the two views. Examples include neuroimaging studies, where thousands of imaging-derived features are related to a relatively small number of cognitive, behavioral, or clinical measurements \citep{smith2015positive,zhuang2020technical}; climate applications, where high-dimensional spatial fields are associated with a small number of climate indices or forecasting targets \citep{barnett1987origins,bretherton1992intercomparison}; and genomics studies, where high-dimensional molecular profiles are linked to a limited number of phenotypic or clinical outcomes \citep{chi2013, jiang2023}. Such dimensional imbalance is particularly prevalent in neuroscience applications, where modern imaging technologies routinely generate tens of thousands of features from a relatively modest number of subjects and phenotypic observations; see Table~\ref{tab:survey} in Appendix~\ref{app:review} for representative examples. In these settings, regularization is primarily needed on the high-dimensional side, whereas imposing sparsity on the low-dimensional view may offer limited benefit while increasing both statistical and computational complexity. This observation suggests that the symmetric regularization strategies adopted by existing sparse CCA methods may not be optimal in many practical settings and motivates the development of methods that explicitly exploit this dimensional imbalance to improve scalability while retaining strong statistical performance.
}


 {
Our approach adopts this perspective by imposing structural constraints on only one of the two canonical loading matrices. By regularizing only the high-dimensional view, we substantially reduce the computational burden associated with theoretically grounded sparse CCA methods while retaining statistical guarantees. At the same time, this asymmetric formulation achieves improved estimation accuracy relative to existing heuristic approaches. As a result, our method effectively leverages the imbalance between the two views to improve scalability without sacrificing statistical performance.
}

\medskip

\noindent {
\xhdr{Contributions}
In this paper, we study asymmetric high-dimensional CCA, where one view is structured and high-dimensional while the other remains low-dimensional. Our main contributions are as follows:
}

\begin{enumerate}

\item  {\textbf{A scalable alternative to Fantope-based sparse CCA.}
In Section~\ref{sec:rrr}, we develop a two-step estimation procedure inspired by existing theoretically grounded sparse CCA methods \citep{gao2015minimax, gao2017sparse, gao2021sparse}, while avoiding the computationally intensive Fantope initialization. Specifically, we first estimate the low-rank signal $\bU \bLambda\bV_0^\top$ via a convex reduced-rank regression formulation and subsequently recover the canonical directions $\bU$ and $\bV$. The resulting method achieves substantially improved computational scalability while preserving strong statistical guarantees.}

\item  {\textbf{Theoretical guarantees under asymmetric structural constraints.}
In Section~\ref{sec:high_d}, we establish finite-sample error bounds for the estimated canonical subspaces under structural constraints imposed on only one view (i.e. one of the datasets). Our analysis shows that consistent estimation is possible in this asymmetric setting and provides theoretical guarantees for a broad class of structured regularization schemes.}

\item  {\textbf{A flexible framework for structured canonical directions.}
Our methodology accommodates a broad class of structural assumptions on the high-dimensional view, including sparsity, group sparsity, and graph-based smoothness penalties (Section~\ref{sec:group-sparse-cca}). This allows flexible structure to be incorporated into the estimation procedure. For instance, features may be partitioned into predefined groups (thus motivating group-sparse loadings), or they may form a spatial or network graph (thereby motivating smoothness constraints that encourage neighboring features to have similar loadings). This flexibility can lead to improved and more interpretable canonical directions in applications where the high-dimensional features exhibit known structure. Through simulations and applications to neuroscience and ecological data (Section~\ref{sec:exp}), we show that our proposed estimator recovers interpretable structured loadings, achieve competitive statistical performance compared to existing sparse CCA methods, and are substantially faster to compute.}

\end{enumerate}

\section{Notations \& Parameter Space} 
For any $t\in \mathbb{Z}_+$, $[t]$ denotes the set $\{1, 2, ..., t\}$. 
For any set $S$, $|S|$ denotes its cardinality and $S^c$ its complement.
For any $a,b\in \R$,
$a\vee b = \max(a,b)$ and ${a\wedge b = \min(a,b)}$. For a vector $u$, $\| u\|=\sqrt{\sum_i u_i^2}$, $\| u\|_0=\sum_i\indc{u_i\neq 0}$, and $\| u\|_1=\sum_i|u_i|$.
For any matrix ${\bA=(a_{ij})\in \R^{p\times k}}$ (denoted by a capital letter and bold font), $\bA_{i\cdot}$ denotes its $i$-th row and 
${\supp{(\bA)}=\{i\in[p]: \|\bA_{i\cdot}\|>0\}}$, the index set of nonzero rows, is called its support.
Throughout this paper, $\sigma_i(\bA)$ stands for the $i^{th}$ largest singular value of the matrix $\bA$ and ${\sigma_{\max}(\bA)=\sigma_1(\bA)}$, ${\sigma_{\min}(\bA)=\sigma_{p\wedge \text{rank}(\bA)}(
\bA)}$. The Frobenius norm and the operator norm of $\bA$ are $\fnorm{\bA}=\sqrt{\sum_{i,j}a_{ij}^2}$ and ${\opnorm{\bA}=\sigma_{\max}(\bA)}$, respectively. The $\ell_{21}$ norm and the nuclear norm are  $\big\|\bA\big\|_{21}=\sum_{i \in [p]}\|\bA_{i\cdot}\|_2$ and ${\nnorm{\bA}=\sum_i\sigma_i(\bA)}$, respectively. 
For any positive semi-definite matrix $\bA$, $\bA^{1/2}$ denotes its principal square root, which is positive semi-definite and satisfies $\bA^{1/2} \bA^{1/2} = \bA$. 
The trace inner product of two matrices $\bA,\bB\in\mathbb{R}^{p \times k}$ is denoted by $\iprod{\bA}{\bB}=\Tr(\bA^\top \bB)$. The notation $\bA^{\dagger}$ is used here to denote the pseudo-inverse of the matrix $\bA$.

We consider here the canonical pair model of \cite{gao2017sparse}, where the observed $n$ pairs of measurement vectors $( x_i,  y_i)_{i=1}^n$ are i.i.d.~from a multivariate Gaussian distribution $\mathcal{N}_{p+q}(0, \bS),$ where
$\bS=\begin{psmallmatrix} \bS_{X} & \bS_{XY} \\ \bS_{YX} & \bS_{Y} \end{psmallmatrix}$
with the cross-covariance matrix $\bS_{XY}$ satisfying \eqref{eq:cca}.
We are interested in the situation where the leading canonical coefficient vectors of $\bU$ are sparse, without assuming any particular structure on $\bV$. Sparsity is here understood in terms of the support of $\bU$. More specifically, we define 
$\mathcal{F}(s_u,p,q,r,\lambda;M)$ to be the collection of all covariance matrices $\bS$ satisfying (\ref{eq:cca}) and 
{\allowdisplaybreaks
\begin{align}
\text{(a)}&~~ \mbox{$\bU\in\mathbb{R}^{p\times r}$ and $\bV\in\mathbb{R}^{q\times r}$ with $|\supp({\bU})|\leq s_u$;}
\label{eq:para}\nonumber\\
\text{(b)}&~~ \mbox{$\sigma_{\min}(\bS_X)\wedge\sigma_{\min}(\bS_Y)\geq \frac1M$ and $\sigma_{\max}(\bS_X)\vee\sigma_{\max}(\bS_Y)\leq M$;}\\
\text{(c)}&~~ \mbox{$\lambda_r\geq\lambda$ and $\lambda_1\leq 1-\frac 1M$\nonumber.}
\end{align}
}
The probability space that we consider throughout this paper is thus
\begin{equation}
\label{eq:para-space}
\big\{( X_i, Y_i)\stackrel{i.i.d.}{\sim} \mathcal{N}_{p+q}(0,\bS)
\quad\text{with }\bS\in \mathcal{F}(s_u ,p,q,r,\lambda;M)\big\}.
\end{equation}
 We shall allow $s_u,p,\lambda$ to vary with $n$, while $M > 1$ is restricted to be an absolute constant. We will also assume that $q$ and $r$ are fixed. Note that this latter assumption is in contrast to the approach by \cite{gao2017sparse} that supposes that $q$ and $r$ can also vary with $n$.


\section{CCA via reduced rank regression} \label{sec:rrr}

In this section, we begin by highlighting a link between reduced rank regression (RRR) and CCA. As we will see in Section~\ref{sec:high_d}, this link will then allow us to formulate a theoretically sound and practical solution for canonical correlation analysis in high dimensions. 
We note here that, while this connection has already been recognized in the existing literature \citep{izenman1975, torre2012}, this formulation has not been previously used in the context of high-dimensional CCA. 

Here we assume that $p, q \leq n$ are fixed, as $n$ is allowed to grow, so that $\widehat\bS_X$ and $\widehat\bS_Y$ are invertible and consistent estimators of $\bS_X$ and $\bS_Y$, respectively. We first consider the CCA formulation as a prediction problem, as noted by \cite{gao2017sparse}:
\begin{equation}
\begin{split}
    \minimize_{\bU\in\R^{p\times r},~\bV\in\R^{q\times r}} \left\|\bY\bV  - \bX\bU\right\|_F^2
    \quad \mbox{subject to}   \quad  \bU^\top\wh\bS_{X} \bU = \bV^\top\wh\bS_{Y} \bV  = \bI_r.
\label{eq:CCA_pred}
\end{split}
\end{equation}

\begin{lemma} 
Letting $\bY_0 = \bY \wh\bS_{Y}^{-1/2}$, the estimators for the first $r$ canonical directions can be recovered as $\wh {\bU}$ and $\wh \bV = \wh\bS_{Y}^{-1/2}\wh {\bV}_0$ from the solution of the following problem:
\begin{equation}
   \wh {\bU}, \wh {\bV}_0 =  \operatorname*{argmin}_{\substack{\bU\in\R^{p\times r},~\bV_0\in\R^{q\times r}\\ 
   \bU^\top\wh\bS_{X} \bU=\bV_0^\top \bV_0= \mathbf I_r}} \|\bY_0 - \bX\bU\bV_0^\top\|_F^2 
\label{rrr:op1}
\end{equation}

\label{lemma:cca:rrr}
\end{lemma}
The proof of this statement can be found in Appendix~\ref{proof:lemma:cca:rrr}. Using this objective function, it is evident that the recovery of the matrix $\bU \bV_0^{\top}$ in (\ref{rrr:op1}) is a constrained instance of the more general reduced-rank regression problem:
\begin{equation}
\begin{split}
    \minimize_{\bB_0\in\R^{p\times q}}\quad \| {\bY_0} - \bX\bB_0\|_F^2
    \quad\mbox{subject to} \quad \rank(\bB_0) =  r,
\label{eq:rrr}
\end{split}
\end{equation}
where $\bB_0$ must have the additional constraint that ${\bB_0}^\top\widehat{\bS}_{X}\bB_0$ and $\widehat{\bS}_X^{1/2}\bB_0{\bB_0}^\top\widehat{\bS}_X^{1/2}$ are projection matrices. The following theorem allows us to tie the solution of the reduced rank regression problem (\ref{eq:rrr}) to the solution of the CCA problem (\ref{eq:CCA_pred}).

\begin{theorem}[Low-dimensional CCAR$^3$ subspace consistency]\label{theorem:cca:rrr}
Let
$\mathbf Y_0=\mathbf Y\widehat{\mathbf\Sigma}_Y^{-1/2},$
where \(\widehat{\mathbf\Sigma}_Y\) is the sample covariance matrix of \(\mathbf Y\).
Assume that the cross-covariance matrix follows the canonical pair model
\[
\mathbf\Sigma_{XY}
=
\mathbf\Sigma_X\mathbf U\mathbf\Lambda \mathbf V^\top \mathbf\Sigma_Y,
\qquad
\mathbf U^\top\mathbf\Sigma_X\mathbf U
=
\mathbf V^\top\mathbf\Sigma_Y\mathbf V
=
\mathbf I_r,
\]
with \(\lambda_r>0\). Let
\[
\widehat{\mathbf B}
=
\operatorname*{argmin}_{\mathbf B\in\mathbb R^{p\times q}}
\|\mathbf Y_0-\mathbf X\mathbf B\|_F^2
=
\widehat{\mathbf\Sigma}_X^{-1}
\widehat{\mathbf\Sigma}_{XY}
\widehat{\mathbf\Sigma}_Y^{-1/2}
\]
and let
$\operatorname{svd}_r(\widehat{\mathbf B})
=
\widehat{\mathbf U}_{\hat B}
\widehat{\mathbf \Lambda}_{\hat B}
\widehat{\mathbf V}_{\hat B}^\top
$
denote the rank-\(r\) truncated singular value decomposition of
\(\widehat{\mathbf B}\), where \(\widehat{\mathbf U}_{\hat B}\) and
\(\widehat{\mathbf V}_{\hat B}\) contain the leading \(r\) left and right singular vectors of $\widehat{\mathbf B}$.
Define
\[
\widehat{\mathbf U}
=
\widehat{\mathbf U}_{\hat B}
\left(
\widehat{\mathbf U}_{\hat B}^\top
\widehat{\mathbf\Sigma}_X
\widehat{\mathbf U}_{\hat B}
\right)^{-1/2},
\qquad
\widehat{\mathbf V}
=
\widehat{\mathbf\Sigma}_Y^{-1/2}\widehat{\mathbf V}_{\hat B} .
\]
Then \(\widehat{\mathbf U}\) and \(\widehat{\mathbf V}\) are consistent estimators of
the canonical subspaces spanned by \(\mathbf U\) and \(\mathbf V\), respectively:
\[
\min_{\mathbf O\in\mathcal O_r}
\|
\widehat{\mathbf U}-\mathbf U\mathbf O
\|_F
\xrightarrow{a.s.}0,
\qquad
\min_{\widetilde{\mathbf O}\in\mathcal O_r}
\|
\widehat{\mathbf V}-\mathbf V\widetilde{\mathbf O}
\|_F
\xrightarrow{a.s.}0.
\]
\end{theorem}
The proof of this theorem is provided in Appendix~\ref{proof:theorem:cca:rrr}.

Lemma~\ref{lemma:cca:rrr} and Theorem~\ref{theorem:cca:rrr} imply that canonical subspaces can be estimated via a simple two-step procedure. First, compute the ordinary least squares (OLS) estimator $\widehat{\bB}$. Second, obtain the leading $r$ left and right singular vectors of $\widehat{\bB}$ and normalize them appropriately to recover estimates of the canonical directions. Under the conditions of Theorem~\ref{theorem:cca:rrr}, these estimators consistently recover the population canonical subspaces.
The full procedure, which we will refer to as CCAR$^3$, is outlined in Algorithm~\ref{alg:rrr}.

\begin{algorithm}
\caption{CCAR$^3$}
\label{alg:rrr}
\textit{Input:} $\bX \in \R^{n \times p}$,  $\bY \in \R^{n \times q}$  with $p, q \leq n;$ number of components $r\leq \min(p,q).$ 
\begin{algorithmic}[1] 
\State Normalize $\bY_0 = \bY \widehat{\bS}_{Y}^{-\frac 12}.$
\State Compute the OLS solution
${\widehat \bB}  = \operatorname*{argmin}_{\bB \in \R^{p\times q}} \|  \bY_0  - \bX\bB\|_F^2.$
{ 
\State Compute the singular value decomposition $ \operatorname{svd}_r(\widehat \bB) = \widehat \bU_{\widehat B} \widehat \bLambda_{\widehat B} \widehat \bV_{\widehat B}^\top.$ 
}
 {\State Normalize directions
$\wh \bU = \widehat \bU_{\widehat B} (  \widehat \bU_{\widehat B}^\top {\wh\bS}_{X}  \widehat \bU_{\widehat B})^{-1/2}$ and $\widehat \bV  = \widehat{\bS}_Y^{-1/2} \widehat \bV_{\widehat B}.$
}
\end{algorithmic}
\textit{Output:} CCA directions $\widehat  \bU $ and  $\widehat  \bV$.

\end{algorithm}

The proposed two-step approach differs from traditional CCA solutions in low dimensions. Initially, given the original data matrix $\bX$ and the normalized matrix $\bY_0 = \bY\widehat{\bS}_{Y}^{-1/2}$, we identify the joint subspace $\wh \bB = \wh\bU\wh\bLambda\wh\bV_0^\top$. This is done by casting the estimation of $\bB$ as a convex optimization problem, much in the same spirit as \cite{gao2017sparse} and \cite{gao2021sparse}. At the same time, this optimization problem is considerably easier than the initialization procedure of \cite{gao2017sparse}. 
We then use the solution ${\widehat \bB}$ to further estimate the canonical directions. Compared to alternative approaches, this method is therefore less prone to initialization issues. 


As dimensions increase, particularly when \( p \) increases and exceeds \( n \), this approach becomes infeasible. The inverse of the matrix \( \widehat \bS_{X} \) ceases to be well-defined, rendering the OLS solution to problem (\ref{eq:CCA_pred}) is non-unique and inconsistent. It is important to remember that in our discussion, \( q \) is considered low-dimensional, so calculating \( \widehat \bS_{Y} \) remains manageable and consistent for the estimation of $\bS_Y$. Unlike \( \bX \), computing \( \bY_0 = \bY \widehat\bS_{Y}^{-1/2} \) is still well-defined.  In the following sections, we will show, nonetheless, that this formulation of the CCA problem is easily extendable to the high-dimensional case.


\section{CCA in high dimensions}
\label{sec:high_d}

As emphasized in the previous section, in high dimensions (more specifically, when $ p$ is on the order of or exceeds $n$), our proposed CCA estimator breaks down as 
the solution of the OLS problem is not necessarily unique.   Imposing some low-dimensional structure on the estimators thus becomes indispensable to reliably estimate the CCA directions.  
A number of such constraints can  be stated in a general form as
$\text{Pen}(\bU)\leq s $, where $\text{Pen}$ is a penalty function on the matrix $\bU$, thus transforming the original CCA objective into the following constrained problem:
\begin{equation}\label{eq:cca_reg_high_di}
    \begin{split}
    \operatorname*{maximize}_{\bU\in\R^{p\times r}, \bV\in\R^{q\times r}} \Tr(\bU^\top \widehat{\bS}_{XY}\bV)  
   \mbox{~subject to~} \bU^\top \widehat{\bS}_{X} \bU = \bV^\top \widehat{\bS}_{Y} \bV  = \mathbf I_r \mbox{~and~}  \operatorname{Pen}(\bU) \leq s.
    \end{split}
\end{equation}
Note that in the traditional CCA model, the directions are obtained up to rotations of the columns. Consequently, the penalty used here is always a row-penalty. Below we specialize our analysis to the different types of constraints that can be imposed on $\bU$:  sparsity, group-sparsity, and graph-sparsity. Importantly, we will highlight how our formulation of the CCA problem enables the flexible adoption of a wide variety of constraints.

\subsection{CCA with sparsity} 
\label{sec:sparse-cca}
As outlined in the introduction, one strategy for estimating the  canonical directions in a high-dimensional context consists in imposing sparsity. The usual hypothesis consists of assuming that $\bU$ is row sparse, indicating that only a limited number of covariates of $\bX$ are involved in determining the canonical directions. Mathematically, this concept is often expressed as \( |\operatorname{supp}(\bU)| \leq s \), which constrains the number of nonzero rows in $\bU$ to be smaller than $s$. 
This immediately implies that \( \bB = \bU\bLambda\bV_0^{\top} \) also exhibits row sparsity: $|\operatorname{supp}(\bB)| \leq s.$ 

To allow consistent estimation of $\bU$ and $\bV$, we propose replacing the OLS step in Algorithm~\ref{alg:rrr} by
\begin{equation}
    {\widehat \bB}  = \operatorname*{argmin}_{\bB \in \R^{p\times q}} \frac{1}{n}\|  \bY_0  - \bX\bB\|_F^2 + \rho \| \bB \|_{21}.
\label{eq:ols:pen}
\end{equation}
Note here that (\ref{eq:ols:pen}) is a convex relaxation of the OLS problem with sparsity constraint $ {|\operatorname{supp}(\bB)| \leq s.}$ 
This ensures the tractability of the objective
and returns a solution that is row-sparse, i.e., for which only a subset $I$ of the rows of ${\widehat \bB}$ are nonzero. We can then resume the algorithm proposed in the previous section. The complete procedure is described in Algorithm \ref{alg:rrr_high_d}. In practice, to efficiently solve the problem (\ref{eq:ols:pen}) we propose a simple implementation relying on the Alternating Direction Method of Multipliers (ADMM; \citealp{boyd2011}), which we describe explicitly in Appendix \ref{algo:spr3-cca}.

\begin{algorithm}
\caption{Sparse CCAR$^3$ in high dimensions}
\label{alg:rrr_high_d}
\textit{Input:} $\bX \in \R^{n \times p}$,  $\bY \in \R^{n \times q}$ with $q \leq n$ and $p\gg n$; number of components $r\leq \min(p,q).$
\begin{algorithmic}[1] 
\State Normalize $\bY_0 = \bY \widehat{\bS}_{Y}^{-1/2}.$
\State Solve the penalized least squares using ADMM:
$${\widehat \bB}  = \operatorname*{argmin}_{\bB \in \R^{p\times q}} {\frac 1n} \|  \bY_0  - \bX\bB\|_F^2 + \rho \| \bB\|_{21}. $$
{ 
\State Compute the singular value decomposition
$ \operatorname{svd}_r({\widehat \bB}) = \widehat \bU_{\wh B} \widehat \bLambda_{\wh B} \widehat \bV^\top_{\wh B}.$ 
\State Normalize directions
$ \widehat  \bU = \wh{\bU}_{\widehat B} (\wh\bU_{\widehat B}^\top \wh\bS_X \wh\bU_{\widehat B})^{-1/2}$ and $\widehat  \bV = {\bf \widehat \Sigma}^{-1/2}_Y\wh\bV_{\widehat B}.$
}
\end{algorithmic}
\textit{Output:} CCA directions $\widehat  \bU $ and
$\widehat  \bV$.

\label{algo:sparse_CCA}
\end{algorithm}

It is possible to show that the addition of the $\ell_{21}$-penalty ensures that the learned $\widehat{\bB}$ is not too far from its population version $\bB^* = \bU\bLambda \bV^\top \bS_Y^{1/2}$ (Theorem~\ref{theorem:rrr_ols}). This allows in turn to ensure that the estimates $\widehat{\bU}$ and $\widehat{\bV}$ are good approximations of $\bU$ and $\bV$, respectively (Theorem~\ref{theorem:rrr_ols_2}).
To keep this manuscript concise, all proofs are deferred to Appendix~\ref{appendix:proof_sparse}. 

\begin{theorem}
\label{theorem:rrr_ols}
 Consider the family $\mathcal{F}(s_u, p, q, r, \lambda, M)$ of covariance matrices satisfying assumptions~(\ref{eq:para}).
Assume $({{q}+s_u\log({ep}/{s_u})})/{n}\leq c_0$
   for some sufficiently small constant ${c_0\in(0,1)}$. We write $\mathbf{\Delta} = \widehat{\bB} - {\bB}^*$ with  ${\bB}^*  = \bU\bLambda \bV^\top\bS_{Y}^{1/2} $, and choose $\rho \geq C_u M^2\sqrt{({q+\log p})/{n}}$ for some large constant $C_u$. Then the solution $\wh \bB$  of problem (\ref{eq:ols:pen}) is such that, for any $C'>0$, there exist constants $C_1$ and $C_2$ that solely depend on $C'$ and $c_0$ such that:
\begin{equation}\label{eq:reg_bound_app}
 \Big\|\widehat{\bS}_X^{\frac 12} \mathbf{\Delta} \Big\|_F^2 \leq   C_1 M^5 s_u \frac{q+\log p}{n} \mbox{\quad and \quad}               \|\widehat \bB-\bB^*\|_F^2\lesssim M^6s_u(q+\log p)/n
\end{equation}
with probability at least 
${1 -\exp(-C' q)-\exp(-C' s_u\log({ep}/{s_u}))} -\exp(-C' (q +\log p))-\exp(- C'(r +\log p)).$
 Moreover, with the same probability, the size of the support of $\wh \bB$ is of the order of $s_u$, that is $|\supp({\wh \bB})| \lesssim M^2 s_u.$
 \end{theorem}

Theorem~\ref{theorem:rrr_ols} implies that, under certain regularity conditions for the matrix $\bS_X$ (more specifically, that it is well conditioned and $\sigma_{\max}(\bS_X) \leq M$), we know that the support of the solution $\wh \bB$ is roughly of the order of the actual support of ${\bB}^*$. This means in particular that the number of discoveries --- and in particular, false discoveries, where a false discovery is understood as a nonzero row in $\widehat{\bB}$ that is not in the support of ${\bB}^*$ --- is of constant order and does not grow with $n$.

 To put this bound into perspective, we compare it with the results of \cite{gao2017sparse} (Theorem 4.1), who show that the error in the joint estimation of the subspace $\bU\bV^\top$ from their Fantope initialization procedure is
$ \| \wh \bA - \bU\bV^\top \|_F^2 \leq  C s_u s_v { \log(p+q)}/{n},$
where $s_v$ is the number of non-zero rows of $\bV$. The error bound of our procedure for estimating the matrix $ \bU \bLambda  \bV^\top\bS_Y^{1/2}$ is therefore of the same order, since we do not assume any sparsity structure on $\bV$ (i.e. $s_v=q$). Moreover, since we do not have to identify the rows of $\bV$ relevant to the canonical directions, we do not have to pay the price of the additional logarithmic factor of $q$ that the setting considered by \cite{gao2017sparse} incurs. 
Theorem~\ref{theorem:rrr_ols_2} provides the analysis of the error bounds on both $\wh {\bU}$ and $\wh {\bV}$ from Algorithm~\ref{algo:sparse_CCA}.
{ 
\begin{theorem}
\label{theorem:rrr_ols_2}
Assume the conditions of Theorem~\ref{theorem:rrr_ols}. Then, for any $C'>0$, the procedure outlined in Algorithm~\ref{alg:rrr_high_d} yields normalized bases $\widehat\bU$ and $\widehat\bV$ such that there exist orthogonal matrices $\bO, \widetilde\bO\in\R^{r\times r}$ with
\begin{equation}
    \begin{split}
\big\| \widehat{\bU}  - \bU \widetilde{\bO} \big\|_F &\leq C\frac{M^{6}}{\lambda} \sqrt{rs_u\frac{ (q + \log p)}{n}}~\text{ and }~
\big\| \widehat{\bV} - \bV \bO\big\|_F  \leq   C\frac{M^{\frac{11}{2}}}{\lambda} \sqrt{rs_u\frac{ (q + \log p)}{n}} 
    \end{split}
\end{equation}
with probability at least 
$1 -\exp(-C' q)-\exp(-C' s_u\log({ep}/{s_u}))-\exp(-C' (q+\log p ))-\exp(-C' (r+\log p )).$
Here $C$ is a constant that depends solely on $C'$ and $c_0$. 
\end{theorem}

}

We note that if the optional $r\times r$ rotation in Algorithm~\ref{alg:rrr_high_d} is applied and the canonical correlations are separated, ordered-direction consistency holds up to diagonal sign changes, with the displayed rate multiplied by a constant depending on the inverse canonical-correlation eigengap.

\xhdr{Observation 1: comparison with existing results} These results have to be compared to the theoretical results obtained by \cite{gao2017sparse}, where the authors derive the following bound for the error of the second step of their procedure:
$$  \min_{\bO \in \mathcal{P}}\Big\|\bS_{X}^{1/2} (\widehat{\bU}  - \bU \bO)\Big\|_F^2 \leq   C \frac{s_u({ r + \log p})}{\lambda^2{n}}.$$
One of the remarkable properties of this bound is that it decouples the dimensions of $\bS_X$ and $\bS_Y$: the bound on $\wh {\bU}$ does not depend on $q$. In our approach, by contrast, the estimation is performed jointly, and the estimation errors of $\wh {\bU}$ and $\wh {\bV}$ are therefore coupled. However, contrary to the method proposed by \cite{gao2017sparse}, our approach does not require splitting the data into 3 folds, and is able to operate on the entire dataset at once --- a situation, that, in practice, can be more efficient from an estimation and computational standpoint (see details in Appendix \ref{algo:comp}). 
Our bound is closer to that for the joint estimation procedure of \cite{gao2021sparse}, which is shown to be upper bounded as
$$ \min_{\bO\in\mathcal{P}} \big\|\wh \bV - \bV\bO\big\|_F \vee  \min_{ {\widetilde\bO\in\mathcal{P}}} \big\|\wh \bU - \bU\wt\bO\big\|_F \leq C\left(\frac{s'}{s_u+q}\right)^{\frac32} \frac{(3 +  \frac{2}{M^2})}{\lambda}\sqrt{\frac{{r(s_u + q)\log(p + q)}}{n}}. $$
Here $C$ is a constant that depends on the spectrum of $\bS_{XY}$, and $s'$ is a parameter controlling the hard row-sparsity of their initial estimator of the matrix $\bA = (\bU^\top | \bV^\top)$. This bound is more efficient than ours for the estimation of $\bV$, as ours scales with the product $q s_u$, rather than the sum $q + s_u$. 

\xhdr{Observation 2: Computational efficiency} 
 {The method of \cite{gao2021sparse} relies on a Fantope-based initialization, which, as discussed in the introduction, scales as $O\big((p+q)^3\big)$ and can incur substantial computation times. By contrast, the computational cost of our procedure when $n << p$ is $O(n^2q+npq)$, which is linear in the larger dimension (see Appendix~\ref{algo:spr3-cca} for more details). 
While an ADMM formulation is used for both CCAR$^3$ and Fantope-initialization approaches, each ADMM step in our method involves only simple matrix multiplications and shrinkage operations. In comparison, each ADMM step for the Fantope projection requires solving a large Lasso problem and computing the SVD of a $(p+q) \times (p+q)$ matrix (see Appendix~\ref{algo:comp}). 
Consequently, our approach provides an efficient alternative that substantially reduces computation while retaining theoretical guarantees.}

\xhdr{Simulation experiments} 
We generate a set of synthetic experiments that allow us to test the behavior of the different methods in the high-dimensional setting. 
To this end, we first construct a block diagonal covariance matrix as 
$\bS_{X} = \begin{psmallmatrix}    
        \bU_{X}\bU_{X}^\top & \mathbf{0}\\
        \mathbf{0} &  \mathbf I\\
    \end{psmallmatrix},$
where $\bU_{X} \in \R^{p_1 \times r_{pca}}$ 
is a randomly generated matrix from orthonormal columns, whose diagonal we then set to all ones. This creates a covariance structure for $\bX$ that is not trivial, but such that the first $r_{pca}$ principal components are sparse and are only spanned by $p_1$ out of $p$ rows. In our experiments, {we set $p_1 = 20$.} 
   We generate a covariance matrix for $\bY$ in a similar fashion as $\bS_{Y}=\bU_Y \bU_Y^\top$. Next, we construct the cross-covariance matrix $\bS_{XY}$ as $\bS_{XY} = \bS_{X} \bU \bLambda \bV^\top \bS_{Y}$. Here $\bU\in\R^{p\times r},~\bV\in\R^{q\times r}$ are sampled uniformly at random and chosen to have support of size $n_{nnz}$ with $n_{nnz}<p$ (row-sparse), and $q$ respectively. Both $\bU$ and $\bV$ are then normalized so that ${\bU^\top \bS_{X} \bU = \bV^\top\bS_{Y} \bV=\mathbf I_r}$. Further, $\bLambda$ is a diagonal matrix with $r$ entries $(\lambda_i)_{i=1}^r$ equally spaced in the following intervals, depending on the desired signal strength: (0.75, 0.9) for ``high'', (0.55, 0.7) for ``medium'', and (0.35, 0.5) for a ``low'' signal strength.
Finally, we generate the observations $( x_i,  y_i)_{i=1}^n$ by sampling from the normal joint distribution $\mathcal{N}_{p+q}\left(0, \begin{psmallmatrix}
    \bS_{X} & \bS_{XY}\\
    \bS_{YX} & \bS_Y
\end{psmallmatrix}\right).$

\begin{figure}
    \centering
\includegraphics[width=0.8\textwidth]{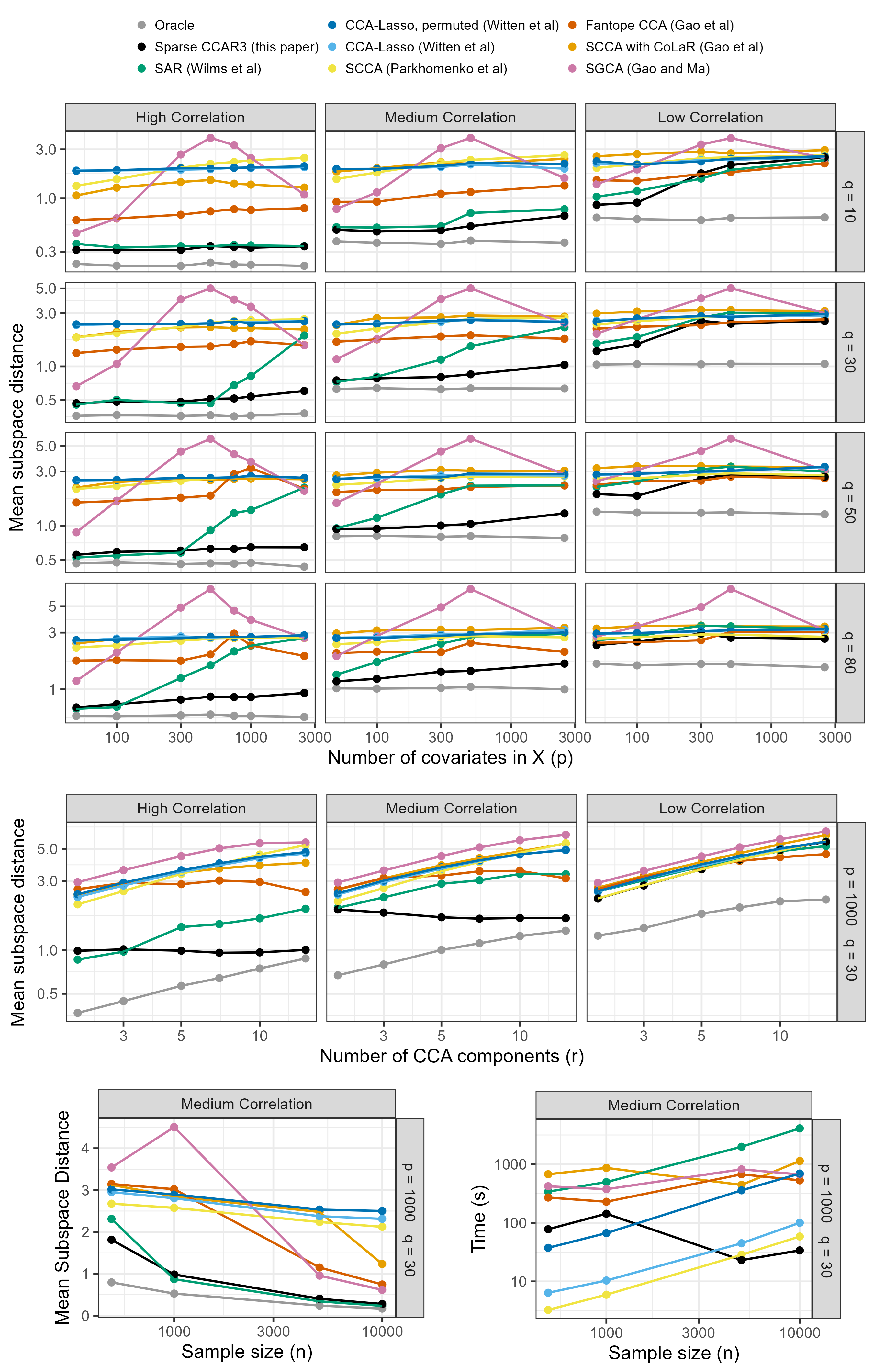}
\caption{Simulation results for sparse model with $r_{pca} = 5$. Top panel: subspace distance as a function of $p$, with $n=500$, $r = 3$ and $n_{nnz}=6$.  Middle panel: subspace distance as a function of the rank $r$, with $n=500,\quad  p=1,000, \quad q=30$ and $n_{nnz}=20$. Bottom panel: subspace distance (left) and computation time (right) as a function of $n$, with $q = 30$, $p=1000$, $r = 3$ and $n_{nnz}=20$. Results are averaged over 100 simulations.
}
    \label{fig:results_sparse}
\end{figure}

{We use the synthetic data to compare the performance of our sparse $\text{CCAR}^3$ method with  six alternative approaches of CCA with sparsity. These include a set of ``heuristic-based'' methods: SAR \citep{wilms2015sparse}, CCA-Lasso \citep{witten2009extensions}, and the SCCA approach of \cite{parkhomenko2009sparse}. We also compare our method with the ``theory-grounded'' approaches: Fantope CCA (the initialization step of the algorithm from \cite{gao2017sparse} and \cite{gao2021sparse}), SCCA with CoLaR (the full algorithm from \cite{gao2017sparse}), and SGCA from \cite{gao2021sparse}}. 
We compare all methods to an ``oracle" solution, which is simply the estimated CCA directions (using a standard low-dimensional CCA solver) on the true support of $\bU$. In our method, we select an optimal regularization parameter $\rho$ through 5-fold cross validation. To robustify the estimates of the canonical variates, we further shrink $\wh \bS_Y$ using its Ledoit-Wolf estimate \citep{ledoit2022power}. {Similarly, we select the regularization parameters in all heuristic-based approaches using either 5-fold cross validation, or the permutation approach proposed by \cite{witten2009extensions} for their method (``permuted'' in Figure~\ref{fig:results_sparse}). For the Fantope initialization of the methods in \cite{gao2017sparse} and \cite{gao2021sparse}, we used the theoretical values for $\rho$ as proposed in the respective papers. However, in the second stage of the CCA with CoLaR, which involves solving a group-lasso problem, we applied cross-validation.}  
To evaluate the resulting CCA reconstructions, we measure the distances between subspaces spanning canonical variates, which can be done by computing the principal angle \citep{bjork1973} {between the subspaces associated with the matrices $ \begin{psmallmatrix}
    \widehat \bU\\
    \widehat \bV
\end{psmallmatrix}$ and $\begin{psmallmatrix}
    \bU\\
    \bV
\end{psmallmatrix}.$}

We examine the dependence of the subspace distance between the estimated and true canonical subspaces on the number of covariates \(p\), the number of canonical components \(r\), and the sample size \(n\) (Figure~\ref{fig:results_sparse}). Results are averaged over 100 simulation replications.

The top panel of Figure~\ref{fig:results_sparse} investigates the effect of increasing dimensionality \(p\). Across the medium- and high-correlation settings, sparse CCAR$^3$ consistently achieves the smallest subspace distance among the competing methods. The advantage becomes increasingly pronounced as \(p\) grows, demonstrating the ability of sparse CCAR$^3$ to accurately recover the canonical subspace in high-dimensional regimes. The only method exhibiting comparable performance is SAR~\citep{wilms2015sparse}, particularly when \(p\) is small and the problem remains effectively low-dimensional. However, the performance gap widens substantially as \(p\) increases. In the low-correlation setting, sparse CCAR$^3$ remains competitive and typically outperforms the remaining benchmarks, with only a few isolated cases in which Fantope CCA attains slightly lower subspace distance.

The middle panel of Figure~\ref{fig:results_sparse} examines the effect of the number of canonical components \(r\) in a representative high-dimensional setting (\(p=1000\), \(q=30\)). Sparse CCAR$^3$ exhibits remarkable stability in high and medium correlation settings as \(r\) increases, maintaining nearly constant subspace distance across all values of \(r\). By contrast, most other baselines' performances deteriorate  with increasing rank. If SAR achieves similar performance for small \(r\), sparse CCAR$^3$ generally provides the most accurate recovery of the canonical subspace over the full range of ranks considered. In low-correlation signals, however, the performance is very similar across  competitors.

The bottom-left panel studies the effect of sample size \(n\). As expected, the subspace distance decreases for all methods as more observations become available. Sparse CCAR$^3$ displays one of the fastest convergence rates. SAR exhibits a similar rate of improvement, whereas the remaining methods improve more gradually and remain substantially farther from the CCAR$^3$ solution even for large sample sizes.

 {
To further assess practical applicability, the bottom-right panel of Figure~\ref{fig:results_sparse} compares computational costs. To provide an idea of the applicability of our method in real-life settings, we used parallelization across 5 workers to perform the cross-validation methods.  We used the efficient code for CCA-Lasso \citep{witten2009extensions} in the \texttt{PMA} package. However, the SAR method by \cite{wilms2015sparse} was not parallelized, as the code for this was not optimized by the authors. 
}

 {
We observe that sparse CCAR$^3$ is slower than the heuristic methods CCA-Lasso and SCCA when $n$ is moderate. However, these methods are consistently less accurate than sparse CCAR$^3$ across all simulation settings considered above. Furthermore, while the computational cost of competing heuristic methods appears to increase exponentially with the sample size $n$, the running time of sparse CCAR$^3$ is largely insensitive to $n$ and, in fact, decreases slightly for larger sample sizes as the ADMM algorithm converges more rapidly. Consequently, for large sample sizes, the computational cost of sparse CCAR$^3$ becomes comparable to CCA-Lasso and SCCA.
}

 {
Among all competing methods, SAR is the closest competitor in terms of accuracy. Nevertheless, its computational cost is substantially higher. For example, when $n=10{,}000$, SAR requires approximately 4,000 seconds, whereas sparse CCAR$^3$ requires only 30 seconds. Even if SAR were fully parallelized, our method would remain approximately 25 times faster. We also note that fitting sparse CCAR$^3$ on a single cross-validation fold is considerably less expensive than solving the optimization problem underlying Fantope CCA: sparse CCAR$^3$ achieves roughly a 100-fold reduction in computational time relative to Fantope CCA, despite the fact that Fantope CCA does not require cross-validation and is evaluated using the theoretically optimal tuning parameter. These results demonstrate that sparse CCAR$^3$ effectively balances statistical and computational efficiency, combining state-of-the-art estimation accuracy with computational scalability.
}

\subsection{CCA with structured penalties}
\label{sec:group-sparse-cca}

The row-sparse formulation from Section~\ref{sec:sparse-cca} can be extended when prior structural information on the covariates of $\bX$ is available. We consider the estimator
\begin{align}
    \widehat{\bB}
    = \operatorname*{argmin}_{\bB \in \R^{p\times q}}
    \frac{1}{n}\|\bY_0-\bX\bB\|_F^2 + \rho\,\operatorname{Pen}(\bB).
    \label{eq:ols:pen:group}
\end{align}
Two choices of penalty are particularly useful.

\paragraph{Group penalty.}
Suppose that $\bX$ covariates are partitioned into known non-overlapping groups
\[
\mathcal G=\{g_1,\ldots,g_M\}, 
\qquad 
\bigcup_{m=1}^M g_m=[p], 
\qquad 
g_m\cap g_{m'}=\varnothing \ \text{for } m\neq m'.
\]
Writing $\bB_g$ for the block of $\bB$ corresponding to the rows in $g$, we take
\[
\operatorname{Pen}(\bB)
=
\sum_{g\in\mathcal G}\sqrt{|g|}\,\|\bB_g\|_F.
\]
This encourages the joint selection of entire groups of covariates. In neuroimaging, for example, the groups may correspond to regions defined by a reference atlas. The row-sparse penalty is recovered as the special case $\mathcal G=\{\{1\},\ldots,\{p\}\}$.

\paragraph{Graph penalty.}
Suppose instead that the covariates lie on a known graph with vertex set $[p]$ and edge set $E$. Let $\bGamma\in\R^{m\times p}$ be the associated edge-incidence matrix, where $m=|E|$. For an edge $(k,k')\in E$, the corresponding row of $\bGamma$ has entries $1$ and $-1$ in columns $k$ and $k'$, and zeros elsewhere. We then take
\begin{equation}\label{eq:graph_penalty}
    \operatorname{Pen}(\bB)=\|\bGamma\bB\|_{21}.
\end{equation}
This is an instance of the multivariate total-variation penalty \citep{hutter2016optimal}. It encourages neighboring covariates to have similar coefficients while still allowing a small number of jumps across edges, i.e. graph sparsity \citep{tran2022generalized}. This is again natural in neuroimaging, where nearby voxels are expected to exhibit similar activation patterns.

Both versions of \eqref{eq:ols:pen:group} can be solved efficiently using the same ADMM framework as in the sparse case. The updates for the group and graph-penalized version are described in Appendix~\ref{algo:spr3-cca}.

\paragraph{Theoretical results.}
For the group penalty, let
$m_u=\big|\{g\in\mathcal G:\|\bU_g\|_F\neq 0\}\big|$
denote the number of active groups, and let
$s_u=\sum_{g:\|\bU_g\|_F\neq 0}|g|$ be their total size. The proof of Theorem~\ref{theorem:rrr_ols} extends directly to this setting with the row-sparse complexity $s_u(q+\log p)$ replaced by the group complexity $s_u q + m_u\log |\mathcal G|$, up to constants and group-size factors when the groups are balanced. Consequently, the corresponding direction-subspace error scales as
\[
\frac{\sqrt r}{\lambda}\sqrt{\frac{s_u q + m_u\log |\mathcal G|}{n}}
\]
up to powers of the conditioning constant $M$.

For the graph penalty, the relevant structural assumption is that $\bGamma\bU$ is row-sparse. Let $s_u=|\supp(\bGamma\bU)|$, let $n_c$ be the number of connected components, let $\bL=\bGamma^\top\bGamma$ be the graph Laplacian, and let $\kappa_2$ denote the smallest nonzero eigenvalue of $\bL$. The graph direction bounds depend on the rate
\[
\delta_G=
\sqrt{n_c\frac{q+\log n_c}{n}}
+
\left(1+\frac{1}{\kappa_2}\right)
\sqrt{\sigma_{\max}(\bL)s_u\frac{q+\log m}{n}}.
\]
The dependence on $n_c$ reflects the fact that $\|\bGamma\bU\|_{21}$ does not penalize componentwise constants, so each connected component contributes one unpenalized degree of freedom. The dependence on the smallest nonzero Laplacian eigenvalue $\kappa_2$ reflects connectivity: smaller $\kappa_2$ corresponds to a more weakly connected graph \citep{chung1996lectures}, and therefore to weaker regularization for a fixed total-variation budget. The rigorous theorem statements, along with their proofs, are provided in Appendices~\ref{appendix:proof_sparse_group} and~\ref{appendix:proof_graph}.

\medskip

\xhdr{Simulation experiments} 
We adopt a data generation mechanism similar to that of the previous subsection, modifying only the generation of $\bU$. Specifically, we generate $\bU$ either according to a known group-sparse structure or according to a graph-sparse structure in $\bX$. Additional details on the simulation setup are provided in Appendix~\ref{appendix:plots:simulation}.

The results for the data with group and graph structure are presented in Figures \ref{fig:simu:group:sd} and \ref{fig:simu:graph:sd} of the Appendix~\ref{appendix:plots:simulation}. We observe that both modifications perform well in terms of the resulting subspace distances between the estimated canonical directions and their population counterpart. SAR is once again the best competitor, achieving better results compared to our methods in settings where $p$ is small ($p\ll n$) and the signal is high. However, as soon as $p$ increases, our methods outperform all competitors. These results are consistent across different values of $q$.

\section{Real-World Experiments}
\label{sec:exp}

In this section, we underscore the advantages of our approach by conducting a comparative analysis of our method against established methods on three real-world datasets.\footnote{ {Our method is available on CRAN under the package \hyperlink{https://cran.r-project.org/web/packages/ccar3/index.html}{ccar3}. The code for the simulations and real-world examples can be found on \hyperlink{https://github.com/donnate/CCAR3code}{https://github.com/donnate/CCAR3code}.}}  

\subsection{The Nutrimouse data}
\label{sec:nutrimouse}

The Nutrimouse dataset is a common benchmark in multivariate analysis \citep{gonzalez2012, rodosthenous2020} that contains genomic and nutritional data from a 2007 study on 40 mice \citep{martin2007novel}. This dataset includes the expression profiles of 120 pre-selected genes in the liver, as well as measurements of 21 hepatic fatty acid concentrations (expressed as percentages) observed under various dietary treatments. We denote the genomic and fatty acid measurements by $\bX$ and $\bY$, respectively.
Each of the 40 mice was assigned to one of five diets, and its genotype was classified as either wildtype or genetically modified (PPAR). We set the number of canonical directions to $r=5$ upon examining the scree plot of the singular values of ~$\widehat \bS_{XY}$, and compare our method's performance to the benchmarks introduced in the previous section.

To evaluate the performance of the methods, we divide the data into eight folds (5 mice per fold) at random, and use seven folds for training, and keep one for testing.
For {sparse CCAR$^3$}, the optimal $\rho$ is chosen using cross-validation on the training data, based on the mean squared error (MSE) between the canonical variates $\bX\widehat \bU$ and $\bY\widehat \bV$ as per equation (\ref{eq:CCA_pred}).  For the other methods, we use their default implementations to determine the appropriate amount of regularization (e.g. using cross-validation, BIC or a permutation-based analysis, depending on the methods).
Performance across methods is measured on the test folds using the  MSE between $\bX\widehat \bU$ and $\bY\widehat \bV$, and the average correlation between directions.  

\medskip

\xhdr{Results} The results, averaged over 10 random 8-fold splits of the data, are presented in Table~\ref{tab:nutrimouse}. Sparse CCAR$^3$ achieves the lowest test MSE and the highest test correlation among all competing methods. In particular, it reduces the average test MSE by approximately 36\% relative to SAR~\citep{wilms2015sparse}, 38--41\% relative to the CCA-Lasso methods of \cite{witten2009extensions}, and 39\% relative to SCCA with CoLaR~\citep{gao2021sparse}. The improvement is also reflected in the substantially higher test correlation, with sparse CCAR$^3$ achieving a correlation of \(0.62\), compared to at most \(0.53\) for the competing methods.

Sparse CCAR$^3$ is also substantially more computationally efficient than existing theoretically grounded sparse CCA methods. For instance, fitting the full sparse CCAR$^3$ model requires only 35 seconds, including 8-fold cross-validation over 30 candidate values of $\rho$, compared to 777 seconds for Fantope CCA with no cross-validation. Sparse CCAR$^3$ thus achieves better predictive performance while requiring only a small fraction of the computational effort of existing theory-based approaches.


\begin{table}[h!]
    \centering
\begin{scriptsize}
\resizebox{\textwidth}{!}{
\begin{tabular}{|c||c|c|c|c|c|}
\hline
         & Test & Test & Diet Label & Genotype Label / Cluster & Running  \\
      & MSE & Correlation & Accuracy (LDA) & Accuracy (LDA / cluster) & Time (s) \\ \hline
\begin{tabular}[c]{@{}c@{}}\textit{sparse CCAR$^3$} \\ (this paper)\end{tabular} & \textbf{0.57 $\pm$ 0.10} & \textbf{0.62 $\pm$ 0.04} & \textbf{0.9} & 1 / 0.575 & 34.63 \\ \hline
\begin{tabular}[c]{@{}c@{}}\textit{SAR} \\ (Wilms et al.)\end{tabular} & 0.89 $\pm$ 0.11 & 0.53 $\pm$ 0.07 & 0.6 & 1 / 0.6 & 6.02 \\ \hline
\begin{tabular}[c]{@{}c@{}}\textit{CCA-Lasso} \\ with CV (Witten et al.)\end{tabular} & 0.96 $\pm$ 0.12 & 0.41 $\pm$ 0.08 & 0.7 & 1 / 0.575 & 2.86 \\ \hline
\begin{tabular}[c]{@{}c@{}}\textit{CCA-Lasso} \\ permuted (Witten et al.)\end{tabular} & 0.92 $\pm$ 0.09 & 0.43 $\pm$ 0.06 & 0.7 & 1 / 0.575 & 2.44 \\ \hline
\begin{tabular}[c]{@{}c@{}}\textit{SCCA} \\ (Parkhomenko et al.)\end{tabular} & 0.86 $\pm$ 0.12 & 0.48 $\pm$ 0.08 & 0.8 & 1 / 0.625 & 0.23 \\ \hline
\begin{tabular}[c]{@{}c@{}}\textit{Fantope CCA} \\ (Gao et al.)\end{tabular} & 2.37 $\pm$ 2.15 & 0.41 $\pm$ 0.07 & 0.7 & 1 / 0.5 & 777 \\ \hline
\begin{tabular}[c]{@{}c@{}}\textit{SCCA with CoLaR} \\ (Gao et al.)\end{tabular} & 0.94 $\pm$ 0.14 & 0.49 $\pm$ 0.04 & 0.7 & 1 / 0.7 & 801 \\ \hline
\begin{tabular}[c]{@{}c@{}}\textit{SGCA} \\ (Gao \& Ma)\end{tabular} & 1.98 $\pm$ 1.18 & 0.38 $\pm$ 0.07 & 0.7 & 1 / 0.675 & 1,356 \\ \hline
\end{tabular}
}
\end{scriptsize}
    \caption{Results of the 8-fold cross-validation on the Nutrimouse dataset. Reported values correspond to the average test MSE and correlation over 10 random splits of the data in 8 folds. The ability of CCA methods to yield directions that can separate genotypes is evaluated through both LDA, and, to allow more discrimination between methods, Gaussian clustering. }
 \label{tab:nutrimouse}
\end{table}

\begin{figure}[p]
\centering
\begin{subfigure}{0.7\linewidth}
\includegraphics[trim={0 8.5cm 0 0}, clip, width=\textwidth]{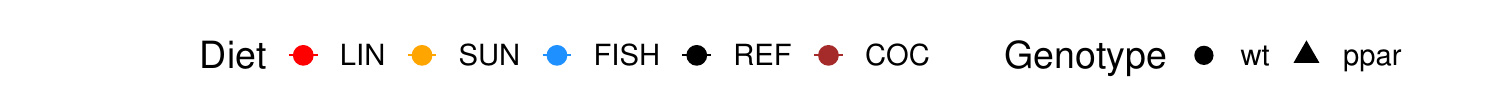}
	\end{subfigure}
 \vfill
 \begin{subfigure}{0.9\linewidth}
 \includegraphics[width=0.3\textwidth]{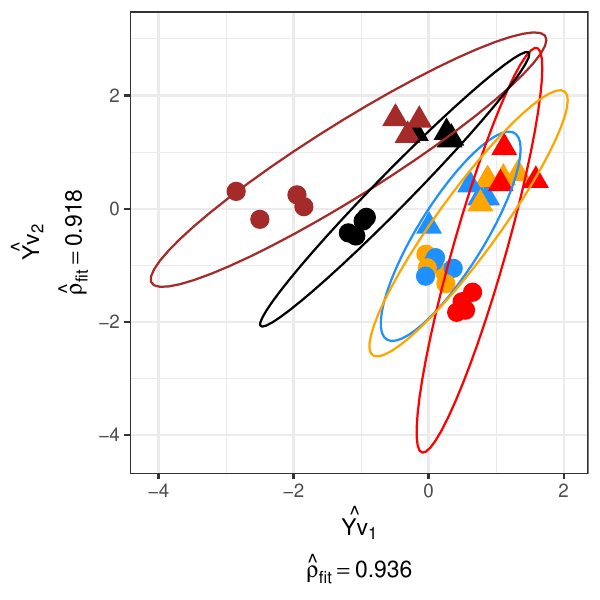}
\includegraphics[width=0.3\textwidth]{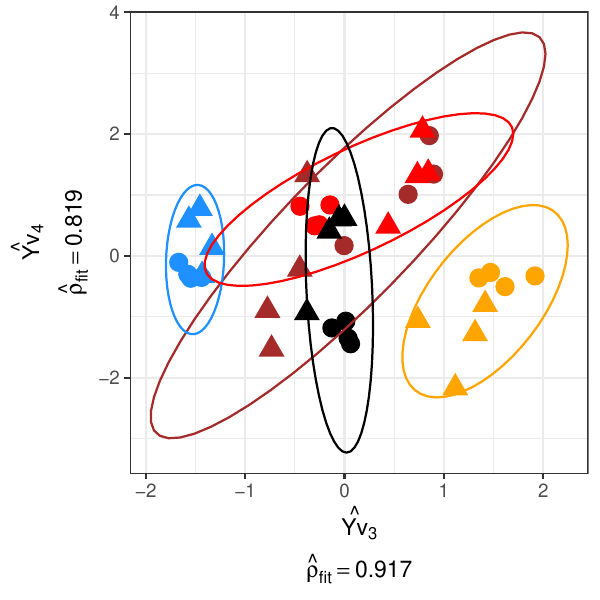}
\includegraphics[width=0.3\textwidth]{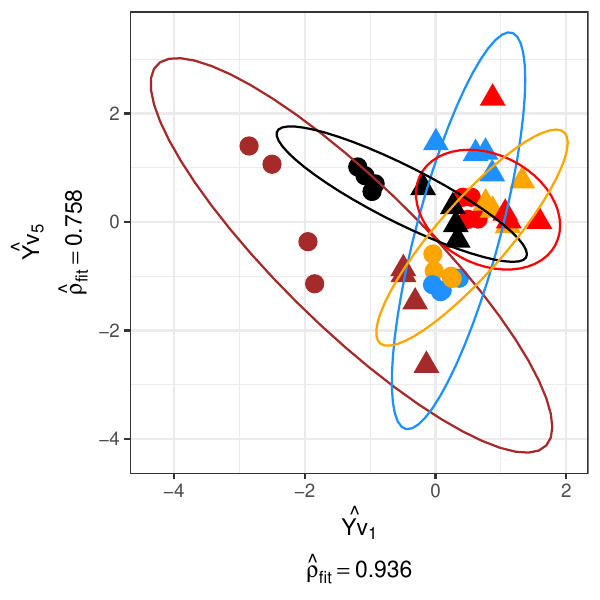}
		\caption{sparse CCAR$^3$ (our method)}
		\label{fig:nutriXU_ours}
	\end{subfigure}
\vfill
\begin{subfigure}{0.9\linewidth}
\includegraphics[width=0.3\textwidth]{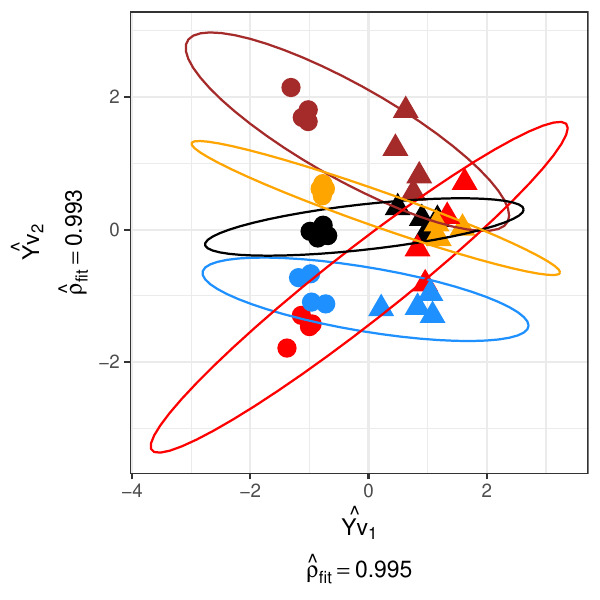}
\includegraphics[width=0.3\textwidth]{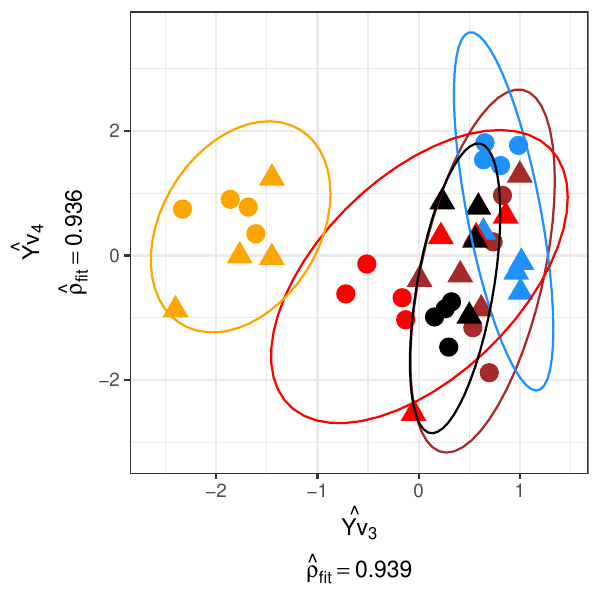}
\includegraphics[width=0.3\textwidth]{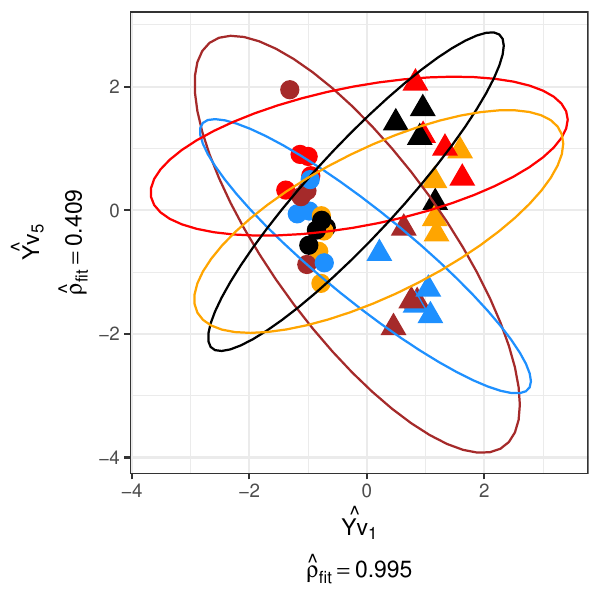}
\caption{SAR \citep{wilms2015sparse}}
		\label{fig:nutriXU_fitsar}
	\end{subfigure}
\vfill\begin{subfigure}{0.9\linewidth}
 \includegraphics[width=0.3\textwidth]{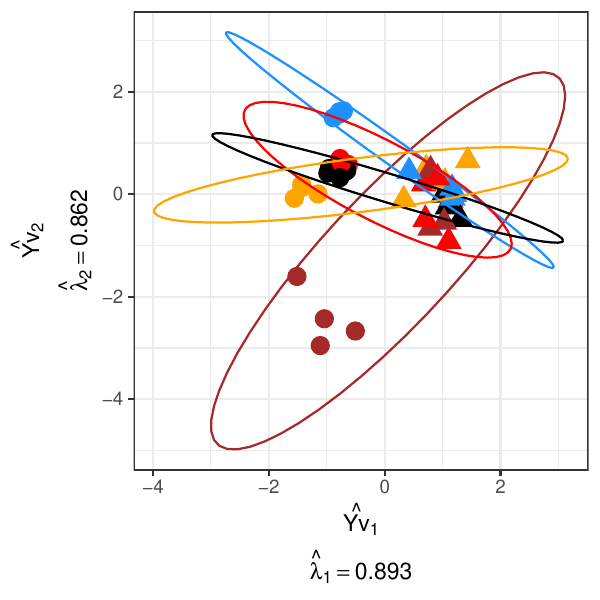}
\includegraphics[width=0.3\textwidth]{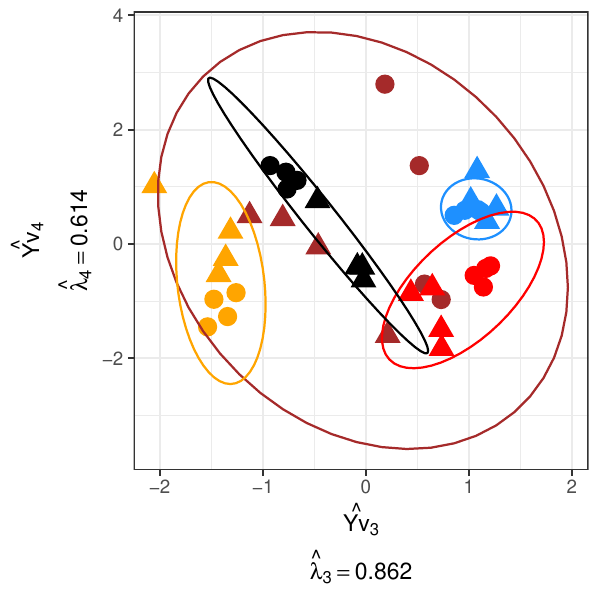}
\includegraphics[width=0.3\textwidth]{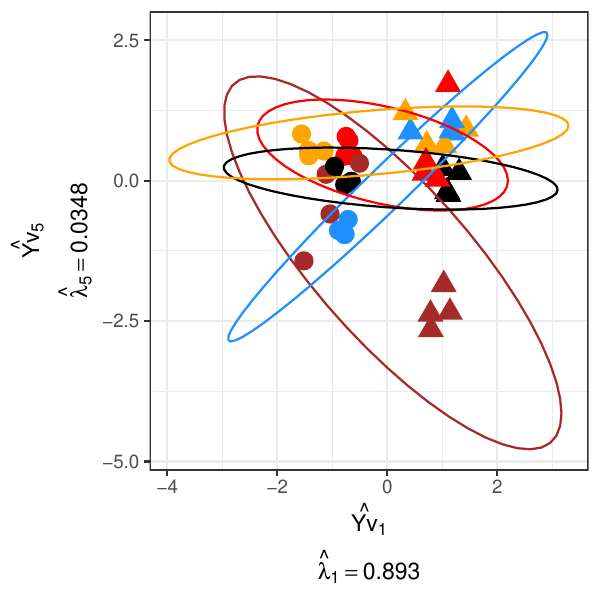}
		\caption{SCCA with CoLaR \citep{gao2017sparse}}
		\label{fig:nutriXU_colar}
	\end{subfigure}
 \vfill
\caption{Scatter plots for canonical variates produced for the Nutrimouse dataset: $1^{st}$ vs  $2^{nd}$ (left panel), $3^{rd}$ vs $4^{th}$ (middle panel), and $1^{st}$ vs  $5^{th}$ (right panel). Sample points are colored by diet, different shapes correspond to the genotype. The plots include 95\% confidence ellipses to emphasize diet clustering.}
	\label{fig:nutriXU}
\end{figure}

We further evaluate the quality of the estimated canonical directions by examining whether the resulting canonical variates separate the different diet groups and genotypes. The canonical variates obtained by SAR \citep{wilms2015sparse}, SCCA with CoLaR \citep{gao2017sparse}, and sparse CCAR$^3$ are displayed in Figure~\ref{fig:nutriXU}. All three methods successfully separate the two genotypes using the first two canonical directions (Figure~\ref{fig:nutriXU}, left panels). However, differences emerge when considering the dietary groups. While SAR identifies the COC and SUN diets as distinct clusters (Figure~\ref{fig:nutriXU_fitsar}), both SCCA with CoLaR and sparse CCAR$^3$ achieve substantially better separation among the dietary groups. In particular, sparse CCAR$^3$ provides the clearest separation of the COC cluster, which appears less distinct under SCCA with CoLaR (Figures~\ref{fig:nutriXU_ours} and \ref{fig:nutriXU_colar}).

To quantify these observations, we classify samples into the five diet groups using the estimated canonical variates \((\bX\widehat{u}_1,\ldots,\bX\widehat{u}_5)\). Classification is performed using linear discriminant analysis (LDA), with \(3/4\) of the data used for training and \(1/4\) for testing, and the results averaged across the four folds. As reported in Table~\ref{tab:nutrimouse}, sparse CCAR$^3$ achieves the highest classification accuracy, improving diet classification accuracy by 50\% relative to SAR and by 28\% relative to SCCA with CoLaR. 


Turning to the loadings, we investigate whether the canonical directions recovered by sparse CCAR$^3$ are consistent with the biological findings of \citet{martin2007novel} (see Figure~\ref{fig:nutrimouse_UV_rrr} in Appendix~\ref{appendix:plots:nutrimouse}). For loadings obtained by competing methods, we refer the interested reader to the online supplement\footnote{\url{https://github.com/donnate/CCAR3code/blob/main/additional_plots}.}

The first canonical variate separates mice fed the coconut-oil diet from those fed the fish- and linseed-oil diets (Figure~\ref{fig:nutriXU_ours}, left panel). The dominant loadings involve fatty acids C20:3n6, C20:3n9, C16:0, C16:1n9, and C18:2n6 together with genes \textit{CAR1}, \textit{CYP3A11}, \textit{PLTP}, \textit{AOX}, and \textit{CYP2c29}. Notably, \textit{AOX}, \textit{CYP3A11}, \textit{CYP2c29}, and \textit{PLTP} were among the variables discussed by \citet{martin2007novel} as contributing to differences across dietary interventions. The recovery of these variables suggests that the first canonical direction captures the primary diet-related variation in the data.

The second canonical variate separates wild-type and PPAR$\alpha$-deficient mice and is characterized by large loadings on long-chain polyunsaturated fatty acids, including C22:5n6, C22:5n3, and C20:5n3 (EPA), together with genes \textit{ACAT2}, \textit{Lpin}, \textit{PLTP}, \textit{CYP3A11}, and \textit{CYP2c29}. Several of these variables, including \textit{Lpin}, \textit{PLTP}, \textit{CYP3A11}, and \textit{CYP2c29}, were highlighted by \citet{martin2007novel} as being associated with PPAR$\alpha$ activity. Likewise, EPA and related long-chain polyunsaturated fatty acids were among the lipid species exhibiting substantial variation across genotypes in the original study. The recovery of these variables suggests that the second canonical direction captures the primary genotype-related variation in the data.

The third canonical variate separates the reference diet from the fish- and sunflower-oil diets, indicating an additional dietary effect beyond the contrast represented by the first canonical variate. The corresponding loadings involve DHA (C22:6n3), C22:4n6, C22:5n6, and C20:3n9 together with genes \textit{CYP2c29}, \textit{ACAT2}, \textit{PLTP}, and \textit{CYP3A11}. Notably, both DHA and the C22 polyunsaturated fatty acids featured prominently in the lipid profiles analyzed by \citet{martin2007novel}, while \textit{PLTP}, \textit{CYP3A11}, and \textit{CYP2c29} were among the genes reported to vary across dietary and genotype conditions. The recovery of these variables suggests that the third canonical direction captures an additional dietary signal present in the data. Overall, the recovered canonical directions closely align with the two principal sources of variation identified by \citet{martin2007novel}: dietary fatty-acid composition and PPAR$\alpha$ activity.

\subsection{Neuroscience data: grouping brain regions with group-sparse CCAR$^3$}

In this section, we re-investigate the dataset from \cite{tozzi2021relating}. The matrix $\bX$ consists of the fMRI brain activations from 229 distinct brain regions of interest (ROI) for 153 individuals in response to monetary gains versus losses during a gambling task. These regions include the 210 cortical areas as identified by the Brainetome atlas and the 19 subcortical areas of the Desikan-Kellamy atlas. The matrix $\bY$ contains self-reported measures on nine variables: drive, fun seeking, reward responsiveness, and overall behavioral inhibition (evaluated via the Behavioral Approach System/Behavioral Inhibition Scale), as well as distress, anhedonia, and anxious arousal (as defined through the Mood and Anxiety Symptom Questionnaire), and positive and negative emotional states (derived from the Positive and Negative Affect Schedule). Both datasets $\bX$ and $\bY$ were adjusted for sex differences and standardized across each column. We refer the reader to the original papers by \cite{tozzi2021relating,tuzhilina2021canonical} for more details on the data collection and preprocessing of these datasets.

We propose a joint analysis of the brain activation and questionnaire datasets by calculating the first two canonical directions based on an initial assessment of the singular values from the cross-covariance matrix. We evaluate sparse and group-sparse penalty types for CCAR$^3$ discussed in Section~\ref{sec:high_d}. 
For the group-sparse method, we assign each of the 229 ROIs to 29 groups of brain regions (which we further split into two, depending on whether they are in the right or left hemisphere) corresponding to the main gyri or subcortical regions, as well as the brain stem. 
We employ the same cross-validation approach as described in the previous section (using 20 folds here). Table~\ref{tab:neuro_agg} summarizes the performance of our two CCAR$^3$ variants on the validation folds, along with that of existing benchmarks. Overall, we observe that group CCAR$^3$ outperforms all other benchmarks, attaining the lowest MSE and highest correlation. We also note that our group modification reduces by 8.4\% the MSE of the sparse CCAR$^3$, the second best performing method.


\begin{table}[h]
    \centering
\begin{scriptsize}
\begin{tabular}{|c||c|c|}
 \hline
 & Test MSE & Test Correlation\\
 \hline \hline
\textit{sparse CCAR$^3$} (this paper) & 1.67 $\pm$ 0.10 & 0.05 $\pm$ 0.03\\
\hline
\textit{group CCAR$^3$} (59 L/R groups; this paper) & \textbf{1.53 $\pm$ 0.03} & \textbf{0.11 $\pm$ 0.01}\\
\hline
\textit{SAR} (Wilms et al.) & 2.03 $\pm$ 0.08 & 0.05 $\pm$ 0.05\\
\hline
\textit{CCA-Lasso} (Witten et al.) & 2.02 $\pm$ 0.13 & -0.01 $\pm$ 0.06\\
\hline
\textit{SCCA} (Parkhomenko et al.) & 1.75 $\pm$ 0.07 & 0.05 $\pm$ 0.02\\
\hline
\textit{Fantope CCA} (Gao et al.) & 2.15 $\pm$ 0.10 & 0.02 $\pm$ 0.04\\
\hline
\textit{SCCA with CoLaR} (Gao et al.) & 1.90 $\pm$ 0.13 & 0.02 $\pm$ 0.03\\
\hline
\textit{SGCA} (Gao \& Ma) & 1.81 $\pm$ 0.03 & 0.05 $\pm$ 0.01\\
\hline
\end{tabular}
\end{scriptsize}
    \caption{Results of the 20-fold cross-validation on the Neuroscience dataset with rank $r=2$. Values correspond to the test MSE over the test folds, averaged over  five random splits of the data into 20 folds, $\pm$ standard deviation.}
    \label{tab:neuro_agg}
\end{table}

\begin{figure}[p]
\centering
\begin{subfigure}{0.63\linewidth}
\includegraphics[width=\textwidth]{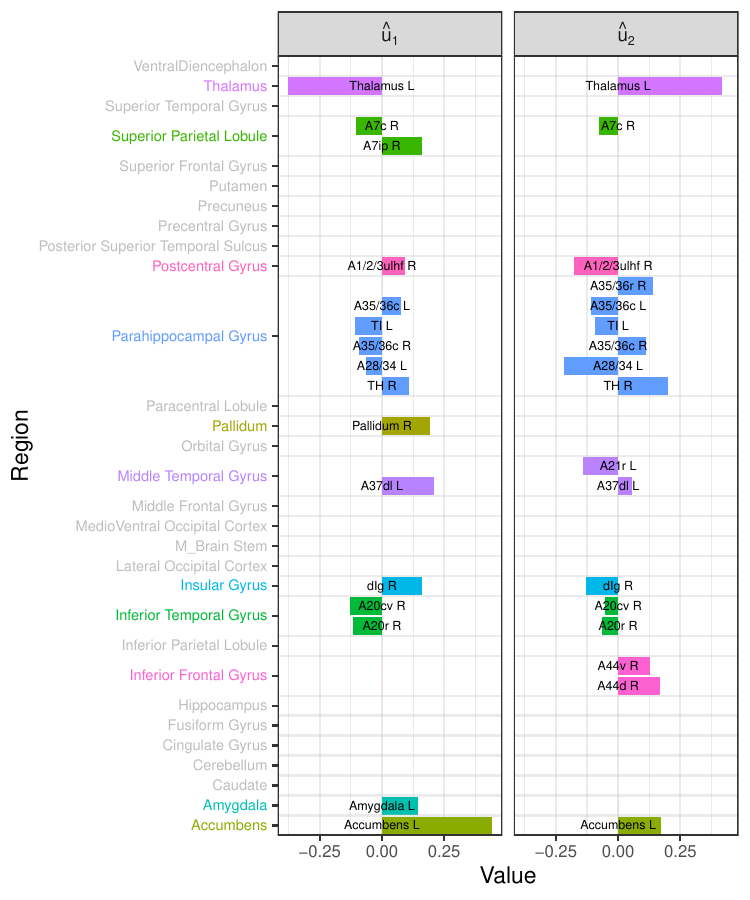}
\end{subfigure}
\begin{subfigure}{0.36\linewidth}
\includegraphics[width=\textwidth]{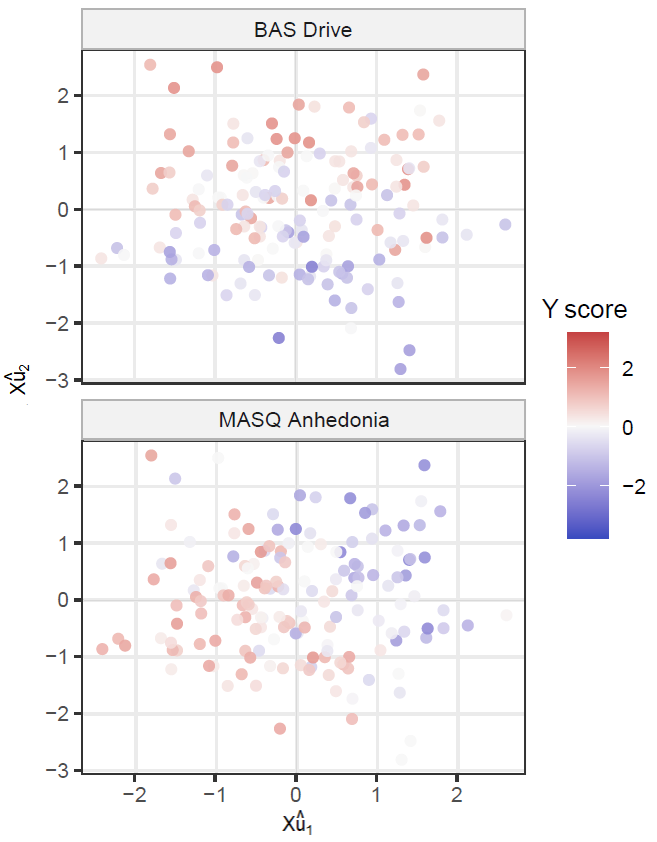}
\end{subfigure}
\includegraphics[width=0.55\textwidth]{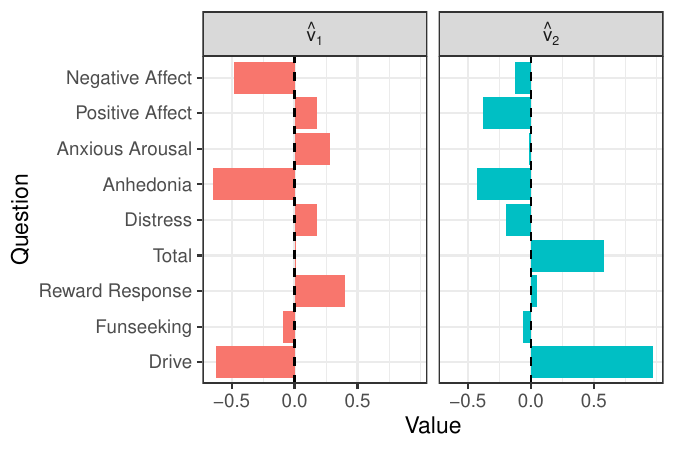}

\caption{Results on the Neuroscience dataset using group sparse CCAR$^3$. Bar plots: canonical directions $(\widehat{{v}}_i)_{i=1}^2$ representing the questionnaire and top 20 brain activation directions $(\widehat{{u}}_i)_{i=1}^2$ grouped along the y-axis and colored according to the brain gyri. Scatter plots: canonical variates $(\bX\widehat{{u}}_i)_{i=1}^2$, colored by drive and anhedonia scores. }
\label{fig:neuro:group}
\end{figure}

In Figure~\ref{fig:neuro:group}, we present the estimated canonical variates and directions obtained using group CCAR$^3$. Corresponding results for sparse CCAR$^3$, SCCA \citep{parkhomenko2009sparse}, and SGCA \citep{gao2021sparse} are provided in the online supplement.\footnote{\url{https://github.com/donnate/CCAR3code/tree/main/additional_plots}.} As expected, the group-sparse formulation yields a more parsimonious solution for $(\widehat{u}_i)_{i=1}^2$, selecting only 11 of the 30 brain regions, compared with a larger number selected by the sparse method.

The bottom bar plot of Figure~\ref{fig:neuro:group} shows that the first canonical direction identified by group CCAR$^3$ is significantly associated with anhedonia, whereas drive contributes strongly to both the first and second canonical directions. These behavioral dimensions induce a clear structure in the canonical variate space $\bX\widehat{\bU}$, where subjects with different levels of anhedonia and drive occupy distinct regions (see the scatter plot of Figure~\ref{fig:neuro:group}).

To assess the extent to which the estimated canonical variates capture variation in individual questionnaire measures, we regress each of the nine behavioral outcomes in $\bY$ on $\bX\widehat{\bU}$. The adjusted $R^2$ values are reported in Table~\ref{tab:corr_Y}. Group CCAR$^3$ achieves the strongest associations with anhedonia and drive, indicating that these variables are the dominant behavioral correlates of the brain connectivity patterns extracted by the method.

\begin{table}[t]
\resizebox{\textwidth}{!}{
\begin{tabular}{|c||c|c|c|c|c|c|c|c|c|}
\hline
 & Drive & Funseeking & Reward Resp. & Total & Distress & Anhedonia & Anx. Arousal & Pos. Affect & Neg. Affect \\ \hline \hline
\textit{sparse CCA$\text{R}^3$} & 0.10 & \textbf{0.41} & 0.05 & 0.14 & 0.11 & 0.15 & 0.06 & 0.15 & 0.12 \\
\hline
\textit{group CCA$\text{R}^3$} & \textbf{0.38} & 0.08 & \textbf{0.28} & 0.04 & 0.22 & \textbf{0.46} & 0.01 & 0.29 & 0.22 \\
\hline
\textit{SAR} & 0.10 & 0.28 & 0.00 & 0.04 & 0.00 & 0.02 & 0.04 & -0.01 & 0.03 \\
\hline
\textit{CCA-Lasso} & 0.22 & 0.07 & 0.10 & 0.10 & 0.12 & 0.13 & 0.10 & 0.12 & 0.16 \\
\hline
\textit{SCCA} & 0.11 & 0.07 & 0.11 & 0.12 & 0.23 & 0.38 & 0.08 & 0.28 & 0.14 \\
\hline
\textit{Fantope CCA} & -0.01 & 0.00 & -0.01 & -0.01 & -0.01 & -0.01 & -0.01 & -0.01 & -0.01 \\
\hline
\textit{SGCA} & 0.07 & 0.07 & 0.06 & \textbf{0.36} & \textbf{0.52} & 0.45 & \textbf{0.35} & \textbf{0.32} & \textbf{0.51} \\
\hline
\end{tabular}
}
\caption{Adjusted $R^2$ for the regression of $\bX\widehat \bU$ onto each of the columns of $\bY$.}
\label{tab:corr_Y}
\end{table}

From a neuroscientific perspective, group CCAR$^3$ identifies the thalamus as an important contributor to the canonical directions. The thalamus plays a central role in arousal, attention, and consciousness, making its involvement plausible in the context of a reward-based gambling task. In addition, group CCAR$^3$ assigns large weight to the nucleus accumbens, a key component of the brain's reward circuitry that has been strongly linked to reward anticipation and processing \citep{day2007}.
Interestingly, the Parahippocampal Gyrus is also highlighted by our method. While the Parahippocampal Gyrus is primarily known for its role in memory encoding and contextual associative processing, it has also been implicated in the integration of contextual information during value-based learning and decision making. This suggests that contextual and memory-related processes may contribute to the neural mechanisms underlying reward-related behavior in this task.

\subsection{Climate data: exploring spatial structure using the graph-sparse model}

 {In this section, we illustrate the proposed graph CCAR$^3$ method using a climate dataset. The predictor variables $\bX$ consist of sea surface temperature (SST) measurements from the Extended Reconstructed Sea Surface Temperature dataset, Version~5 (ERSSTv5), produced by the National Oceanic and Atmospheric Administration (NOAA). The dataset provides monthly SST observations on a $2^\circ \times 2^\circ$ latitude--longitude grid covering the global ocean, with records available since January 1854. We focus on the equatorial Pacific region, selecting grid points with longitudes between $120^\circ$E and $280^\circ$E and latitudes between $20^\circ$S and $20^\circ$N. To reduce spatial resolution, the grid is coarsened to $4^\circ$, resulting in a time-by-space matrix $\bX$ where each row corresponds to a monthly observation and each column to a spatial grid location. The seasonal cycle is removed prior to analysis.}

 {The response variables $\bY$ comprise several large-scale climate indices obtained from NOAA datasets, including the Arctic Oscillation (AO), North Atlantic Oscillation (NAO), North Pacific index (NP), outgoing longwave radiation (OLR), Pacific--North American pattern (PNA), and the Southern Oscillation Index (SOI). These indices summarize major modes of atmospheric variability and are widely used to characterize large-scale climate dynamics.}

 {After removing missing values, the dataset contains $n = 589$ observations, $p = 441$ predictors, and $q = 6$ response variables. We construct train--test--validation splits using consecutive time points: 250 for training, 100 for validation, and 100 for testing. To reduce temporal dependence between the sets, a gap of 25 time points is introduced between consecutive segments. By shifting the starting time point forward by 25 at each iteration, we obtain four distinct train--test--validation splits.}

 {We fit sparse CCA methods with $r = 3$ components. The graph CCAR$^3$ method uses the spatial locations of features in $\bX$ to construct a graph by connecting neighboring grid points within a $4^\circ$ distance. After selecting the optimal penalty parameter for each method using the test set, we evaluate their performance on the test set.}

 {Sparse CCAR$^3$ achieves a validation correlation of $0.383$ with an MSE of $1.09$. In contrast, graph CCAR$^3$ performs better, attaining a substantially higher test correlation of $0.474$ and a lower MSE of $0.946$. By comparison, SAR achieves a correlation of $0.370$ with an MSE of $1.27$, and CCA-Lasso attains a correlation of $0.420$ with an MSE of $1.08$, both underperforming relative to the graph-based approach. Notably, CCA-Lasso fails to recover more than two components for most penalty choices.}

 {In terms of computational efficiency, both CCAR$^3$ variants are substantially faster than SAR: training graph CCAR$^3$ and sparse CCAR$^3$ requires, on average, $2.30$ and $1.70$ seconds, respectively, compared to $25.7$ seconds for SAR, while CCA-Lasso is the fastest at $0.0583$ seconds.}

\begin{figure}[h!]
\centering
\begin{subfigure}{0.95\linewidth}
    \centering
    \includegraphics[width=0.72\textwidth]{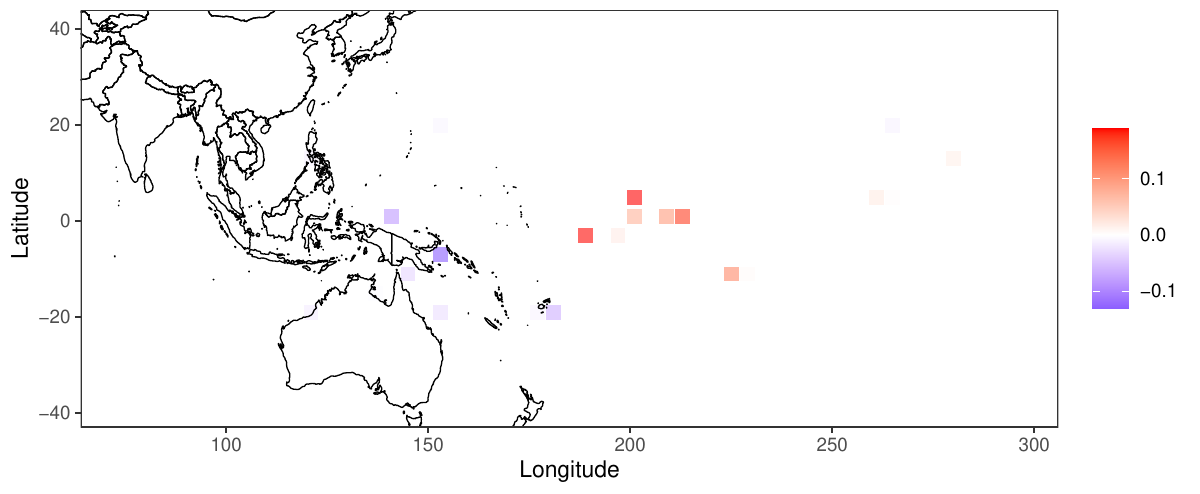}
    \includegraphics[width=0.27\textwidth]{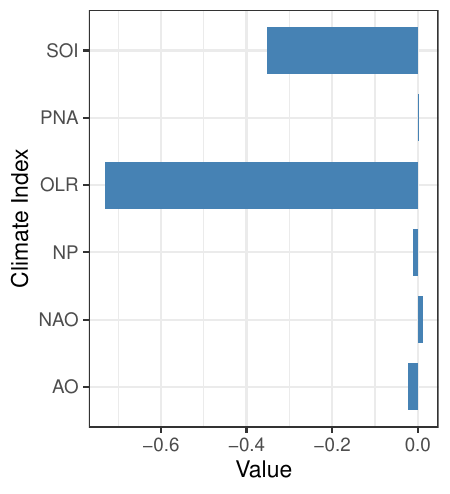}
    \caption{sparse CCAR$^3$}
    \label{fig:clim_sparse}
\end{subfigure}
\vspace{0.5cm} 
\begin{subfigure}{0.95\linewidth}
    \centering
    \includegraphics[width=0.72\textwidth]{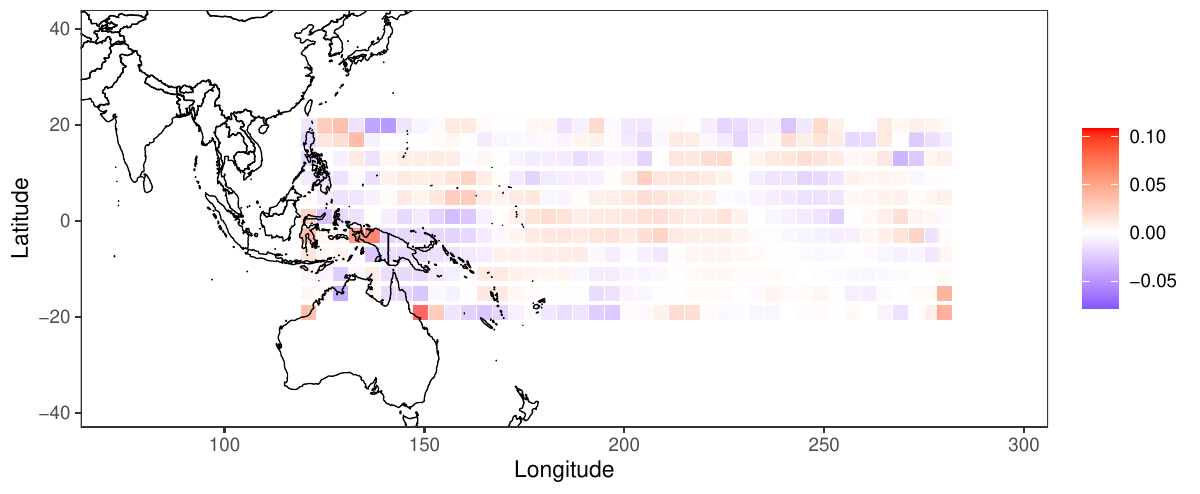}
    \includegraphics[width=0.27\textwidth]{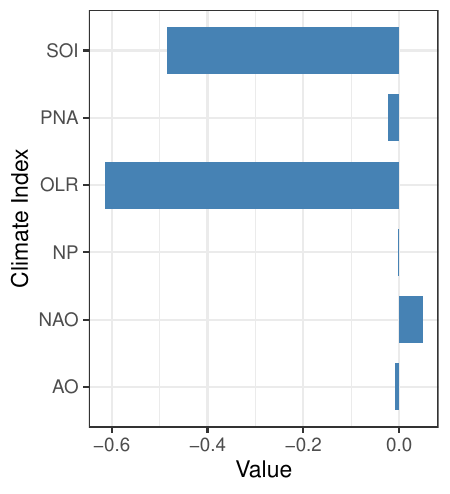}
    \caption{graph CCAR$^3$}
    \label{fig:clim_graph}
\end{subfigure}
\caption{Comparison of the first canonical directions obtained by sparse and graph CCAR$^3$. Left panels represent the canonical directions $\widehat{u}_1$ corresponding to SST features. Right panels represent the canonical directions $\widehat{v}_1$ corresponding to climate indices.}
\label{fig:climate}
\end{figure}

 {Figure~\ref{fig:climate} compares the first canonical directions obtained by sparse and graph CCAR$^3$. The $\widehat{v}_1$ vectors exhibit similar patterns, highlighting the influence of SOI and OLR. In contrast, the $\widehat{u}_1$ vectors differ substantially: the sparse method identifies a few isolated locations with non-zero values (positive near longitude $200^\circ$E and latitude $0^\circ$, and negative near longitude $150^\circ$E and latitude $-10^\circ$), whereas the graph-based method reveals clusters of significant regions, with similar overall positions of positive and negative important features.}

 {These positions correspond to physically meaningful SST regions: the central/eastern equatorial Pacific (around $200^\circ$E) and the western Pacific  warm pool  (around $150^\circ$E) are key areas influencing El Ni{\~n}-Southern Oscillation variability, consistent with  Southern Oscillation Index (SOI) and outgoing longwave radiation (OLR) patterns. The graph-regularization captures spatially coherent SST structures, reflecting connected regions that jointly influence large-scale climate indices, whereas the sparse method highlights only individual grid points, potentially missing broader regional effects. Overall, the graph-based approach provides a more interpretable and physically meaningful representation of ocean-atmosphere interactions.}

\section{Related Works}\label{sec:related_works}

To overcome the challenge posed by the high dimensionality of the data, several adaptations of the original CCA problem have recently been developed. All of these modifications incorporate some form of regularization or constraint on the estimators of $\bU$ and $\bV$, and can be classified into one of two categories depending on the type of regularization they employ.

{\textit{Ridge-regularized CCA methods}} attempt to resolve the issue of the ill-conditioning of $\widehat{\bS}_{X}$ and $\widehat{\bS}_{Y}$
by adjusting their diagonals \citep{leurgans1993}. In this setting, the CCA problem becomes:
\begin{equation}\label{eq:cca_reg}
    \begin{split}
       \maximize_{\bU\in\R^{p\times r},~\bV\in\R^{q\times r}} \quad & \Tr\Big(\bU^\top(\widehat{\bS}_{X} + \rho_1 \mathbf I)^{-1}\widehat{\bS}_{XY}(\widehat{\bS}_{Y}+\rho_2 \mathbf I)^{-1} \bV \Big)  \\
   \text{ subject to }    \quad & \bU^\top \widehat{\bS}_{X} \bU =  \bV^\top \widehat{\bS}_{Y} \bV  =\mathbf I_r.
    \end{split}
\end{equation}
While $\ell_2$-regularized CCA methods have demonstrated some empirical success, to the best of our knowledge, there are no theoretical guarantees as to the accuracy of the proposed estimators of $\bU$~and~$\bV$. Moreover, while these methods shrink coefficients, they often yield dense solutions that do not necessarily lend themselves well to subsequent interpretation and analysis.

{\textit{Sparse CCA methods}}, on the other hand,  suggest imposing sparsity constraints on the CCA directions. This is typically done by introducing a variant of the $\ell_1$ penalty, which
 effectively reduces the complexity of the model by identifying and using only the most informative variables in $\bX$ and $\bY$ to estimate the directions $\bU$ and $\bV$. 
Building on the success of the lasso in the regression setting \citep{tibshirani1996regression}, initial approaches to sparse canonical correlation analysis focused primarily on imposing a sparsity penalty on the individual canonical directions. 
For instance, \cite{witten2009extensions} and \cite{parkhomenko2009sparse} propose algorithms imposing a lasso penalty on each of the canonical directions for the case where $r=1$. However, their proposed methods are not straightforwardly applicable to the estimation of several directions, particularly when the covariance matrices of $\bX$ and $\bY$ are not diagonal and it is no longer possible to neglect the normalization constraints $\bU^\top\widehat\bS_{X}\bU=\mathbf I_r$ and $\bV^\top \widehat{\bS}_{Y} \bV  = \mathbf I_r$. As demonstrated by our experiments in Section~\ref{sec:sparse-cca}, in this case, these methods struggle to estimate the canonical directions even when $n$ is large. To allow the estimation of multiple canonical directions, \cite{wilms2015sparse} propose an iterative algorithm based on an alternating minimization scheme to solve for each pair $({u}_i, {v}_i)$. This method allows the computation of more than one pair of canonical directions by progressively deflating the matrices $\bX$ and~$\bY$. Since the sparsity of canonical directions obtained on the deflated matrices does not necessarily translate into sparse canonical directions expressed in terms of the original data matrices $\bX$ and $\bY$, the authors propose adding a postprocessing step that sparsifies their estimated canonical directions in the original covariate basis. Although the proposed technique is efficient, the alternating nature of this method makes it sensitive to initialization. Moreover, while each of the resulting CCA directions $({u}_i)_{i=1}^r$ and $({v}_i)_{i=1}^r$ are typically sparse, the matrices $\bU$ and $\bV$ are not necessarily row-sparse, which may lead to non-sparse canonical subspaces (defined by the span of the columns of $\bU$~or~$\bV$).

While sparse CCA variants have proliferated over the years, one of the few theoretical treatments is provided by Gao and coauthors \citep{chen2013sparse, gao2015minimax, gao2017sparse}.  
More specifically, \cite{gao2015minimax} consider the setting where the support sizes, i.e., the number of nonzero rows of $\bU$ and $\bV$, are small: $|\text{supp}(\bU)| \leq s_u$ and $|\text{supp}(\bV)| \leq s_v$ for some $s_u \leq p$, $s_v \leq q$. Under this assumption, they show that the minimax rate for estimators of $\bU$ and $\bV$ under the joint loss 
\(\| \widehat{\bU} \widehat{\bV}^\top - \bU\bV^\top \|_F^2\) is
\[
    \frac{1}{n \lambda_r^2} \Big( r(s_u + s_v) + s_u \log({ep}/{s_u}) + s_v \log({eq}/{s_v}) \Big),
\]
where $\lambda_r$ denotes the $r^{\text{th}}$ canonical correlation.

Since the estimator proposed by \cite{gao2015minimax} is not tractable in practice, \cite{gao2017sparse} introduce a three-step algorithm that achieves the minimax rate. This procedure requires splitting the data into three batches. In the first step, the cross-product \(\mathbf F = \bU \bV^\top\) is estimated from the first batch, transforming the original non-convex CCA problem into a convex optimization problem:
\begin{equation}\label{eq:fantope_revised}
\begin{split}
\maximize_{\mathbf F \in \R^{p\times q}}~ \langle \widehat\bS_{XY}, \mathbf F \rangle - \rho \| \mathbf F \|_{11} ~~\text{subject to} ~ \| \widehat\bS_X^{1/2} \mathbf F \widehat\bS_Y^{1/2} \|_* \le r, ~ 
\| \widehat\bS_X^{1/2} \mathbf F \widehat\bS_Y^{1/2} \|_{\rm op} \le 1,
\end{split}
\end{equation}
where \(\rho\) is a sparsity-controlling penalty parameter, and \(\|\cdot\|_{11}\), \(\|\cdot\|_*\), and \(\|\cdot\|_{\rm op}\) denote the usual matrix \(\ell_1\), nuclear, and operator norms, respectively. This step provides initial estimates of the canonical directions \(\widehat{\bU}\) and \(\widehat{\bV}\). In the second step, using the second batch of data, the estimators are refined by solving separate group-lasso regression problems for \(\bU\) and \(\bV\), enforcing sparsity via the \(\ell_{21}\) penalty. The third step uses the third batch to normalize the resulting estimates.  

While theoretically sound, this procedure has important practical limitations. The first step relies on an ADMM algorithm that requires repeated SVDs of $p \times q$ matrices and solving lasso problems with $pq$ parameters at each iteration, resulting in substantial computational costs. Moreover, splitting the data into three batches can be problematic in small-sample settings, further limiting the method's practical applicability.

Building on this work, \cite{gao2021sparse} propose an alternative two-step procedure. The first step serves as an initialization for 
$\mathbf{F}  = \begin{psmallmatrix}
\bU \\
\bV
\end{psmallmatrix} 
 \begin{psmallmatrix}
\bU^\top & 
\bV^\top
\end{psmallmatrix} 
\in \mathbb{R}^{(p+q)\times(p+q)}
$
and is based on a Fantope projection problem:
\[
\minimize_{\mathbf{F} \in \mathcal F} ~  - \langle \widehat{\mathbf{\Sigma}}, \mathbf{F}  \rangle + \rho \| \mathbf{F}  \|_1 
\quad\text{subject to} \quad  \widehat{\mathbf{\Sigma}}_0^{1/2} \mathbf{F}  \widehat{\mathbf{\Sigma}}_0^{1/2} \in \mathcal F,
\]
where 
\[
\mathcal F = 
\left\{
\mathbf{X} \in \mathbb{R}^{(p+q)\times(p+q)} : 0 \preceq \mathbf{X} \preceq \mathbf{I}_{p+q}, \ \Tr(\mathbf{X}) = r
\right\}.
\]
The second step refines the estimates using a thresholded gradient descent algorithm. While this approach improves estimation, the initialization still relies on an ADMM algorithm that involves repeated SVDs of a $(p+q) \times (p+q)$ matrix, which scales poorly with dimension and becomes computationally expensive for large datasets.

While these methods achieve minimax-optimal statistical guarantees, the quality of the estimators critically depends on the initializer obtained in the first step, which is computationally expensive (see Appendix~\ref{algo:comp} for details). Moreover, selecting the regularization parameter \(\rho\) via cross-validation requires repeatedly solving this initialization problem, further amplifying the computational burden.

\section{Conclusion} \label{sec:conclusion}

In this paper we present CCAR$^3$, a novel technique for estimating canonical directions based on the link between CCA and reduced rank regression when one of the datasets is high-dimensional while the dimension of the other remains small.
Our method can easily incorporate different types of penalties, allowing a good compromise between computational efficiency, statistical accuracy, and practical flexibility in a wide range of practical settings. 
A potential direction for future research consists in extending CCAR$^3$ to settings where both $p$ and $q$ are large. This will require further investigation of computationally efficient alternatives to the Fantope optimization problem, which is both the crux of the success and the computational bottleneck of \cite{gao2017sparse} and \cite{gao2021sparse}. 

\section*{Acknowledgments}
C.D. was supported by the National Science Foundation under Award Number 2238616, as well as the resources provided by the University of Chicago's Research Computing Center.
E.T. was supported by the Natural Sciences and Engineering Research Council of Canada under grant RGPIN-2023-04727; the University of Toronto Data Science Institute Catalyst grant; and the University of Toronto McLaughlin Center under grant MC-2023-05. 
This research was supported in part by grants from the NSF (DMS-2235451) and Simons Foundation (MPS-NITMB-00005320) to the NSF-Simons National Institute for Theory and Mathematics in Biology (NITMB).

\bibliography{references.bib}

\appendix

\newpage 

\section{Additional plots and tables} \label{appendix:tabplot}
\subsection{A review of CCA in recent applications}
\label{app:review}

\begin{table}[h!]
    \centering
    \resizebox{\textwidth}{!}{%
    \begin{tabular}{|c|c|c|c|c|}\hline 
       Study  & Objective & Number of Observations & Size of dataset 1 & Size of dataset 2 \\ \hline \hline
           \cite{wang2023improved} & $\bX$:  features extracted from EEG data & $n=15$& $p=310$ & $q=3$\\
    &  $\bY$: Emotion Data    & & &\\ \hline
               \cite{looden2022patterns} & $\bX$:  Connectivity features extracted from fMRI & $n=299$& $p=30,3081$ & $q=3$\\
    &  $\bY$: Behavioral Measurements    & & &\\ \hline
               \cite{luobrain} & $\bX$:  Connectivity features extracted from fMRI measurements & $n=122$& $p=25,651$ & $q \in [30, 50]$\\
    &  $\bY$: Clinical and Behavioural Measurements    & & &\\ \hline
    \cite{tozzi2021relating} & $\bX$: fMRI activation of each voxel in the brain & $n=269$& $p=90,000$ & $q=89$\\
    &  $\bY$: Questionnaire Data    & & &\\ \hline
    \cite{pigeau2022impact} & $\bX$: vertex-level cortical thickness measurements & $n=25,043$& $p=38,561$ & $q=52$\\
    &  $\bY$: Behavioural Data    & & &\\
        \hline
    \end{tabular}
    }
    \caption{Short survey of recent applications of high-dimensional CCA.}
    \label{tab:survey}
\end{table}

\subsection{Simulations}
 \label{appendix:plots:simulation}

\xhdr{Group Sparsity} 
We simulate here a setting where we know the groups. We adopt a similar data generation mechanism as in the previous subsection, adapting only the procedure for generating~$\bU$. In particular, here, we split the $p$ covariates into groups of size 10, and select 5 of these groups to be nonzero. 
We then select at random each of the entries of these groups and normalize as for the simulations in the previous section. 

\begin{figure}[h!]
    \centering
    \includegraphics[width = 0.7\textwidth]{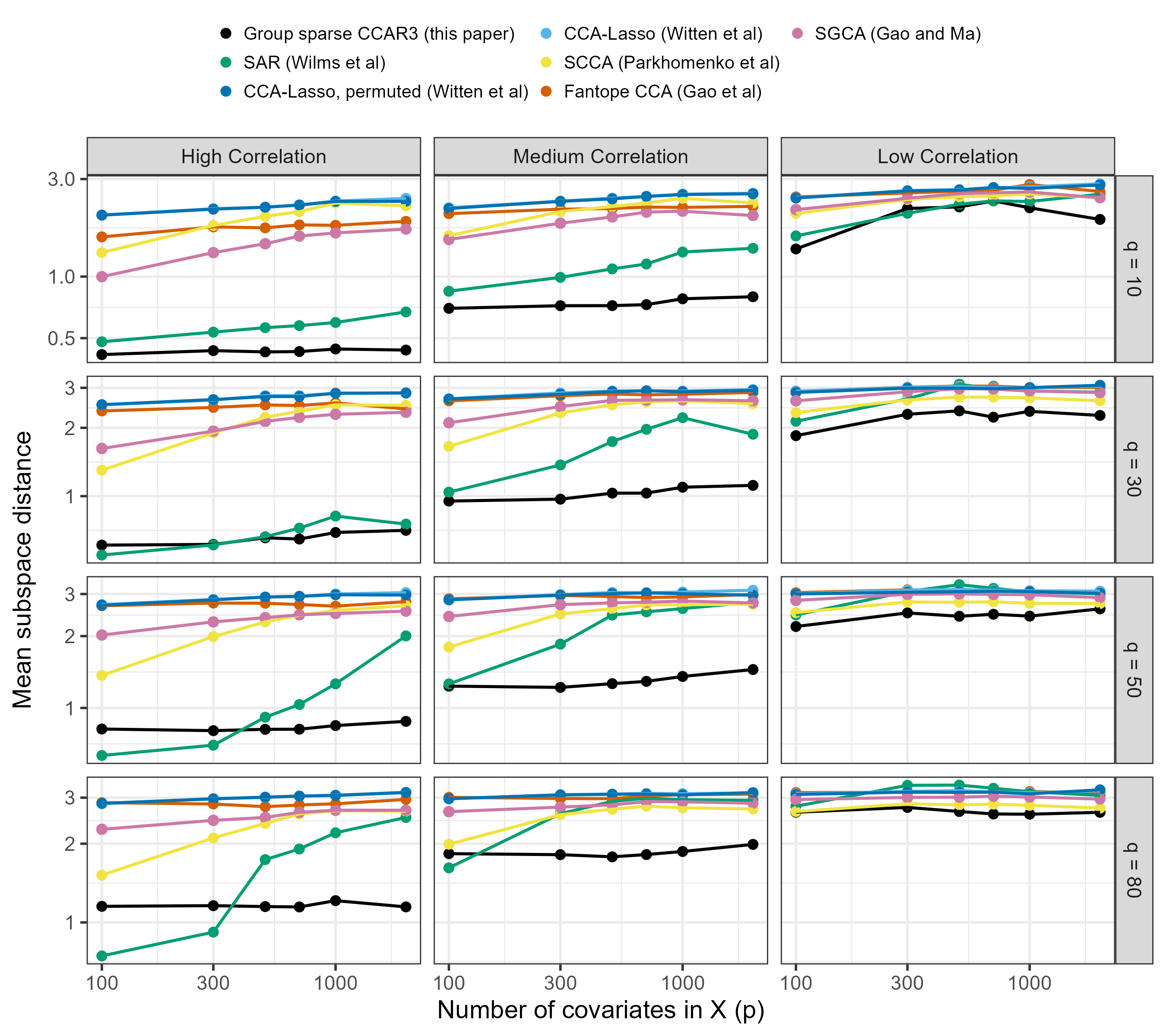}
    \caption{Simulation results for group model with $r_{pca} = 5$,  $5$ nonzero groups and $n=500$. Subspace distance as a function of $p$. Results are averaged over 100 simulations.}
    \label{fig:simu:group:sd}
\end{figure}

\clearpage

\begin{figure}[h!]
    \centering
    \includegraphics[width = 0.8\textwidth]{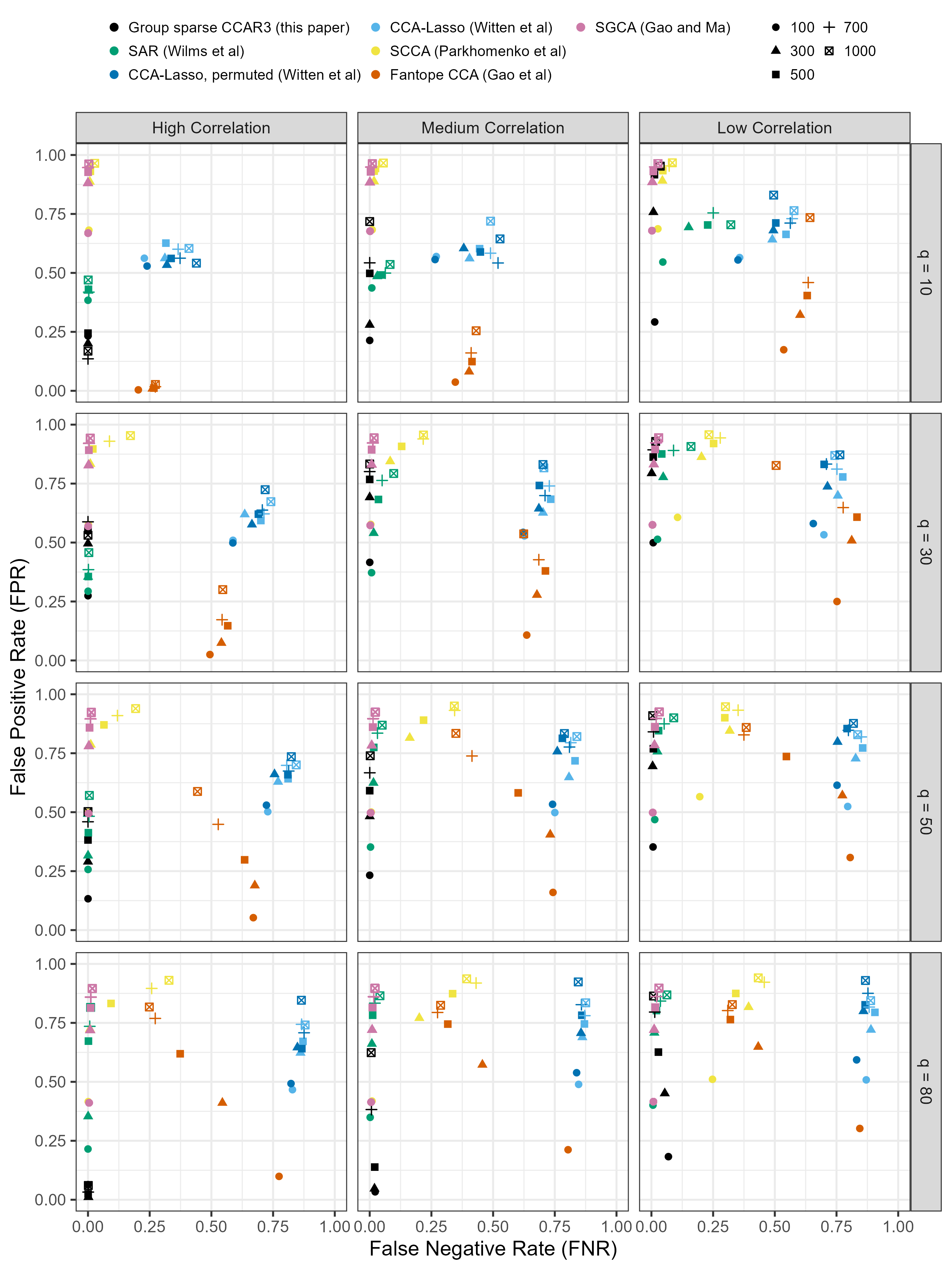}
    \caption{Simulation results for sparse model with $r_{pca} = 5$. False Negative Rate vs False Positive Rate with $n=500$, $r = 3$ and $n_{nnz}=6$. The shape of each point represents $X$ data dimensionality ($p$). Results are averaged over 100 simulations.}
    \label{fig:simu:group:fr}
\end{figure}

In Figure~\ref{fig:simu:group:fr} the true discovery is understood as an overlap between the supports of the estimator and the true support of $U^*$, and a false discovery is understood as the complement of this intersection in the support of $\wh U$.

\clearpage

\xhdr{Graph Sparsity} 
We reiterate the previous set of synthetic experiments, changing only the data generation mechanism for $\bU$ to accommodate a graph-structure. More specifically, we first generate a graph (here a 2D grid) with $p$ nodes and $m$ edges. We sample $\widetilde{\bU} \in \R^{m \times r}$ in the same way as in our sparse CCA experiments. We then apply a transformation to $\widetilde \bU$ as $\bU = \mathbf \Gamma^{\dagger} \widetilde{\bU}$, where $\bm\Gamma^{\dagger}$ denotes the pseudoinverse of $\mathbf\Gamma$. This results in a $\bU$ such that $\bm \Gamma \bU$ is sparse and $\bm\Pi \bm U=0$, where $\bm\Pi = \bm I_{p} - \bm\Gamma^{\dagger} \bm\Gamma$. Similar to the previous experiments, we vary the signal strength and sizes of $p$ and $q$.

\begin{figure}[h!]
    \centering
    \includegraphics[width = 0.7\textwidth]{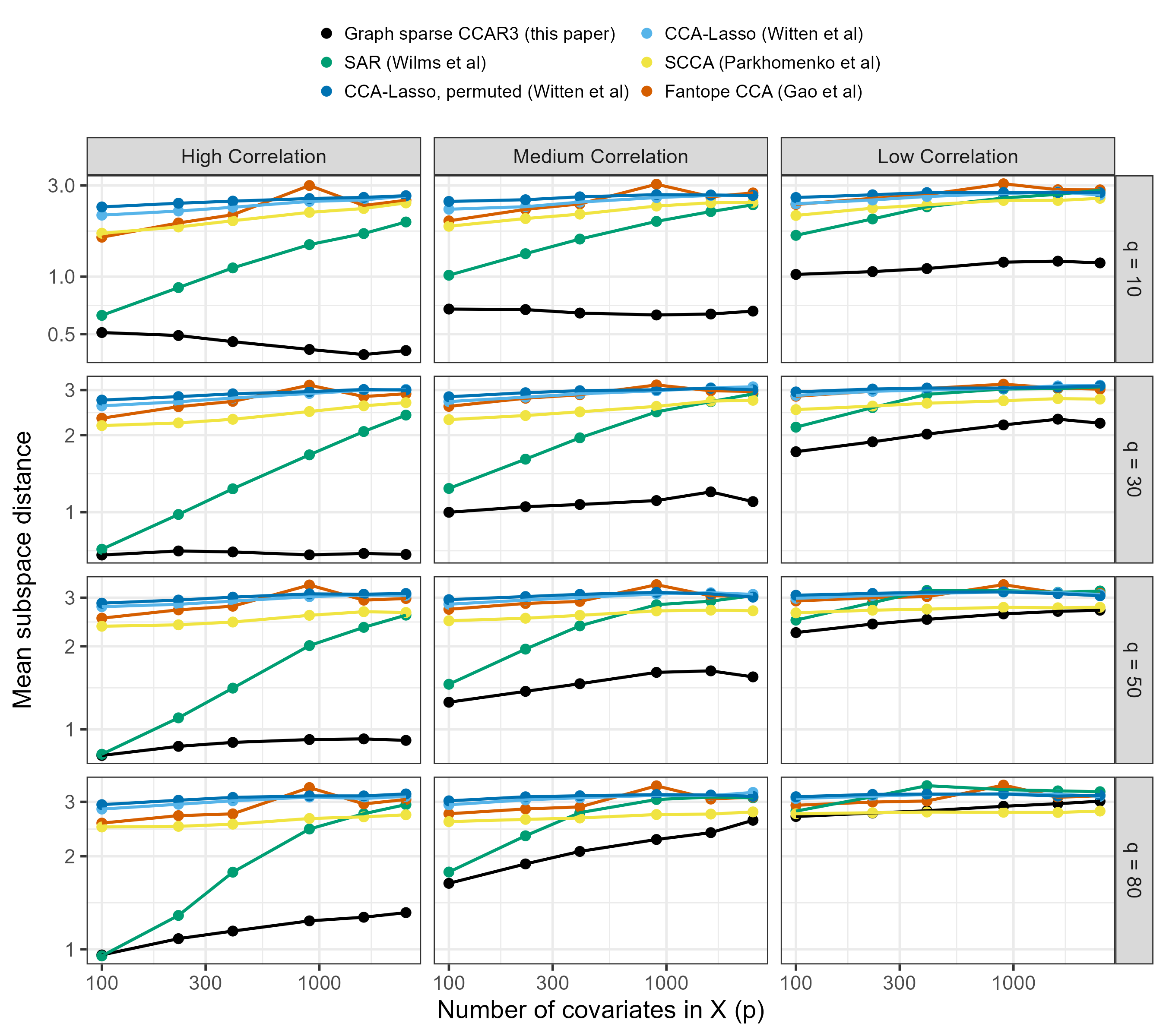}
    \caption{Simulation results for graph model with $r_{pca} = 5$,  $\supp(\bGamma\bB)=5$ and $n=500$. Subspace distance as a function of $p$. Results are averaged over 100 simulations.}
    \label{fig:simu:graph:sd}
\end{figure}

\clearpage

\subsection{Nutrimouse dataset}
\label{appendix:plots:nutrimouse}

Below, we present the loading vectors recovered by sparse CCAR$^3$.

\begin{figure}[h!]
    \centering
    \begin{subfigure}[b]{\linewidth}
        \includegraphics[width=\linewidth]{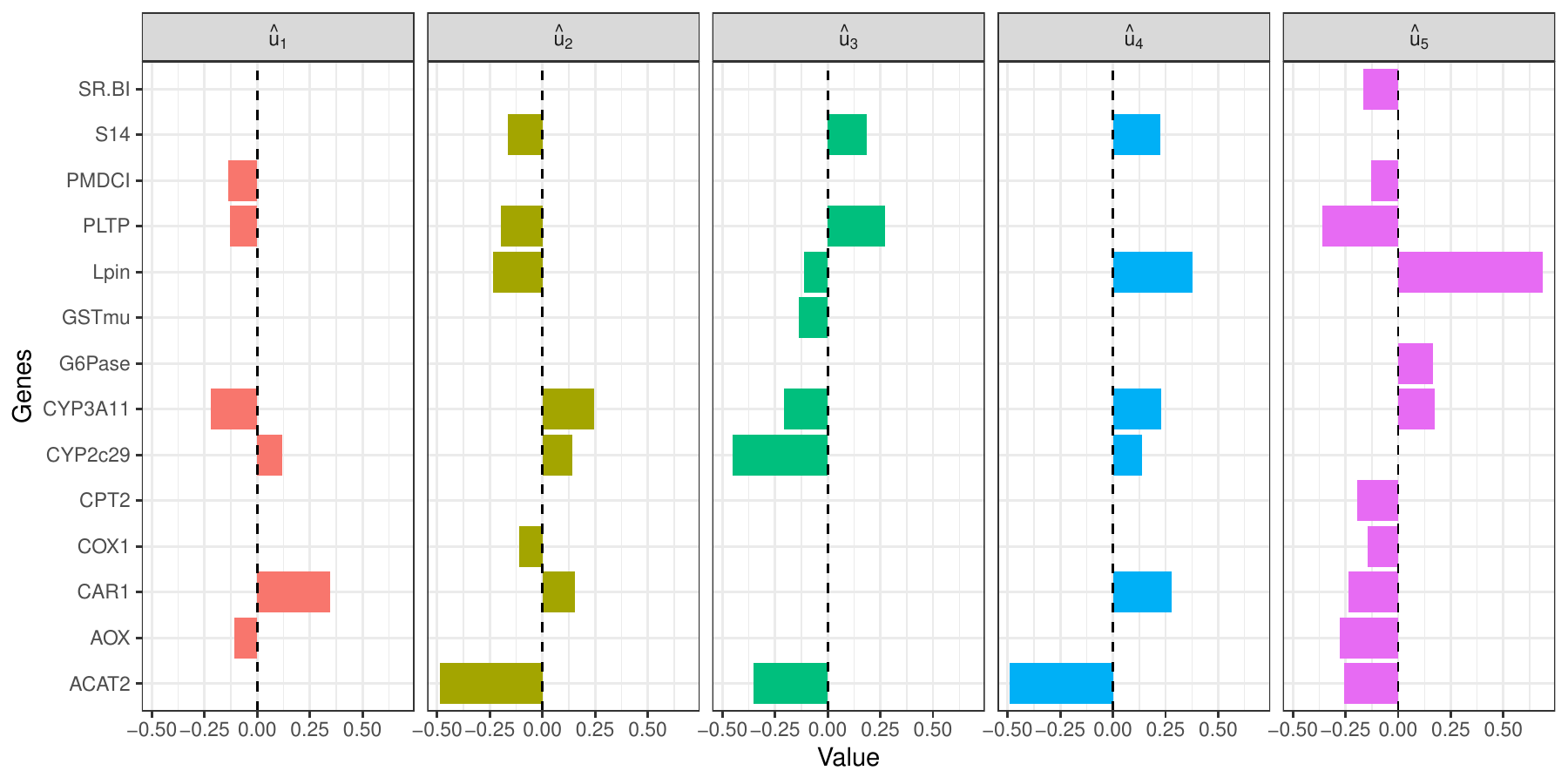}
        \includegraphics[width=\linewidth]{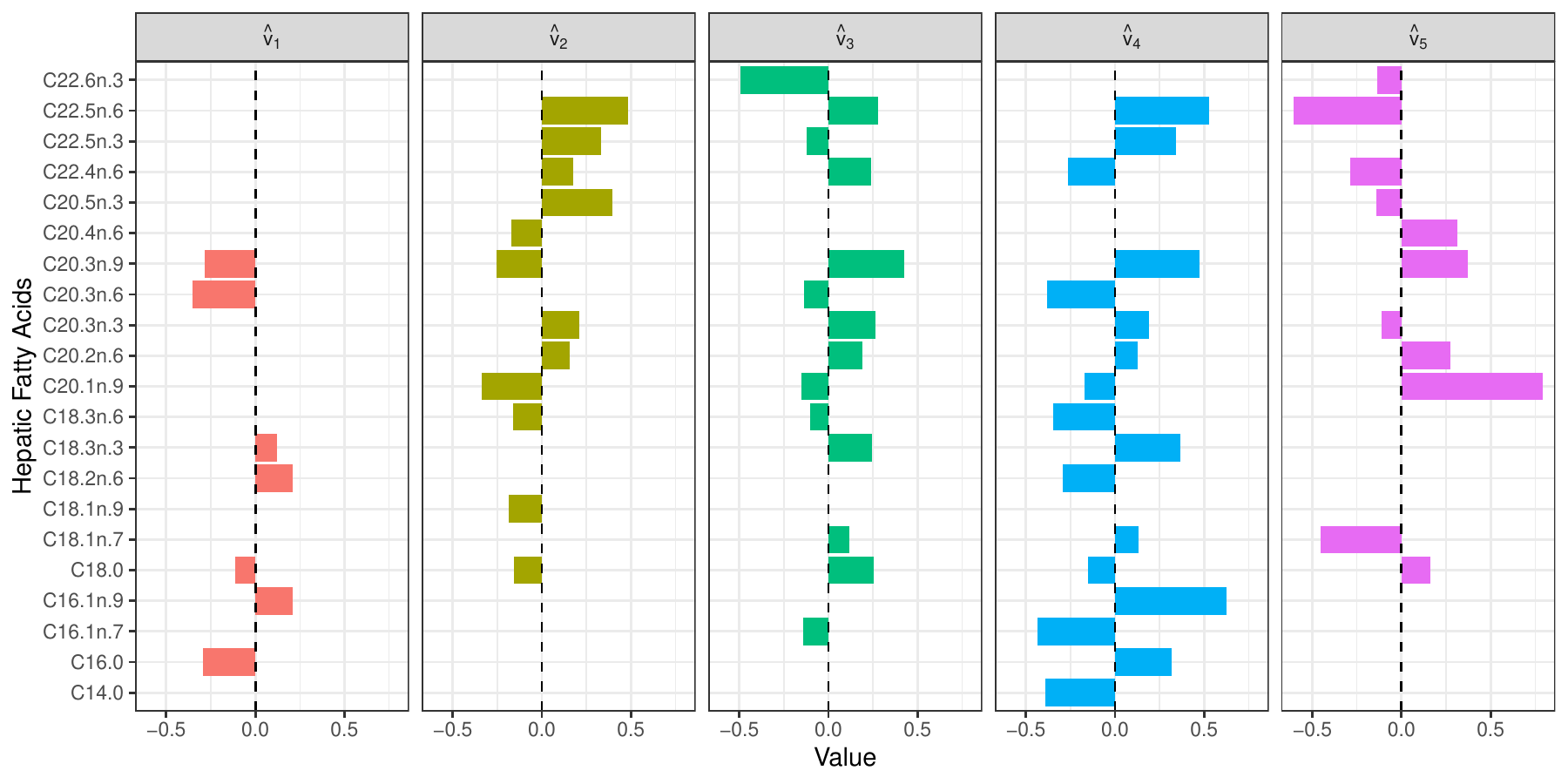}
    \end{subfigure}
    \caption{Sparse CCAR$^3$ presented in this paper. Five canonical directions corresponding to genes, i.e. $(\widehat{\bm{u}}_i)_{i=1}^5$, and fatty acids, i.e. $(\widehat{\bm{v}}_i)_{i=1}^5$, produced for the Nutrimouse dataset. For the convenience of visualization, the plot represents only values greater than the $0.1$ threshold.}
\label{fig:nutrimouse_UV_rrr}
 \end{figure}

\section{Algorithms}\label{appendix:algo}
\subsection{Computational complexity of theoretical methods}
\label{algo:comp}
We review the computational complexity of the three-stage procedure in
\emph{Sparse CCA: Adaptive Estimation and Computational Barriers} by
\cite{gao2017sparse} and the two-stage procedure in the
\emph{Thresholded Gradient Descent method for Sparse GCA} by
\cite{gao2021sparse}. In both methods, the main computational burden arises
from the convex ADMM-based initialization. The subsequent refinement stages are
significantly cheaper when $r \ll \min(p,q)$.

\paragraph{Sparse CCA by \cite{gao2017sparse}}

The data are split into three independent batches indexed by $0,1,2$.

\medskip

\textit{Stage 1: Initialization.} The initialization solves
$$
\max_{\mathbf{F}\in\mathbb{R}^{p\times q}}
\langle \widehat{\mathbf{\Sigma}}_{XY}^{(0)}, \mathbf{F}\rangle
-\rho\|\mathbf{F}\|_1
\quad
\text{s.t.}
\quad
\| (\widehat{\mathbf{\Sigma}}_X^{(0)})^{1/2}\mathbf{F}(\widehat{\mathbf{\Sigma}}_Y^{(0)})^{1/2}\|_* \le r,
\quad
\| (\widehat{\mathbf{\Sigma}}_X^{(0)})^{1/2}\mathbf{F}(\widehat{\mathbf{\Sigma}}_Y^{(0)})^{1/2}\|_{op} \le 1.
$$
The optimization variable $\mathbf{F}$ is a $p\times q$ matrix, so the problem
has $pq$ parameters. The algorithm is based on ADMM and each iteration consists
of three main steps.

\begin{enumerate}
\item \textit{Lasso.} Solve an $\ell_1$-penalized quadratic problem with $pq$ parameters.
\item \textit{Singular value capped soft-thresholding.} This requires computing the SVD of a $p\times q$ matrix.
\item \textit{Matrix multiplications.} Multiply $p\times p$, $p\times q$, and $q\times q$ matrices.
\end{enumerate}

\medskip

\textit{Stage 2: Group Lasso Refinement.} Let
$\mathbf{V}^{(0)}\in\mathbb{R}^{q\times r}$ store the top $r$ right singular
vectors of the solution $\widehat{\mathbf{F}}$ from Stage~1. The second stage
solves a group lasso problem with $pr$ parameters to obtain
$\mathbf{U}^{(1)}$. The analogous update for $\mathbf V$ uses the estimated
left singular vectors and has $qr$ parameters.

\medskip

\textit{Stage 3: Normalization.}
The final estimators are normalized on the third data split, e.g.
$
\widehat{\mathbf{U}}
=
\widehat{\mathbf{U}}^{(1)}
\left(
\widehat{\mathbf{U}}^{(1)\top}
\widehat{\mathbf{\Sigma}}_X^{(2)}
\widehat{\mathbf{U}}^{(1)}
\right)^{-1/2}
$
and analogously for $\widehat{\mathbf V}$.

\medskip

Overall, the computational bottleneck of Sparse CCA is the convex ADMM initialization,
dominated by repeated lasso solves and SVDs of $p\times q$ matrices. Additionally,
this approach requires splitting data into three batches, which is not ideal for many
applications with limited data.

\bigskip

\paragraph{Sparse GCA by \cite{gao2021sparse}} This approach requires two stages.

\textit{Stage 1: Initialization via Fantope Relaxation.}
Let
$
\mathbf{F}=\mathbf{A}\mathbf{A}^\top\in\mathbb{R}^{(p+q)\times(p+q)},
$
where
$
\mathbf{A}=
\begin{pmatrix}
\mathbf{U}\\
\mathbf{V}
\end{pmatrix}
\in\mathbb{R}^{(p+q)\times r}.
$
The initialization is based on a Fantope projection problem, which solves
$$
\begin{aligned}
\min_{\mathbf{F} \in \mathcal F}
\quad & -\langle \widehat{\mathbf{\Sigma}}, \mathbf{F} \rangle
+ \rho \|\mathbf{F}\|_1
\quad \text{subject to}\quad
& \widehat{\mathbf{\Sigma}}_0^{1/2} \mathbf{F} \widehat{\mathbf{\Sigma}}_0^{1/2} \in \mathcal F,
\end{aligned}
$$
where
$
\mathcal F
=
\left\{
\mathbf{X} \in \mathbb{R}^{(p+q)\times(p+q)}
:
0 \preceq \mathbf{X} \preceq I_{p+q},
\ \operatorname{Tr}(\mathbf{X})=r
\right\}.
$
This problem is also solved using ADMM. Like in \cite{gao2017sparse}, it requires
computing an SVD at each iteration, but the matrix size is now $(p+q)\times(p+q)$.
The per-iteration SVD cost is therefore $O((p+q)^3)$.

\medskip

\textit{Stage 2: Thresholded Gradient Descent.}
After initialization, refinement is performed via thresholded gradient descent.
The per-iteration complexity is
$
O\!\left(n(p+q)r + (p+q)\log(p+q)\right).
$
The first term corresponds to gradient computation, and the second to thresholding/sorting.

\bigskip

Overall, the bottleneck of SGCA is the repeated SVDs of $(p+q)\times(p+q)$ matrices.

\subsection{Algorithm for CCAR$^3$}
\label{algo:spr3-cca}

The sparse and group-sparse versions of Step 2, and the profiled graph-sparse version
described below, can be written as
\begin{equation}
\label{eq:structured-regression}
\widehat{\bTheta}
=
\operatorname*{argmin}_{\bTheta\in\mathcal C}
\frac{1}{n}\|\widetilde{\bY}_0-\widetilde{\bX}\bTheta\|_F^2
+
\rho \sum_{m=1}^M w_m \|\bTheta_{\mathcal G_m\cdot}\|_F,
\end{equation}
where $\mathcal G_1,\ldots,\mathcal G_M$ form a partition of $[\tilde p]$,
$w_m>0$ are prespecified weights, and $\mathcal C\subseteq\mathbb R^{\tilde p\times q}$
is a closed linear subspace. For the sparse and group-sparse cases,
$\mathcal C=\mathbb R^{\tilde p\times q}$. For the graph case, $\mathcal C$ enforces
that the edge variable belongs to the column space of the incidence matrix.
The three special cases are as follows.

\paragraph{Sparse CCAR$^3$.}
Take
$
\tilde p=p,
\quad
\widetilde{\bX}=\bX,
\quad
\widetilde{\bY}_0=\bY_0,
\quad
\mathcal C=\mathbb R^{p\times q},
\quad
\mathcal G_m=\{m\},
\quad
{w_m=1}.
$
Then $\widehat{\bB}=\widehat{\bTheta}$.

\paragraph{Group-sparse CCAR$^3$.}
Take
$
\tilde p=p,
\quad
\widetilde{\bX}=\bX,
\quad
\widetilde{\bY}_0=\bY_0,
\quad
\mathcal C=\mathbb R^{p\times q},
\quad
\mathcal G_m=g_m,
\quad
w_m=\sqrt{|g_m|}.
$
Then $\widehat{\bB}=\widehat{\bTheta}$.

\paragraph{Graph-sparse CCAR$^3$.}
Let $\bGamma\in\mathbb{R}^{m\times p}$ be the graph incidence matrix, where
$m=|E|$, and let
$
\mathbf\Pi=\mathbf I_p-\bGamma^\dagger\bGamma
$
be the orthogonal projector onto $\ker(\bGamma)$. Any coefficient matrix can be
written as
$
\bB=\bB_{\mathrm{null}}+\bGamma^\dagger\bTheta,
$
where $\bB_{\mathrm{null}}\in\ker(\bGamma)$ and
$\bTheta=\bGamma\bB\in\operatorname{col}(\bGamma)^{\oplus q}$. Define the projection
onto the fitted null-space design by
\[
\mathbf P_\Pi
=
\bX\mathbf\Pi
\bigl\{(\bX\mathbf\Pi)^\top(\bX\mathbf\Pi)\bigr\}^\dagger
(\bX\mathbf\Pi)^\top .
\]
Profiling out the null-space component gives the exact profiled problem
\eqref{eq:structured-regression} with
\[
\tilde p=m,
\qquad
\widetilde{\bX}=(\mathbf I_n-\mathbf P_\Pi)\bX\bGamma^\dagger,
\qquad
\widetilde{\bY}_0=(\mathbf I_n-\mathbf P_\Pi)\bY_0,
\qquad
\mathcal C=\{\bTheta:(\bGamma\bGamma^\dagger)\bTheta=\bTheta\},
\]
and singleton edge groups $\mathcal G_m=\{m\}$ with $w_m=1$. After obtaining
$\widehat{\bTheta}$, the profiled null-space component is
\[
\widehat{\bB}_{\mathrm{null}}
=
\mathbf\Pi
\bigl\{(\bX\mathbf\Pi)^\top(\bX\mathbf\Pi)\bigr\}^\dagger
(\bX\mathbf\Pi)^\top
\left(\bY_0-\bX\bGamma^\dagger\widehat{\bTheta}\right),
\]
and the final estimator is
\[
\widehat{\bB}
=
\widehat{\bB}_{\mathrm{null}}+\bGamma^\dagger\widehat{\bTheta}.
\]
The constraint $\widehat{\bTheta}\in\operatorname{col}(\bGamma)^{\oplus q}$ is part of
the exact equivalence with the original penalty $\|\bGamma\bB\|_{21}$.

To solve \eqref{eq:structured-regression}, introduce a splitting variable
$\mathbf Z\in\mathbb{R}^{\tilde p\times q}$ and consider the constrained problem
$$
\operatorname*{minimize}_{\bTheta\in\mathcal C,\,\mathbf Z}
\frac{1}{n}\|\widetilde{\bY}_0-\widetilde{\bX}\bTheta\|_F^2
+
\rho \sum_{m=1}^M w_m \|\mathbf Z_{\mathcal G_m\cdot}\|_F
\quad\text{subject to}\quad
\bTheta=\mathbf Z.
$$
Let $\mathbf P_{\mathcal C}$ denote the orthogonal projector onto $\mathcal C$.
Thus $\mathbf P_{\mathcal C}=\mathbf I_{\tilde p}$ for the sparse and group-sparse
cases, while $\mathbf P_{\mathcal C}\bTheta=(\bGamma\bGamma^\dagger)\bTheta$ for
the graph-sparse case. Using the scaled augmented Lagrangian with scaled dual
variable $\mathbf D$,
$$
\mathcal L_\delta(\bTheta,\mathbf Z,\mathbf D)
=
\frac{1}{n}\|\widetilde{\bY}_0-\widetilde{\bX}\bTheta\|_F^2
+
\rho \sum_{m=1}^M w_m \|\mathbf Z_{\mathcal G_m\cdot}\|_F
+
\frac{\delta}{2}\|\bTheta-\mathbf Z+\mathbf D\|_F^2
-
\frac{\delta}{2}\|\mathbf D\|_F^2,
$$
we obtain the ADMM updates given in Algorithm~\ref{alg:structured-admm}.

\begin{algorithm}[h]
\caption{ADMM for \eqref{eq:structured-regression}}
\label{alg:structured-admm}

\textit{Input:}
$\widetilde{\bX}\in\mathbb{R}^{n\times \tilde p}$,
$\widetilde{\bY}_0\in\mathbb{R}^{n\times q}$,
groups $\mathcal G_1,\ldots,\mathcal G_M$,
weights $w_1,\ldots,w_M$,
regularization parameter $\rho>0$,
ADMM parameter $\delta>0$,
projector $\mathbf P_{\mathcal C}$,
stopping tolerances $\epsilon_{\mathrm{pri}},\epsilon_{\mathrm{dual}}>0$.

\textit{Initialize:}
$\bTheta^{(0)},\mathbf Z^{(0)},\mathbf D^{(0)}\in\mathbb{R}^{\tilde p\times q}$.

\textit{Precompute:}
$
\mathbf K_\delta=\frac{2}{n}\widetilde{\bX}^\top\widetilde{\bX}+\delta \mathbf I_{\tilde p}
$
and
$
\mathbf C=\frac{2}{n}\widetilde{\bX}^\top\widetilde{\bY}_0.
$
When $\tilde p$ is moderate, factorize $\mathbf K_\delta$ once and reuse the
factorization throughout; when $\tilde p\gg n$, the Woodbury identity described
below is more efficient.

\textit{Repeat until convergence:}
\begin{algorithmic}[1]
\State Update $\bTheta$ by solving
$
\mathbf K_\delta \bTheta^{(t+1)}
=
\mathbf P_{\mathcal C}\!\left\{
\mathbf C+
\delta\bigl(\mathbf Z^{(t)}-\mathbf D^{(t)}\bigr)
\right\}.
$
\For{$m=1,\ldots,M$}
\State Update the block $\mathbf Z_{\mathcal G_m\cdot}$ via group soft-thresholding
$
\mathbf Z^{(t+1)}_{\mathcal G_m\cdot}
=
\mathcal S_{\rho w_m/\delta}
\!\left(
\bTheta^{(t+1)}_{\mathcal G_m\cdot}
+
\mathbf D^{(t)}_{\mathcal G_m\cdot}
\right),
$
where
$
\mathcal S_{\tau}(\mathbf V)
=
\left(1-\frac{\tau}{\|\mathbf V\|_F}\right)_+\mathbf V
$
with the convention $\mathcal S_\tau(0)=0$.
\EndFor
\State Update the scaled dual variable:
$
\mathbf D^{(t+1)}
=
\mathbf D^{(t)}+\bTheta^{(t+1)}-\mathbf Z^{(t+1)}.
$
\end{algorithmic}

\textit{Output:}
$\widehat{\bTheta}$, and hence $\widehat{\bB}$ through the case-specific reconstruction above.
\end{algorithm}
Define the primal and dual residuals by
$
\mathbf r^{(t+1)}=\bTheta^{(t+1)}-\mathbf Z^{(t+1)}
$
and
$
\mathbf s^{(t+1)}=\delta\bigl(\mathbf Z^{(t+1)}-\mathbf Z^{(t)}\bigr).
$
We terminate the iterations when
$
\|\mathbf r^{(t+1)}\|_F\le \epsilon_{\mathrm{pri}}
$
and
$
\|\mathbf s^{(t+1)}\|_F\le \epsilon_{\mathrm{dual}}.
$

\paragraph{Choice of $\delta$.}
The ADMM parameter $\delta > 0$ does not affect the optimizer of
\eqref{eq:structured-regression}, but it influences convergence speed and numerical
stability. In the $\bTheta$-update, a larger $\delta$ improves conditioning of
$\mathbf K_\delta$, which is beneficial when $\widetilde{\bX}^\top\widetilde{\bX}$
is ill-conditioned. In the $\mathbf Z$-update, a larger $\delta$ gives a smaller
proximal threshold $\rho w_m/\delta$ for that iteration, but this changes only the
ADMM trajectory, not the limiting optimizer. Overall, $\delta$ controls the balance
between the primal and dual residuals. In all experiments, we use $\delta=1$, though
adaptive residual-balancing schemes can lead to faster convergence.

\paragraph{Computational cost.}
For fixed $\delta$, the matrices $\mathbf K_\delta$ and $\mathbf C$ are constant
across iterations and may be precomputed once. With a dense factorization of
$\mathbf K_\delta$, the one-time cost is
$
O(n\tilde p^2+\tilde p^3+n\tilde p q),
$
and each subsequent $\bTheta$-update costs $O(\tilde p^2q)$. The $\mathbf Z$- and
$\mathbf D$-updates cost $O(\tilde p q)$, so the overall per-iteration cost is
$O(\tilde p^2q)$ after factorization.

When $\tilde p\gg n$, the Woodbury identity gives
$$
\mathbf K_\delta^{-1}
=
\frac{1}{\delta}\mathbf I_{\tilde p}
-
\frac{2}{n\delta^2}\widetilde{\bX}^\top
\left(
\mathbf I_n+\frac{2}{n\delta}\widetilde{\bX}\widetilde{\bX}^\top
\right)^{-1}
\widetilde{\bX}.
$$
Thus it suffices to factorize an $n\times n$ matrix. In that case the
precomputation cost is
$O(n^2\tilde p+n^3+n\tilde p q)$,
and each $\bTheta$-update costs
$O(n^2q+n\tilde p q)$.
This is preferable, for example, in graph-sparse CCAR$^3$ when $\tilde p=|E|\gg n$.

\section{Proofs: ordinary CCAR$^3$}\label{appendix:proof}
\subsection{Proof of Lemma~\ref{lemma:cca:rrr}}
\label{proof:lemma:cca:rrr}


\begin{proof}
Using the change of variables $\mathbf{V}_0 = \wh{\mathbf{\Sigma}}_Y^{\frac 12} \mathbf{V}$ and $\mathbf{Y}_0 = \mathbf{Y} \wh{\mathbf{\Sigma}}_{Y}^{-\frac 12}$, and the orthonormality of the columns of $\mathbf{V}_0$ (i.e. $\mathbf{V}_0^\top \mathbf{V}_0 = \mathbf{I}_r$), we transform the objective in Equation \eqref{eq:CCA_pred} into
\begin{equation*}
\begin{split}
\| \mathbf{Y}\mathbf{V} - \mathbf{X}\mathbf{U} \|_F^2
&= \| \mathbf{Y}_0 \mathbf{V}_0 - \mathbf{X}\mathbf{U} \|_F^2  \\
&= \underbrace{\Tr (\mathbf{V}_0^\top \mathbf{Y}_0^\top \mathbf{Y}_0 \mathbf{V}_0)}_{=rn}
   -2 \Tr(\mathbf{U}^\top \mathbf{X}^\top \mathbf{Y}_0 \mathbf{V}_0)
   + \underbrace{\Tr (\mathbf{U}^\top \mathbf{X}^\top \mathbf{X}\mathbf{U})}_{=rn}  \\
&=\underbrace{\Tr (\mathbf{Y}_0^\top\mathbf{Y}_0)}_{=qn}
   -2 \Tr (\mathbf{V}_0 \mathbf{U}^\top \mathbf{X}^\top\mathbf{Y}_0 )
   +  \underbrace{\Tr (\mathbf{V}_0\mathbf{U}^\top \mathbf{X}^\top \mathbf{X}\mathbf{U}\mathbf{V}_0^\top)}_{=rn}
   + (r-q)n\\
&=\| \mathbf{Y}_0 - \mathbf{X}\mathbf{U}\mathbf{V}_0^\top \|_F^2+ (r-q)n .
\end{split}
\end{equation*}
The additive constant $(r-q)n$ does not affect the minimizer, which proves the claim.
\end{proof}

\subsection{Proof of Theorem~\ref{theorem:cca:rrr}}
\label{proof:theorem:cca:rrr}

\begin{proof}
By the Law of Large Numbers, since \(p\) and \(q\) are fixed,
\[
\widehat{\mathbf\Sigma}_X\xrightarrow{a.s.}\mathbf\Sigma_X,
\qquad
\widehat{\mathbf\Sigma}_Y\xrightarrow{a.s.}\mathbf\Sigma_Y,
\qquad
\widehat{\mathbf\Sigma}_{XY}\xrightarrow{a.s.}\mathbf\Sigma_{XY}.
\]
Because \(\mathbf\Sigma_X\) and \(\mathbf\Sigma_Y\) are positive definite, the sample
covariance matrices are positive definite for all sufficiently large \(n\), almost surely.
Moreover, by continuity of the inverse and inverse-square-root maps on the positive
definite cone,
\[
\widehat{\mathbf\Sigma}_X^{-1}
\xrightarrow{a.s.}
\mathbf\Sigma_X^{-1},
\qquad
\widehat{\mathbf\Sigma}_Y^{-1/2}
\xrightarrow{a.s.}
\mathbf\Sigma_Y^{-1/2}.
\]
Therefore,
\[
\widehat{\mathbf B}
=
\widehat{\mathbf\Sigma}_X^{-1}
\widehat{\mathbf\Sigma}_{XY}
\widehat{\mathbf\Sigma}_Y^{-1/2}
\xrightarrow{a.s.}
\mathbf B^\star
:=
\mathbf\Sigma_X^{-1}
\mathbf\Sigma_{XY}
\mathbf\Sigma_Y^{-1/2}.
\]
Under the canonical pair model,
\[
\mathbf B^\star
=
\mathbf\Sigma_X^{-1}
\mathbf\Sigma_X\mathbf U\mathbf\Lambda
\mathbf V^\top\mathbf\Sigma_Y
\mathbf\Sigma_Y^{-1/2}
=
\mathbf U\mathbf\Lambda \mathbf V_0^\top,
\qquad
\mathbf V_0:=\mathbf\Sigma_Y^{1/2}\mathbf V .
\]
Since
\[
\mathbf V_0^\top\mathbf V_0
=
\mathbf V^\top\mathbf\Sigma_Y\mathbf V
=
\mathbf I_r,
\]
the columns of \(\mathbf V_0\) are orthonormal. We also note that \(\mathbf U\) has full column rank
because \(\mathbf U^\top\mathbf\Sigma_X\mathbf U=\mathbf I_r\). Hence
\[
\operatorname{rank}(\mathbf B^\star)=r.
\]
Moreover,
\[
\sigma_r(\mathbf B^\star)
=
\sigma_r(\mathbf U\mathbf\Lambda\mathbf V_0^\top)
\ge
\sigma_r(\mathbf U)\lambda_r
>
0.
\]
Thus the \(r\)-dimensional left and right singular subspaces of \(\mathbf B^\star\) are
separated from the null singular subspace by a positive gap.

Let
\(
\mathbf B^\star
=
\mathbf U_B\mathbf \Lambda_B\mathbf V_B^\top
\)
be the rank-$r$ singular value decomposition of \(\mathbf B^\star\), where
\(\mathbf U_B\in\mathbb R^{p\times r}\) and
\(\mathbf V_B\in\mathbb R^{q\times r}\) have orthonormal columns. Since
\[
\mathbf B^\star=\mathbf U\mathbf\Lambda\mathbf V_0^\top,
\]
we have
\[
\operatorname{span}(\mathbf U_B)
=
\operatorname{span}(\mathbf U),
\qquad
\operatorname{span}(\mathbf V_B)
=
\operatorname{span}(\mathbf V_0).
\]
Because both \(\mathbf V_B\) and \(\mathbf V_0\) have orthonormal columns and span the same
subspace, there exists an orthogonal matrix \(\mathbf Q_Y\in\mathcal O_r\) such that
\[
\mathbf V_B=\mathbf V_0\mathbf Q_Y.
\]
Similarly, define
\[
\mathbf U_\diamond
:=
\mathbf U_B
\left(
\mathbf U_B^\top\mathbf\Sigma_X\mathbf U_B
\right)^{-1/2}.
\]
Then
\[
\mathbf U_\diamond^\top\mathbf\Sigma_X\mathbf U_\diamond
=
\mathbf I_r,
\]
and \(\operatorname{span}(\mathbf U_\diamond)=\operatorname{span}(\mathbf U)\).
Since \(\mathbf U\) is also \(\mathbf\Sigma_X\)-orthonormal, there exists
\(\mathbf Q_X\in\mathcal O_r\) such that
\[
\mathbf U_\diamond=\mathbf U\mathbf Q_X.
\]

To prove consistency, we apply a standard Davis--Kahan/Wedin singular-subspace perturbation bound to
\(\widehat{\mathbf B}\) and \(\mathbf B^\star\). Since
\(\widehat{\mathbf B}\to\mathbf B^\star\) in Frobenius norm almost surely and
\(\sigma_r(\mathbf B^\star)>0\), there exist sequences
\(\mathbf O_n,\widetilde{\mathbf O}_n\in\mathcal O_r\) such that
\[
\|
\widehat{\mathbf U}_{\hat B}-\mathbf U_B\mathbf O_n
\|_F
\xrightarrow{a.s.}0,
\qquad
\|
\widehat{\mathbf V}_{\hat B}-\mathbf V_B\widetilde{\mathbf O}_n
\|_F
\xrightarrow{a.s.}0.
\]

We show that this implies the right singular-vector convergence to the estimator
\(\widehat{\mathbf V}\). Since \(\mathbf V_B=\mathbf V_0\mathbf Q_Y\) and
\(\mathbf V_0=\mathbf\Sigma_Y^{1/2}\mathbf V\), we have
\[
\mathbf\Sigma_Y^{-1/2}\mathbf V_B
=
\mathbf V\mathbf Q_Y.
\]
Therefore,
\begin{align*}
\|
\widehat{\mathbf V}
-
\mathbf V\mathbf Q_Y\widetilde{\mathbf O}_n
\|_F
&=
\|
\widehat{\mathbf\Sigma}_Y^{-1/2}\widehat{\mathbf V}_{\hat B}
-
\mathbf\Sigma_Y^{-1/2}\mathbf V_B\widetilde{\mathbf O}_n
\|_F \\
&\le
\|
\widehat{\mathbf\Sigma}_Y^{-1/2}
\|_{op}
\|
\widehat{\mathbf V}_{\hat B}-\mathbf V_B\widetilde{\mathbf O}_n
\|_F \\
&\quad+
\|
\widehat{\mathbf\Sigma}_Y^{-1/2}
-
\mathbf\Sigma_Y^{-1/2}
\|_{op}
\|
\mathbf V_B
\|_F
\xrightarrow{a.s.}0.
\end{align*}
Since \(\mathbf Q_Y\widetilde{\mathbf O}_n\in\mathcal O_r\), this implies
\[
\min_{\widetilde{\mathbf O}\in\mathcal O_r}
\|
\widehat{\mathbf V}-\mathbf V\widetilde{\mathbf O}
\|_F
\xrightarrow{a.s.}0.
\]

We now turn to the proof of consistency for $\mathbf U$. Let
\[
\widehat{\mathbf A}_n
:=
\widehat{\mathbf U}_{\hat B}^\top
\widehat{\mathbf\Sigma}_X
\widehat{\mathbf U}_{\hat B},
\qquad
\mathbf A
:=
\mathbf U_B^\top
\mathbf\Sigma_X
\mathbf U_B.
\]
Since
\[
\widehat{\mathbf U}_{\hat B}-\mathbf U_B\mathbf O_n
\xrightarrow{a.s.}0
\quad\text{and}\quad
\widehat{\mathbf\Sigma}_X\xrightarrow{a.s.}\mathbf\Sigma_X,
\]
we have
\[
\widehat{\mathbf A}_n
-
\mathbf O_n^\top\mathbf A\mathbf O_n
\xrightarrow{a.s.}0.
\]
The matrix \(\mathbf A\) is positive definite. Hence, for all sufficiently large \(n\),
\(\widehat{\mathbf A}_n\) is also positive definite almost surely, and by continuity of the
inverse-square-root map,
\[
\widehat{\mathbf A}_n^{-1/2}
-
\mathbf O_n^\top\mathbf A^{-1/2}\mathbf O_n
\xrightarrow{a.s.}0.
\]
Equivalently,
\[
\mathbf O_n\widehat{\mathbf A}_n^{-1/2}
-
\mathbf A^{-1/2}\mathbf O_n
\xrightarrow{a.s.}0.
\]
Now
\[
\widehat{\mathbf U}
=
\widehat{\mathbf U}_{\hat B}\widehat{\mathbf A}_n^{-1/2},
\qquad
\mathbf U_\diamond
=
\mathbf U_B\mathbf A^{-1/2}.
\]
Thus,
\begin{align*}
\|
\widehat{\mathbf U}
-
\mathbf U_\diamond\mathbf O_n
\|_F
&=
\|
\widehat{\mathbf U}_{\hat B}\widehat{\mathbf A}_n^{-1/2}
-
\mathbf U_B\mathbf A^{-1/2}\mathbf O_n
\|_F \\
&\le
\|
\widehat{\mathbf U}_{\hat B}-\mathbf U_B\mathbf O_n
\|_F
\|
\widehat{\mathbf A}_n^{-1/2}
\|_{op} \\
&\quad+
\|
\mathbf U_B\
\|_{op}
\|
\mathbf O_n\widehat{\mathbf A}_n^{-1/2}
-
\mathbf A^{-1/2}\mathbf O_n
\|_F
\xrightarrow{a.s.}0.
\end{align*}
Since \(\mathbf U_\diamond=\mathbf U\mathbf Q_X\), we obtain
\[
\|
\widehat{\mathbf U}
-
\mathbf U\mathbf Q_X\mathbf O_n
\|_F
\xrightarrow{a.s.}0.
\]
Because \(\mathbf Q_X\mathbf O_n\in\mathcal O_r\), it follows that
\[
\min_{\mathbf O\in\mathcal O_r}
\|
\widehat{\mathbf U}-\mathbf U\mathbf O
\|_F
\xrightarrow{a.s.}0.
\]
This completes the proof.
\end{proof}

\section{Proofs: sparse CCAR$^3$}\label{appendix:proof_sparse}

 \subsection{Useful lemmas}
This subsection provides a list of the technical lemmas adapted from \cite{gao2017sparse} and \cite{gao2015minimax} that we used in this manuscript to derive our theoretical results.

\begin{lemma}[Adaptation of Lemma 3 in \cite{gao2015minimax}]
\label{lem:cov} 
Assume $ \frac{s_u+q}{n} < c_0$ for  $c_0\in (0, 1)$. Consider a deterministic set $T_u \in [p]$ with $|T_u| =  s_u$.
For any constant $C'>0$ and assuming that $c_0$ is small enough, 
there exists a constant $C> 0$ that depends solely on $C'$ and $c_0$, such that:
 \begin{equation}
     \begin{split}
\opnorm{ [\widehat{\mathbf{\Sigma}}_X]_{T_uT_u}  - [\mathbf{\Sigma}_X]_{T_uT_u}}^2 \leq \frac{C M^2 }{n}s_u \\
\opnorm{ [\widehat{\mathbf{\Sigma}}_{X}]^{\frac12}_{T_uT_u}  - [{\mathbf{\Sigma}}_{X}]^{\frac12}_{T_uT_u}}^2 \leq \frac{CM^2  }{n}s_u \\
\opnorm{ \widehat{\mathbf{\Sigma}}_{Y}  - {\mathbf{\Sigma}}_{Y}}^2 \leq \frac{CM^2}{n} q\\
\opnorm{ \widehat{\mathbf{\Sigma}}^{\frac12}_{Y}  - {\mathbf{\Sigma}}^{\frac12}_{Y}}^2 \leq \frac{C M^2  }{n}q \\
     \end{split} 
 \end{equation}
 with probability at least $1 - \exp\{-C's_u \} - \exp\{-C'q\}. $
\end{lemma}
\begin{proof}
    We adapt Lemma 3 of \cite{gao2015minimax} to explicit the dependency on $M$. Note that the set $T_u$  considered here is deterministic. 
    Therefore, by Lemma 6.5 of \cite{wainwright2019high}, there are positive universal constants $\{c_j\}_{j=0}^3$ such that, for any $\delta \in (0,1)$:
    $$ \mathbb{P}\left[\frac{\opnorm{ [\widehat{\mathbf{\Sigma}}_X]_{T_uT_u}  - [\mathbf{\Sigma}_X]_{T_uT_u}}}{\opnorm{\mathbf{\Sigma}_X}} > c_1\Big(\sqrt{\frac{s_u}{n}}+{\frac{s_u}{n}}\Big) + \delta\right] \leq c_2\exp\big\{-c_3n (\delta \wedge \delta^2)\big\}.$$
    Denote $c_2 = e^{s_u \gamma_2}$ and choose $\delta^2 = \frac{s_u}{n} \frac{C' + \gamma_2}{c_3 }$. Assuming that $c_0$ is small enough (i.e. $c_0 \leq \frac{c_3 }{C' + \gamma_2}$) so that $\delta \leq 1$ (and thus, $\delta^2 \leq \delta$)  we have: 
      $$ \mathbb{P}\left[\opnorm{[\widehat{\mathbf{\Sigma}}_X]_{T_uT_u}  - [\mathbf{\Sigma}_X]_{T_uT_u}} > c_1'M\sqrt{\frac{s_u}{n}}\right] \leq \exp\{-Cs_u\}$$
      for a choice of $c_1'=2c_1+\sqrt{\frac{C' + \gamma_2 }{c_3}}\leq 2c_1 + \frac{1}{\sqrt{c_0}}$.
      The same result holds for $\mathbf{\Sigma}_Y$ replacing $s_u$ by $q$.
      Finally, for the operator norm of the square roots, by Lemma 2 of the supplementary material of \cite{gao2015minimax}, we have:
      $$ \opnorm{[\widehat{\mathbf{\Sigma}}_{X}]^{\frac12}_{T_uT_u}  - [{\mathbf{\Sigma}}_{X}]^{\frac12}_{T_uT_u}} \leq 
      \frac{\opnorm{[\widehat{\mathbf{\Sigma}}_X]_{T_uT_u}  - [\mathbf{\Sigma}_X]_{T_uT_u}}}{\sigma_{\min}\Big([\widehat{\mathbf{\Sigma}}_{X}]^{\frac12}_{T_uT_u}\Big) +\sigma_{\min}\Big( [{\mathbf{\Sigma}}_{X}]^{\frac12}_{T_uT_u}\Big)
    }$$
      Let us assume for now  that $\opnorm{\mathbf{\Sigma}_X}=1,$ and take $M=1.$
      Let $\mathcal{A}_X$ denote the event $\mathcal{A}_X =\Big\{ \opnorm{[\widehat{\mathbf{\Sigma}}_X]_{T_uT_u}  - [\mathbf{\Sigma}_X]_{T_uT_u}} \leq  c_1'\sqrt{\frac{s_u}{n}}\Big\}$, with $c'_1$ as previously defined. Then, on $\mathcal{A}_X$, by Weyl's inequality \citep{giraud2021introduction},
$$\sigma_{\min}\Big([\widehat{\mathbf{\Sigma}}_{X}]^{\frac12}_{T_uT_u}\Big) = \sigma_{\min}\Big([\widehat{\mathbf{\Sigma}}_{X}]_{T_uT_u}\Big)^{\frac12}\geq \sqrt{\sigma_{\min}\Big( [{\mathbf{\Sigma}}_{X}]_{T_uT_u}\Big) - c_1' \sqrt{\frac{s_u}{n}}}.$$     
Therefore, as long as $c_1' \sqrt{\frac{s_u}{n}} \leq \frac{3}{4} \sigma_{\min}( \mathbf{\Sigma}_{X})$, which happens as soon as $c_0 \leq \left(\frac{\frac{3}{4}\sigma_{\min}(\mathbf{\Sigma}_X)}{c'_1}\right)^2$, we have:
      $$ \opnorm{[\widehat{\mathbf{\Sigma}}_{X}]^{\frac12}_{T_uT_u}  - [{\mathbf{\Sigma}}_{X}]^{\frac12}_{T_uT_u}}
      \leq \frac{c_1' \sqrt{\frac{s_u}{n}}}{ \frac32\sigma_{\min}\Big( [{\mathbf{\Sigma}}_{X}]^{\frac12}_{T_uT_u}\Big)}.$$
      Scaling the previous inequality by $\opnorm{\mathbf{\Sigma}_X^{\frac12}}$, and {using the fact that $\sigma_{\min}\Big( \mathbf{\Sigma}_{X}^{\frac12}\Big) \geq \frac{1}{\sqrt{M}}$, we have:
      $$ \opnorm{[\widehat{\mathbf{\Sigma}}_{X}]^{\frac12}_{T_uT_u}  - [{\mathbf{\Sigma}}_{X}]^{\frac12}_{T_uT_u}} \leq M{c_1'' \sqrt{\frac{s_u}{n}}}.$$}
      The proof for $\mathbf{\Sigma}_Y^{\frac12}$ is similar and considers instead the event 
      $$\mathcal{A}_Y =\Big\{ \opnorm{\widehat{\mathbf{\Sigma}}_Y -  \mathbf{\Sigma}_Y} > c_1'\sqrt{\frac{q}{n}}\Big\}.$$
\end{proof}
The following lemma is a direct consequence of the previous lemma by applying a union bound.
\begin{lemma}[Adapted from Lemma 12 in \cite{gao2015minimax}]\label{lem:sparsespec}
Let $S_u$ be a set of indices of size $|S_u|=s_u$. 
Assume $\frac{q+s_u\log\left(\frac{ep}{s_u}\right)}{n}  < c_0$ for some constant $c_0\in (0,1)$. For any constant $C'>0$, there exists some constant $C>0$ only depending on $C'$ and $c_0$ such that:
$$\opnorm{ [\wh{\mathbf{\Sigma}}_{X}]_{S_uS_u} - [{\mathbf{\Sigma}}_{X}]_{S_uS_u}}^2 \leq \frac{CM^2}{n} s_u\log\left(\frac{ep}{s_u}\right) $$
$$ \opnorm{ [\wh{\mathbf{\Sigma}}_{X}]^{\frac12}_{S_uS_u} - [{\mathbf{\Sigma}}_{X}]^{\frac12}_{S_uS_u}}^2 \leq \frac{CM^{2}}{n} s_u \log\left(\frac{ep}{s_u}\right)$$
$$ \opnorm{ {\widehat{\mathbf{\Sigma}}_{Y}} - {\mathbf{\Sigma}}_{Y} }^2 \leq \frac{CM^{2}}{n} q$$
$$ \opnorm{ {\widehat{\mathbf{\Sigma}}_{Y}}^{\frac12} - {\mathbf{\Sigma}}^{\frac12}_{Y} }^2 \leq \frac{CM^{2}}{n} q$$
with probability at least $ 1  -\exp\{ -C' q\}  -\exp\big\{ -C' s_u \log\big(\frac{ep}{s_u}\big)\big\}$.
\end{lemma}

\begin{proof}
Bounding over all possible sets $T_u \subset [p]$ with cardinality at most $s_u$, we obtain:
    \begin{equation}
        \begin{split}
    \mathbb{P}\left[\opnorm{ [\wh{\mathbf{\Sigma}}_{X}]_{S_uS_u} - [{\mathbf{\Sigma}}_{X}]_{S_uS_u}} \geq t\right]   
    &\leq   \mathbb{P}\left[\max_{T_u \in [p]:~|T_u| =s_u}\opnorm{ [\wh{\mathbf{\Sigma}}_{X}]_{T_uT_u} - [{\mathbf{\Sigma}}_{X}]_{T_uT_u}} \geq t\right] \\   
     &\leq   \sum_{T_u \in [p]} \mathbb{P}\left[\opnorm{ [\wh{\mathbf{\Sigma}}_{X}]_{T_uT_u} - [{\mathbf{\Sigma}}_{X}]_{T_uT_u}} \geq t\right] \\ 
          &\leq  {p\choose{s_u}}\mathbb{P}\left[\opnorm{ [\wh{\mathbf{\Sigma}}_{X}]_{T_uT_u} - [{\mathbf{\Sigma}}_{X}]_{T_uT_u}} \geq t\right] 
        \end{split}
    \end{equation}
Therefore, noting that ${p\choose{s_u}}\leq \left(\frac{ep}{s_u}\right)^{s_u} $, for any $\delta>0$, by  the proof of Lemma~\ref{lem:cov}:
    \begin{equation}
        \begin{split}
    \mathbb{P}\left[\opnorm{[\wh{\mathbf{\Sigma}}_{X}]_{S_uS_u} - [{\mathbf{\Sigma}}_{X}]_{S_uS_u}} \geq  \opnorm{\mathbf{\Sigma}_X} \Bigg(c_1\Big(\sqrt{\frac{s_u}{n}}+{\frac{s_u}{n}}\Big) + \delta\Bigg)\right] \\
    \leq  c_2\exp\Big\{s_u\log\left(\frac{ep}{s_u}\right)-c_3n(\delta\wedge \delta^2)\Big\} 
        \end{split}
    \end{equation}
    Writing $c_2=e^{\gamma_2 s_u\log\left(\frac{ep}{s_u}\right)}$ for some constant $\gamma_2$, we choose $\delta^2 = \frac{C'+1 +  (\gamma_2)_+}{c_3}\cdot\frac{s_u \log\left(\frac{ep}{s_u}\right)}{n}$ for a constant $C'>0$.
    Assuming that $c_0$ is small enough so that $\delta \leq 1$, which happens as soon as $c_0 \leq \frac{c_3}{C' + 1 + (\gamma_2)_+}$, we have:
    \begin{equation*}
        \begin{split}
    &\mathbb{P}\Bigg[\opnorm{ [\wh{\mathbf{\Sigma}}_{X}]_{S_uS_u} - [{\mathbf{\Sigma}}_{X}]_{S_uS_u}} \geq  \opnorm{\mathbf{\Sigma}_X} \Bigg(c_1\Big(\sqrt{\frac{s_u}{n}}+{\frac{s_u}{n}}\Big) + \sqrt{\frac{c_3+1+( \gamma_2)_+}{c_3}\cdot\frac{s_u \log\big(\frac{ep}{s_u}\big)}{n}}\Bigg)\Bigg]\\  
          &\leq  \exp\Big\{-\big(C'+1+(\gamma_2)_+-1-\gamma_2\big)~s_u\log\left(\frac{ep}{s_u}\right)\Big\}
                  \end{split}
    \end{equation*}
  implying that      
        \begin{equation*}
        \begin{split}
           \mathbb{P}\Bigg[\opnorm{ [\wh{\mathbf{\Sigma}}_{X}]_{S_uS_u} - [{\mathbf{\Sigma}}_{X}]_{S_uS_u}} \geq  c_1'\opnorm{\mathbf{\Sigma}_X}\sqrt{\frac{s_u \log\big(\frac{ep}{s_u}\big)}{n}}\Bigg] 
          &\leq  \exp\Big\{-C's_u\log\left(\frac{ep}{s_u}\right)\Big\} \\ 
        \end{split}
    \end{equation*}
for $c_1' = 2c_1 + \sqrt{\frac{C'+1 + (\gamma_2)_+}{c_3}} \leq 2 c_1 +\frac{1}{\sqrt{c_0}}.$
The proof for the square root matrices is identical to the proof of the previous lemma (Lemma ~\ref{lem:cov}), and also requires that $c_0$ is small enough in the sense that $c_0 \leq \left(\frac{\frac{3}{4}\sigma_{\min}(\mathbf{\Sigma}_X)}{c'_1}\right)^2$.
\end{proof}


\begin{lemma}[Adapted from Lemma 6.7 of \cite{gao2017sparse}] \label{lem:ep}
Assume $\frac{{q}+\log p}{n}\leq c_0$ for some sufficiently small constant $c_0\in(0,1)$. For $C' >0$, there exists a constant $C>0$ only depending on $C'$ and $c_0$ such that
$$\max_{j\in[p]}\Big\|\big[\wh{\mathbf{\Sigma}}_{XY}\wh{\mathbf{\Sigma}}_Y^{-\frac12}-\wh{\mathbf{\Sigma}}_X\mathbf{B}^*\big]_{j\cdot}\Big\|_2\leq C  M^2 \sqrt{\frac{q + \log p}{n} },$$
with probability at least 
$${1 -\exp\{-C' q\}-\exp\{-C' s_u\log(ep/s_u)\} -\exp\{-C' (q +\log p)\}}-\exp\{- C'(r +\log p)\}.$$
\end{lemma}

\begin{proof}
Recall that we defined $\mathbf{B}^*$ as
$ \mathbf{B}^* = \mathbf{U}\mathbf{\Lambda} \mathbf{V}^\top \mathbf{\Sigma}_Y^{\frac12}$. By definition of $\mathbf{B}^*$, we know that ${\mathbf{\Sigma}_X \mathbf{B}^* = \mathbf{\Sigma}_{XY} \mathbf{\Sigma}_Y^{-\frac12}}$. We thus have 
\begin{equation}\label{ineq}
    \begin{split}
\max_{j\in[p]}\Big\|\big[\wh{\mathbf{\Sigma}}_{XY}\wh{\mathbf{\Sigma}}_Y^{-\frac 12}&-\wh{\mathbf{\Sigma}}_X\mathbf{B}^*\big]_{j\cdot}\Big\|\leq \underbrace{\max_{j\in[p]}\Big\|\big[(\wh{\mathbf{\Sigma}}_{XY}-\mathbf{\Sigma}_{XY})\wh{\mathbf{\Sigma}}_Y^{-\frac 12}\big]_{j\cdot}\Big\|}_{(A)}\\&+\underbrace{\max_{j\in[p]}\Big\|\big[({\mathbf{\Sigma}}_{XY}(\wh{\mathbf{\Sigma}}_Y^{-\frac 12}- {\mathbf{\Sigma}}_Y^{-\frac 12})\big]_{j\cdot}\Big\|}_{(B)}
+\underbrace{\max_{j\in[p]}\Big\|\big[\wh{\mathbf{\Sigma}}_X\mathbf{B}^*-\mathbf{\Sigma}_X \mathbf{B}^*\big]_{j\cdot}\Big\|}_{(C)}
\end{split}
\end{equation}
For a given constant $C'>0$, let $\mathcal{A}$ be the event:
  \begin{equation}
     \begin{split}
    \mathcal{A}= \Big\{
\opnorm{[\wh{\mathbf{\Sigma}}_{X}]^{\frac12}_{S_uS_u} - [\mathbf{\Sigma}_{X}]^{\frac12}_{S_uS_u}}^2 &\leq \frac{CM^2}{n} s_u\\
  \opnorm{ {\widehat{\mathbf{\Sigma}}_{Y}}^{\frac12} - {\mathbf{\Sigma}}^{\frac12}_{Y} }^2 &\leq \frac{CM^2}{n} q \Big\}\\
     \end{split} 
 \end{equation}
 where $S_u =\text{supp}(\mathbf{U}),$ and where the constant $C$ was chosen appropriately such that, by Lemma~\ref{lem:cov}, we know that $\mathbb{P}[\mathcal{A}] \geq 1-\exp\{ -C' q\}  -\exp\{ -C' s_u\}$.
 
\xhdr{Bound for (A)}
We note that
\begin{equation*}
    \begin{split}
        \max_{j\in[p]}\Big\|\big[(\wh{\mathbf{\Sigma}}_{XY}-\mathbf{\Sigma}_{XY})\wh{\mathbf{\Sigma}}_{Y}^{-\frac12}\big]_{j\cdot}\Big\| &=   \max_{j\in[p]}\Big\|\big[(\wh{\mathbf{\Sigma}}_{XY}-\mathbf{\Sigma}_{XY})\mathbf{\Sigma}_Y^{-\frac12} \mathbf{\Sigma}_Y^{\frac12}\wh{\mathbf{\Sigma}}_{Y}^{-\frac12}\big]_{j\cdot}\Big\| \\
        &\leq    \max_{j\in[p]}\Big\|\big[(\wh{\mathbf{\Sigma}}_{XY}-\mathbf{\Sigma}_{XY})\mathbf{\Sigma}_Y^{-\frac12}\big]_{j\cdot}\Big\|~ \opnorm{\mathbf{\Sigma}_Y^{\frac12}\wh{\mathbf{\Sigma}}_{Y}^{-\frac12}} \\
    \end{split}
\end{equation*}
By Weyl's theorem \citep{giraud2021introduction}, we know that on the event $\mathcal{A}$:
$$\forall i \in [q], \quad  
\big| \sigma_i\big( \widehat{\mathbf{\Sigma}}^{\frac12}_Y\big) -\sigma_i\big( \mathbf{\Sigma}^{\frac12}_Y\big)\big| \leq \opnorm{ \widehat{\mathbf{\Sigma}}^{\frac12}_Y -  \mathbf{\Sigma}^{\frac12}_Y} \leq \sqrt{ CM^2\frac{q}{n}}  . $$
Therefore, writing $\delta_{q,n} =\sqrt{\frac{q}{n}} $ and assuming $ \sqrt{C}M \delta_{q,n} \leq \frac{1}{2}\sigma_{\min}\big(\mathbf{\Sigma}_Y^{\frac12}\big)$, we have:
\begin{equation}
\label{ineq:sigma_y_sqrt_inv}
    \opnorm{\wh{\mathbf{\Sigma}}^{-\frac12}_Y} = \frac{1}{\sigma_{\min}\big( \widehat{\mathbf{\Sigma}}^{\frac12}_Y\big)} \leq \frac{1}{\sigma_{\min}\big( \mathbf{\Sigma}^{\frac12}_Y\big) -\sqrt{C}M\delta_{q,n}}  \leq \frac{2}{\sigma_{\min}\big( \mathbf{\Sigma}^{\frac12}_Y\big)}.
\end{equation}


\noindent Let $\mathbf{Z}_i \sim \mathcal{N}_p(0, \mathbf{\Sigma}_X)$, $\widetilde{\mathbf{Y}}_i \sim \mathcal{N}_q(0, \mathbf{I}_q)$, and $\mathbf{1}_j^\top \in \R^{1 \times p}$ the vector indicator of row $j$ (i.e. $[\mathbf{1}_j]_i=0$ for all $i\neq j$, and  $[\mathbf{1}_j]_j=1$). Thus, we may write
$$\mathbf{1}_j^\top\big(\wh{\mathbf{\Sigma}}_{XY}\mathbf{\Sigma}_Y^{-\frac12}-\mathbf{\Sigma}_{XY}\mathbf{\Sigma}_Y^{-\frac12}\big)=\frac{1}{n}\sum_{i=1}^n\big(\mathbf{1}_j^\top \mathbf{Z}_i\widetilde{\mathbf{Y}}_i^\top-\mathbb{E}(\mathbf{1}_j^\top \mathbf{Z}_i\widetilde{\mathbf{Y}}_i^\top)\big).$$
Define for each $j \in [p]$ the matrix 
$\mathbf{H}_i^{(j)}=\begin{pmatrix}
\mathbf{1}_j^\top \mathbf{Z}_i \\
\widetilde{\mathbf{Y}}_i
\end{pmatrix}.$
Since $\mathbf{1}_j^\top \mathbf{Z}_i\widetilde{\mathbf{Y}}_i^\top$ is a submatrix of $\mathbf{H}_i^{(j)}(\mathbf{H}_i^{(j)})^{\top}$,   we have:
\begin{equation}
    \begin{split}
        \max_{j\in[p]}  \Big\|\mathbf{1}_j^\top(\wh{\mathbf{\Sigma}}_{XY}\mathbf{\Sigma}_Y^{-\frac12}-\mathbf{\Sigma}_{XY}\mathbf{\Sigma}_Y^{-\frac12})\Big\| &\leq \max_{j\in[p]}\opnorm{\frac{1}{n}\sum_{i=1}^n \big(\mathbf{H}_i^{(j)}\big(\mathbf{H}_i^{(j)}\big)^\top-\mathbb{E}\mathbf{H}_i^{(j)}\big(\mathbf{H}_i^{(j)}\big)^\top\big)}\\
    \end{split}
\end{equation}
Denoting $\mathbf{\mathcal{H}}^{(j)}=\mathbb{E}\mathbf{H}_i^{(j)}\big(\mathbf{H}_i^{(j)}\big)^\top  = \begin{pmatrix}
\mathbf{1}_j^\top  \mathbf{\Sigma}_X \mathbf{1}_j & \mathbf{1}_j^\top  \mathbf{\Sigma}_{XY}\mathbf{\Sigma}_{Y}^{-\frac12}\\
\mathbf{\Sigma}_{Y}^{-\frac12}\mathbf{\Sigma}_{YX} \mathbf{1}_j & \mathbf{I}_q
\end{pmatrix} $, this implies:
\begin{equation}
    \begin{split}
        \opnorm{\mathbf{\mathcal{H}}^{(j)}} & = \sup_{\substack{u_1\in \R,~u_2 \in \R^{q}\\ u_1^2 + \|u_2\|^2=1}} \mathbf{1}_j^\top  \mathbf{\Sigma}_X\mathbf{1}_j u_1^2 + 2 u_1 \mathbf{1}_j^\top  \mathbf{\Sigma}_{XY} \mathbf{\Sigma}_Y^{-\frac12} u_2 + {u_2^\top \mathbf{I}_q u_2}\\
        & \leq \sup_{\substack{u_1\in \R,~u_2 \in \R^{q}\\ u_1^2 + \|u_2\|^2=1}}Mu_1^2 + 2  \lambda_1 \sqrt{M}_1 u_1 \| u_2\|+  \|u_2\|^2\\
        & \leq \sup_{\substack{u_1\in \R,~u_2 \in \R^{q}\\ u_1^2 + \|u_2\|^2=1}} M u_1^2 +   \lambda_1\sqrt{M} (u_1^2  + \| u_2\|^2)+  \|u_2\|^2\\
        &= M(1 + \lambda_1) \leq 2M ,  \\
    \end{split}
\end{equation}
since by assumption $\opnorm{\mathbf{\Sigma}_X}\leq M$ and $\lambda_1 \leq 1.$
By Theorem 6.5 of \cite{wainwright2019high}, there exists universal positive constants $\{c_j\}_{j=1}^3$ such that, for any $t>0$:
$$ \mathbb{P}\Bigg[\opnorm{\frac{1}{n}\sum_{i=1}^n \big(\mathbf{H}_i^{(j)}\big(\mathbf{H}_i^{(j)}\big)^\top -\mathbb{E}\mathbf{H}_i^{(j)}\big(\mathbf{H}_i^{(j)}\big)^\top \big)} \geq  4Mc_1\sqrt{\frac{1+q}{n}} + 2Mt\Bigg] \leq c_2\exp\big\{ -c_3n (t \wedge t^2)\big\}.  $$
Thus, by a simple union bound, we conclude that:
$$ \mathbb{P}\Bigg[\max_{j \in [p]} \opnorm{\frac{1}{n}\sum_{i=1}^n \big(\mathbf{H}_i^{(j)}\big(\mathbf{H}_i^{(j)}\big)^\top -\mathbb{E}\mathbf{H}_i^{(j)}\big(\mathbf{H}_i^{(j)}\big)^\top \big)} \geq  4Mc_1\sqrt{\frac{1+q}{n}} + 2Mt\Bigg] \leq pc_2\exp\big\{ -c_3n (t \wedge t^2)\big\}.$$
We write $c_2 = e^{\gamma_2 (q + \log p)} $, for $\gamma_2$ a constant. 
Let $C'>0$ be a constant and taking ${t^2 = \frac{C'+1+(\gamma_2)_+}{c_3} \cdot \frac{q + \log p}{n}}$, and assuming that $c_0$ is small enough (i.e. $c_0\leq \frac{c_3}{C'+1+(\gamma_2)_+} $)  so that $ t \leq 1$, the previous equation implies
\begin{align}
    \begin{split} 
    \mathbb{P}\Bigg[\max_{j \in [p]} &\opnorm{\frac{1}{n}\sum_{i=1}^n \big(\mathbf{H}_i^{(j)}\big(\mathbf{H}_i^{(j)}\big)^\top -\mathbb{E}\mathbf{H}_i^{(j)}\big(\mathbf{H}_i^{(j)}\big)^\top \big)} \\
    &\geq 4Mc_1\sqrt{\frac{1+q}{n}} +2M\sqrt{\frac{C'+1+(\gamma_2)_+}{c_3} \cdot \frac{q + \log p}{n}}\Bigg] \\
    &\leq \exp\big\{\gamma_2 (q + \log p) + \big(1-C'-1-(\gamma_2)_+\big)\log p  -\big(C'+1 + (\gamma_2)_+\big)q\big\}\\
    &= \exp\big\{ -C'\log p  -(C'+1)q\big\}\\
    &\leq \exp\big\{ -C'(\log p +q))\big\}.\\
            \end{split}
        \end{align}
Therefore, we can choose $c_1' = 4c_1 + 2\sqrt{\frac{C'+1+(\gamma_2)_+}{c_3}}$ such that, as soon as $p\geq3$:
\begin{equation}
    \begin{split} 
    &\mathbb{P}\Bigg[\max_{j \in [p]} \opnorm{\frac{1}{n}\sum_{i=1}^n (\mathbf{H}_i^{(j)}\big(\mathbf{H}_i^{(j)}\big)^\top -\mathbb{E}\mathbf{H}_i^{(j)}\big(\mathbf{H}_i^{(j)}\big)^\top )} \geq c_1'M\sqrt{\frac{q + \log p}{n} }\Bigg] \\
    &\leq\exp\big\{- C'(q + \log p)\big\} 
            \end{split}
        \end{equation}
Let us denote as $\mathcal{A}_1$ the event: 
$$\mathcal{A}_1 = \Bigg\{\max_{j \in [p]} \opnorm{\frac{1}{n}\sum_{i=1}^n (\mathbf{H}_i^{(j)}\big(\mathbf{H}_i^{(j)}\big)^\top -\mathbb{E}\mathbf{H}_i^{(j)}\big(\mathbf{H}_i^{(j)}\big)^\top )} \geq c_1'M\sqrt{\frac{q + \log p}{n} }\Bigg\}.$$
Therefore, with probability at least $1-\exp\{- C'(q + \log p)\}- \exp\{- C'q\}-\exp\{- C's_u\},$ the event $\mathcal{A}\cap \mathcal{A}_1^c$ occurs and
\begin{equation}\label{eq:bound_sigaxy}
    \begin{split}
   \max_{j\in [p]}\Big\|\big[(\wh{\mathbf{\Sigma}}_{XY}-\mathbf{\Sigma}_{XY})\wh{\mathbf{\Sigma}}_{Y}^{-\frac12}\big]_{j\cdot}\Big\| \leq M  c_1' \sqrt{\frac{q + \log p}{n} } \cdot \frac{\sigma_{\max}\big( \mathbf{\Sigma}^{\frac12}_Y\big)}{\sigma_{\min}\big( \mathbf{\Sigma}^{\frac12}_Y\big)}
   \leq M^2  c_1' \sqrt{\frac{q + \log p}{n} }
    \end{split}
    \end{equation}

\xhdr{Bound (B)} We have, on the event $\mathcal{A}$ (and provided that $c_0$ is small enough, in the sense of the proof of Lemma~\ref{lem:ep}):
   \begin{equation}
       \begin{split}
           \max_{j\in [p]}\Big\|\big[({\mathbf{\Sigma}}_{XY}(\wh{\mathbf{\Sigma}}_Y^{-\frac 12}- {\mathbf{\Sigma}}_Y^{-\frac 12})\big]_{j\cdot}\Big\|&\leq \max_{j\in [p]}\Big\|\big[{\mathbf{\Sigma}}_{XY}{\mathbf{\Sigma}}_Y^{-\frac 12}\big]_{j\cdot}\Big\|~  \opnorm{ {\mathbf{\Sigma}}_Y^{\frac 12} \wh{\mathbf{\Sigma}}_Y^{-\frac 12}- \mathbf{I}_q}\\
           &\leq \opnorm{{\mathbf{\Sigma}}_{XY}{\mathbf{\Sigma}}_Y^{-\frac12}}~  \opnorm{ {\mathbf{\Sigma}}_Y^{\frac 12}- \wh{\mathbf{\Sigma}}_Y^{\frac 12}}~\opnorm{\wh{\mathbf{\Sigma}}_Y^{-\frac 12}}.\\
           &\leq 
\sqrt{M}\lambda_1\cdot  \sqrt{C}M   \sqrt{\frac{q}{n}} \cdot \frac{2}{\sigma_{\min}\big( \mathbf{\Sigma}^{\frac12}_Y\big)} \\
  &= 
 2\sqrt{C}  \sqrt{\frac{q}{n}} M^{\frac32}\cdot \frac{\lambda_1}{\sigma_{\min}\big( \mathbf{\Sigma}^{\frac12}_Y\big)} 
       \end{split}
   \end{equation}
 Here we have used Inequality (\ref{ineq:sigma_y_sqrt_inv}) to bound $\opnorm{\widehat{\mathbf{\Sigma}}_Y^{-\frac12}}$ on $\mathcal{A}$.
We thus obtain:
    \begin{equation}
       \begin{split}
           \max_{j\in [p]}\Big\|\big[({\mathbf{\Sigma}}_{XY}(\wh{\mathbf{\Sigma}}_Y^{-\frac 12}- {\mathbf{\Sigma}}_Y^{-\frac 12})\big]_{j\cdot}\Big\|&\leq 
 2M^2\sqrt{C}  \lambda_1 \cdot \sqrt{\frac{q}{n}}  
       \end{split}
   \end{equation}

\xhdr{Bound (C)} By the proof of Lemma 6.7 in \cite{gao2017sparse}:
\begin{equation*}
    \begin{split}
        \max_{j\in [p]}\Big\|\big[(\wh{\mathbf{\Sigma}}_X-\mathbf{\Sigma}_X)\mathbf{B}^*\big]_{j\cdot}\Big\| &= \max_{j\in [p]}\Big\|\big[(\wh{\mathbf{\Sigma}}_X-\mathbf{\Sigma}_X) \mathbf{U}\mathbf{\Lambda} \mathbf{V}^\top  \mathbf{\Sigma}_Y^{\frac 12}\big]_{j\cdot}\Big\| \\
        &\leq  \max_{j\in [p]}\frac{\Big\|\big[(\wh{\mathbf{\Sigma}}_X-\mathbf{\Sigma}_X) \mathbf{U}\big]_{j\cdot}\Big\|}{\opnorm{\mathbf{\Sigma}_X}}\cdot \opnorm{\mathbf{\Sigma}_X}~\opnorm{\mathbf{\Lambda} \mathbf{V}^\top  \mathbf{\Sigma}_Y^{\frac 12}} \\
                &\leq  C_2 \sqrt{\frac{r + \log p}{n}} M\lambda_1  \\
    \end{split}
\end{equation*}
with probability at least $1 -\exp\{- C'(r +\log p)\},$ for $C'$ a constant that can be chosen to be arbitrarily large, and $C_2$ a constant that depends solely on $C'$.
Let 
$$\mathcal{A}_2 = \Bigg\{  \max_{j\in [p]}\Big\|\big[(\wh{\mathbf{\Sigma}}_X-\mathbf{\Sigma}_X)\mathbf{B}^*\big]_{j\cdot}\Big\| \geq  C_2 \lambda_1 M\sqrt{\frac{r + \log p}{n}}\Bigg\}.$$

\xhdr{Combined bound (A) + (B) + (C)}
Therefore, since $r<q$, and $\lambda<1$, on the event $\mathcal{A}\cap \mathcal{A}^c_1 \cap \mathcal{A}_2^c$, Inequality (\ref{ineq}) becomes:
\begin{equation}
    \begin{split}
\max_{j\in [p]}\Big\|\big[\wh{\mathbf{\Sigma}}_{XY}\wh{\mathbf{\Sigma}}_Y^{-\frac 12}-\wh{\mathbf{\Sigma}}_X\mathbf{B}^*\big]_{j\cdot}\Big\|&\leq c_1' M^2 \sqrt{\frac{q + \log p}{n} } +   2\sqrt{C} M^2 \sqrt{\frac{q}{n}}  + {C}_5 M  \sqrt{\frac{r + \log p}{n}}\\
&\leq \widetilde{C} M^2 \sqrt{\frac{q + \log p}{n} }\\
    \end{split}
\end{equation}
 for an appropriate choice of~$\widetilde{C} = 3(c_1' \vee 2\sqrt{C} \vee C_2)$.
 This occurs
with probability at least ${1 -\exp\{-C' q\}-\exp\{-C' s_u\log(ep/s_u)\} -\exp\{-C' (q +\log p)\}}-\exp\{- C'(r +\log p)\}.$  

\end{proof}


\subsection{Proof of Theorem~\ref{theorem:rrr_ols}}
\label{proof:theorem:rrr_ols}


\begin{proof}
The proof {of} Theorem~\ref{theorem:rrr_ols} is based on the same regression argument as before; the only population object that must
be used throughout is
\[
\mathbf B^*=\mathbf U\mathbf\Lambda\mathbf V^\top\mathbf\Sigma_Y^{1/2}
             =\mathbf U\mathbf\Lambda\mathbf V_0^\top,
\]
because $\mathbf Y_0=\mathbf Y\widehat{\mathbf\Sigma}_Y^{-1/2}$. In particular, under our model assumptions:
\[
\mathbf\Sigma_X\mathbf B^*
=\mathbf\Sigma_X\mathbf U\mathbf\Lambda\mathbf V^\top\mathbf\Sigma_Y^{1/2}
=\mathbf\Sigma_{XY}\mathbf\Sigma_Y^{-1/2}.
\]
Let $S_u=\supp(\mathbf B^*)=\supp(\mathbf U)$ and define the high-probability {score} event
\[
\mathcal G=\left\{
\max_{j\in[p]}
\left\|\left[\widehat{\mathbf\Sigma}_{XY}\widehat{\mathbf\Sigma}_Y^{-1/2}
      -\widehat{\mathbf\Sigma}_X\mathbf B^*\right]_{j\cdot}\right\|_2
\le C M^2\sqrt{\frac{q+
\log p}{n}}
\right\}.
\]
 let $\mathcal{A}$ be the event:
  \begin{equation}
     \begin{split}
    \mathcal{A}= \Big\{
{\opnorm{ [\widehat{\mathbf{\Sigma}}_X]_{S_uS_u} - [\mathbf{\Sigma}_X]_{S_uS_u}}^2} &\leq \frac{CM^2}{n} s_u\\
\opnorm{ {\widehat{\mathbf{\Sigma}}_{Y}} - {\mathbf{\Sigma}}_{Y} }^2 &\leq \frac{CM^2}{n} q\\
  \opnorm{ {\widehat{\mathbf{\Sigma}}_{Y}}^{\frac12} - {\mathbf{\Sigma}}^{\frac12}_{Y} }^2 &\leq \frac{CM^2}{n} q \Big\}\\
     \end{split} 
 \end{equation}
{By Lemma~\ref{lem:cov} and the uniform sparse-spectrum event $\mathcal E_{\rm sp}$ of Lemma~\ref{lem:sparsespec} used below, since the set $S_u$ is deterministic, we may work on an event $\mathcal{A}\cap\mathcal E_{\rm sp}$ satisfying
\[
\mathbb{P}[\mathcal{A}\cap\mathcal E_{\rm sp}] \geq 1-\exp\{ -C' q\}  -\exp\{ -C' s_u\log(ep/s_u)\}.
\]}

\xhdr{Bound on ${\fnorm{\widehat{\mathbf{\Sigma}}_X^{\frac 12}\mathbf\Delta}}$ and $\|\mathbf{\Delta}\|_F$}
By the Basic Inequality:
$$\frac{1}{n}\Big\| \mathbf{Y}_0 - \mathbf{X} \widehat{\mathbf{B}}\Big\|_F^2 + \rho \|\widehat{\mathbf{B}}\Big\|_{21} \leq     \frac{1}{n}\Big\| \mathbf{Y}_0 - \mathbf{X} {\mathbf{B}}^* \|_F^2 + \rho \|{\mathbf{B}}^*\|_{21} $$
Therefore, writing $\mathbf{\Delta} = \wh{\mathbf B} - \mathbf{B}^*$:
\begin{equation*}
    \begin{split}
         \frac{1}{n}\big\|\mathbf{X} \mathbf{\Delta} \big\|_F^2 &\leq  \frac{2}{n}\langle \mathbf{X}^\top \mathbf{Y}_0 - \mathbf{X}^\top \mathbf{X} \mathbf{B}^*, \mathbf{\Delta} \rangle + \rho ( \|\mathbf{B}^*\|_{21} - \| \mathbf{\Delta} + \mathbf{B}^* \|_{21}) \\
      \fnorm{{\widehat{\mathbf{\Sigma}}_X^{\frac 12}} \mathbf{\Delta}}^2 &\leq  2 \langle \wh{\mathbf{\Sigma}}_{XY} \wh{\mathbf{\Sigma}}_Y^{-\frac 12} - \wh{\mathbf{\Sigma}}_X {\mathbf{B}}^*, \mathbf{\Delta} \rangle + \rho ( \|\mathbf{B}^*_{S_u\cdot}\|_{21} -  \|\mathbf{B}^*_{S_u\cdot} + \mathbf{\Delta}_{S_u\cdot}\|_{21}  - \| \mathbf{\Delta}_{S_u^c\cdot}\|_{21}) \\
         &\leq  2  \max_{j\in[p]}\Big\| \big[\wh{\mathbf{\Sigma}}_{XY} {\wh{\mathbf{\Sigma}}_Y^{-\frac 12}} - \wh{\mathbf{\Sigma}}_X {\mathbf{B}}^*\big]_{j\cdot}\Big\| \cdot  \sum_{j=1}^p  \|\mathbf{\Delta}_{j\cdot}\| + \rho ( \| \mathbf{\Delta}_{S_u\cdot}\|_{21}  - \| \mathbf{\Delta}_{S_u^c\cdot}\|_{21})  
    \end{split}
\end{equation*}
where the last line follows by an application of H{\"o}lder's inequality to tackle the inner product term, and the reverse triangle inequality for the $\ell_{21}$ terms.
If we select $\rho$ such that
\begin{equation}\label{eq:rho_u}
    \rho > 4  \max_{j\in[p]}\Big\| \big[\widehat{\mathbf{\Sigma}}_{XY} \wh{\mathbf{\Sigma}}_Y^{
      -\frac 12} - \wh{\mathbf{\Sigma}}_X {\mathbf{B}}^*\big]_{j\cdot}\Big\| 
\end{equation}
(and we will provide a value of $\rho$ for which this holds in a later paragraph), then
\begin{equation}\label{eq:cone1}
    \begin{split}
            \fnorm{\widehat{\mathbf{\Sigma}}_X^{\frac 12} \mathbf{\Delta} }^2 &\leq   \frac{3\rho}{2}  \| \mathbf{\Delta}_{S_u\cdot}\|_{21}  - \frac{\rho}{2}\| \mathbf{\Delta}_{S_u^c\cdot}\|_{21} \\  
    \end{split}
\end{equation}
Since $\sum_{j \in S_u}\|\mathbf{\Delta}_{j\cdot}\| \leq \Big(s_u\sum_{j \in S_u}\|\mathbf{\Delta}_{ j\cdot}\|^2\Big)^{\frac 12}$, the upper bound becomes
\begin{equation} \label{eq:Basic3}
    \begin{split}
            \fnorm{\widehat{\mathbf{\Sigma}}_X^{\frac 12} \mathbf{\Delta} }^2 &\leq   \frac{3\rho \sqrt{s_u} }{2}\Bigg(\sum_{j \in S_u}\|\mathbf{\Delta}_{j \cdot}\|^2\Bigg)^{\frac 12}
    \end{split}
\end{equation}
Note that Equation~(\ref{eq:Basic3}) also provides a cone inequality:
\begin{align}
\label{eq:cone}
\sum_{j \in S_u^c}\|\mathbf{\Delta}_{ j \cdot}\| \leq  3\sum_{j \in S_u}\|\mathbf{\Delta}_{ j \cdot}\|    
\end{align}
For an integer $t$, let the set $J_1 = \{j_1, \cdots, j_t\}$ in $S_u^c$ correspond to the $t$ rows with the largest $\ell_2$ norm in $\mathbf{\Delta}$, and define the extended support $\wt{S}_u = S_u \cup J_1$.  
To proceed, we adopt the peeling strategy of \cite{gao2017sparse}.  We partition $\wt{S}_u^c$ into disjoint subsets $J_2,...,J_K$ of size $t$ (with $|J_K|\leq t$), such that $J_k$ is the set of (row) indices corresponding to the {$t$ largest row $\ell_2$ norms} in $\mathbf{\Delta}$ outside $\wt{S}_u \bigcup_{j=2}^{k-1}J_j$. By the triangle inequality,
\begin{align*}
& \fnorm{\wh{\mathbf{\Sigma}}_X^{\frac 12}\mathbf{\Delta}}
\geq \fnorm{\wh{\mathbf{\Sigma}}_X^{\frac 12}\mathbf{\Delta}_{\wt{S}_u\cdot}}-
{\sum_{k=2}^K}\fnorm{\wh{\mathbf{\Sigma}}_X^{\frac 12}\mathbf{\Delta}_{J_k \cdot }} \\
&\geq \sqrt{{\sigma_{\min}^{\wh{\mathbf{\Sigma}}_X}(s_u+t)}}\cdot\fnorm{\mathbf{\Delta}_{\wt{S}_u\cdot}} -  \sqrt{{\sigma_{\max}^{\wh{\mathbf{\Sigma}}_X}(t)}}\cdot
 {\sum_{k=2}^K}\fnorm{\mathbf{\Delta}_{J_k\cdot}}.
\end{align*}
where $\sigma_{\min}^{\wh{\mathbf{\Sigma}}_X}(t+s_u)$ indicates the minimum eigenvalue of $\wh{\mathbf{\Sigma}}_X$ on any set of size at most $s_u +t$, while $\sigma_{\max}^{\wh{\mathbf{\Sigma}}_X}(t)$ denotes the maximum eigenvalue of $\wh{\mathbf{\Sigma}}_X$ on any set of size at most $t$.
By construction, we also have:
\begin{eqnarray}
\label{eq:start}\sum_{k= 2}^K\fnorm{\mathbf{\Delta}_{J_k\cdot}} &\leq& \sqrt{t}\sum_{k= 2}^K\max_{j\in J_k}\|\mathbf{\Delta}_{j\cdot}\| \leq \frac 1{\sqrt{t}}\sum_{k= 2}^K\sum_{j\in J_{k-1}}\|\mathbf{\Delta}_{j\cdot}\| \\
\nonumber &\leq& \frac 1{\sqrt{t}}\sum_{j\in S_u^c}\|\mathbf{\Delta}_{j\cdot}\| \leq \frac 3{\sqrt{t}}\sum_{j\in S_u}\|\mathbf{\Delta}_{j\cdot}\| \\
\label{eq:end}&\leq& 3\sqrt{\frac{s_u}{t}}\Bigg(\sum_{j\in S_u}\|\mathbf{\Delta}_{j\cdot}\|^2\Bigg)^{\frac12}\leq 3\sqrt{\frac{s_u}{t}}\fnorm{\mathbf{\Delta}_{\wt{S}_u\cdot}}.
\end{eqnarray}
In the above derivation, we have used the construction of $J_k$ and the cone condition (\ref{eq:cone}). Hence,
$\fnorm{ \widehat{\mathbf{\Sigma}}_X^{\frac 12}\mathbf{\Delta}}\geq \kappa \fnorm{\mathbf{\Delta}_{\wt{S}_u\cdot}}$
with $\kappa=\sqrt{{\sigma_{\min}^{\wh{\mathbf{\Sigma}}_X}(s_u+t)}}-3\sqrt{\frac{s_u}{t}\sigma_{\max}^{\wh{\mathbf{\Sigma}}_X}(t)}$. 

In view of Lemma {\ref{lem:sparsespec}} {(on the event $\mathcal E_{\rm sp}$)} and by direct application of Weyl's inequality \citep{giraud2021introduction}, taking $t=c_1s_u$ for some sufficiently large constant $c_1$, {on the event $\mathcal E_{\rm sp}$}, $\kappa$ can be lower-bounded by a positive constant $\kappa_0$ only depending on $\sigma_{\min}(\mathbf{\Sigma}_X)$, $c_0$ and~$c_1$:
\begin{equation*}
\begin{split}
    \kappa &\geq  \sqrt{ \frac{1}{M}} - M\sqrt{{C\frac{(c_1+1)s_u\log\big(\frac{ep}{(c_1+1)s_u}\big)}{n}}}-3\sqrt{ \frac{1}{c_1}} \cdot \Bigg(\sqrt{M}+ M\sqrt{C\frac{c_1s_u\log\big(\frac{ep}{c_1s_u}\big)}{n}}\Bigg)\\
    &\geq  \sqrt{ \frac{1}{M}} -3\sqrt{\frac{M}{c_1}} - M\left(\sqrt{{C\frac{(c_1+1)s_u\log\big(\frac{ep}{s_u}\big)}{n}}}+3\sqrt{C\frac{s_u\log\big(\frac{ep}{s_u}\big)}{n}}\right)\\
    &\geq  \frac{1}{2\sqrt{ M}} -M\left(\sqrt{{C\frac{(c_1+1)s_u\log\big(\frac{ep}{s_u}\big)}{n}}}+3\sqrt{C\frac{s_u\log\big(\frac{ep}{s_u}\big)}{n}}\right) \\
    &{\quad \text{choosing $c_1$ sufficiently large depending only on $M$}}\\
    &\geq  \frac{1}{2\sqrt{ M}} {- M\sqrt{Cc_0}\left(\sqrt{c_1+1} +3\right)}\\
     &{\geq   \frac{1}{4\sqrt{ M}}\qquad \text{for $c_0$ sufficiently small, depending only on $M$ and $c_1$.}}  \\
    \end{split}  
\end{equation*}
Now, assuming  that {$c_1$ is chosen as above} and $c_0$ is small enough, 
$$ \kappa \geq \gamma_0 \sqrt{\sigma_{\min}(\mathbf{\Sigma}_X)} \geq \gamma_0 \frac{1}{\sqrt{M}}$$
where $\gamma_0\geq\frac{1}{4}$ is a constant.
Combined with (\ref{eq:Basic3}), this lower bound means that we have:
\begin{equation}
\fnorm{\mathbf{\Delta}_{\wt{S}_u\cdot}}\leq {C\frac{\rho\sqrt{s_u}}{2\kappa_0^2}} \leq {C\frac{\rho\sqrt{s_u}}{2\gamma_0^2 \sigma_{\min}(\mathbf{\Sigma}_X)}}. \label{eq:Sum1}
\end{equation}
By (\ref{eq:start})-(\ref{eq:end}), we have
\begin{equation}
\fnorm{\mathbf{\Delta}_{\wt{S}_u^c\cdot}}\leq 
 {\sum_{k=2}^K}
\fnorm{\mathbf{\Delta}_{J_k\cdot}}\leq 3\sqrt{\frac{s_u}{t}}\fnorm{\mathbf{\Delta}_{\wt{S}_u\cdot}}\leq \frac{3}{\sqrt {c_1}}\fnorm{\mathbf{\Delta}_{\wt{S}_u\cdot}}. \label{eq:Sum2}
\end{equation}
Summing (\ref{eq:Sum1}) and (\ref{eq:Sum2}), we have $\|{\mathbf{\Delta}}\|_F\leq C_0\frac{\sqrt{s_u}\rho}{\sigma_{\min}(\mathbf{\Sigma}_X)}$,
therefore, Equation~(\ref{eq:Basic3}) becomes
\begin{equation}
    \begin{split}
            \fnorm{\widehat{\mathbf{\Sigma}}_X^{\frac 12} \mathbf{\Delta} }^2 &\leq   C_0\frac {3\rho^2s_u}{2\sigma_{\min}(\mathbf{\Sigma}_X)} \leq   C_0\frac {3M\rho^2s_u}{2}
    \end{split}
\end{equation}
where $C_0$ is a generic constant that depends on $\gamma_0$.

{By Lemma \ref{lem:ep}, we may choose $\rho \geq M^2 C_u\sqrt{\frac{q+\log p}{n}}$ for some large $C_u$ so that (\ref{eq:rho_u}) holds on the score event $\mathcal{G}$ defined above. We work on $\mathcal{A}\cap\mathcal E_{\rm sp}\cap\mathcal{G}$; by a union bound, this event has at least the probability stated in the theorem.}
In this case:
\begin{equation}\label{eq:ols_proof_final}
    \begin{split}
            \fnorm{\widehat{\mathbf{\Sigma}}_X^{\frac 12} \mathbf{\Delta} }^2 &\leq   C_0M^5 \frac{3C_u^2 s_u }{2} \cdot \frac{q+\log p}{n}\leq   C'M^5 s_u\cdot \frac{q+\log p}{n}\\
        \|{\mathbf{\Delta}}\|_F&\leq C''M^3 \sqrt{\frac{s_u(q+\log p)}{n}}
    \end{split}
\end{equation}
with high probability, with $C' = \frac{3}{2} C_0 C_u^2$, and $C'' = C_0 C_u$. {The second line implies $\|\mathbf\Delta\|_F^2\leq (C'')^2M^6s_u(q+\log p)/n$.} This concludes the proof of Equation~\ref{eq:reg_bound_app}.

\xhdr{Bound on $|\supp{\wh{\mathbf B}}|$}
We note that, by the KKT conditions for all $j$, it holds
    $$  \frac{2}{n}\big[\mathbf{X}^\top (\mathbf{X}\wh{\mathbf B} - \mathbf{Y}_0) \big]_{j \cdot} {\in -\rho\,\partial \|\wh{\mathbf B}_{j \cdot}\|_2}$$
This implies that:
    \begin{equation}
        \begin{cases}   
        \frac{1}{n}\Big\| \big[\mathbf{X}^\top (\mathbf{X}\wh{\mathbf B} - \mathbf{Y}_0)\big]_{j\cdot}\Big\|
                = \frac{\rho}{2} \qquad \text{ if } \qquad  \big\|\wh{\mathbf B}_{j \cdot}\big\| \neq  0\\
              \frac{1}{n}   \Big\| \big[\mathbf{X}^\top (\mathbf{X}\wh{\mathbf B} - \mathbf{Y}_0)\big]_{j\cdot}\Big\|       \leq \frac{\rho}{2}   \qquad \text{ if }  \qquad \big\|\wh{\mathbf B}_{j \cdot}\big\| = 0\\
            \end{cases}
    \end{equation}
Let $T = \supp{\wh{\mathbf B}}$. The following inequality therefore holds for any $j \in T$ {on $\mathcal G$}:
    \begin{equation}
           \frac{1}{n}  \Big\| \big[\mathbf{X}^\top (\mathbf{X} \mathbf{\Delta})\big]_{j \cdot}\Big\| \geq   \frac{1}{n}\Big\| \big[\mathbf{X}^\top (\mathbf{X} \wh{\mathbf B} -\mathbf{Y}_0)\big]_{j \cdot}\Big\| -  \frac{1}{n}\Big\| \big[\mathbf{X}^\top (\mathbf{Y}_0 - \mathbf{X}\mathbf{B}^*)\big]_{j \cdot}\Big\|\geq  \frac{\rho}{4}
    \end{equation}
    Thus
    $$ \sum_{j \in \supp{\wh{\mathbf B}}}  \frac{1}{n^2}  \Big\| \big[\mathbf{X}^\top (\mathbf{X} \mathbf{\Delta})\big]_{j \cdot}\Big\|^2 
            \geq   \frac{\rho^2}{16}   |T|.$$
Let 
$$\widehat{M} = \max_{J \subset [p]:~|J| = |T|}\opnorm{[\widehat{\mathbf{\Sigma}}_{X}]_{JJ}}.$$  {Let $\mathbf X_T$ denote the submatrix of $\mathbf X$ with columns indexed by $T$.} The previous inequality implies
\begin{equation}
        \begin{split}
            |T| &\leq {\frac{16}{\rho^2}\frac{1}{n^2}\Big\|\mathbf X_T^\top\mathbf X\mathbf\Delta\Big\|_F^2
            \leq \frac{16}{\rho^2}\frac{\opnorm{\mathbf X_T^\top\mathbf X_T}}{n^2}\|\mathbf X\mathbf\Delta\|_F^2}\\
&{\leq \frac{16 \widehat{M}}{\rho^2}  \fnorm{ \wh{\mathbf{\Sigma}}_X^{\frac12} \mathbf{\Delta}}^2
\leq \frac{16 \widehat{M}}{\rho^2}  C_0 \frac{3M\rho^2s_u }{2} \leq C_1 \widehat{M} M s_u}\\
        \end{split}
    \end{equation}
    for a choice of constant {$C_1$} large enough.
{By Weyl's inequality, combined with the uniform sparse-spectrum event in Lemma \ref{lem:sparsespec}, we have, simultaneously for all relevant sets $J$,
$$\widehat{M} \leq  M + CM \sqrt{ |T| \frac{\log\big(\frac{ep}{|T|}\big)}{{n}}}.$$}
This yields
$${     |T| \leq C_1 M^2 s_u + C_1 M^2s_u \sqrt{    |T|\frac{\log\big(\frac{ep}{|T|}\big)}{n}}.}$$
{Solving this one-dimensional inequality on the same sparse-spectrum event gives
$$     |T| \leq C M^2s_u,$$
provided $c_0$ in $s_u\log(ep/s_u)/n\le c_0$ is sufficiently small. Using that $x/\log(ep/x)$ is increasing on the relevant range, if $|T|>2C_1M^2s_u$, then the preceding display implies
$$|T|\leq 4C_1^2M^4s_u^2\frac{\log(ep/|T|)}{n},$$
which contradicts $|T|>CM^2s_u$ for a sufficiently large constant $C$ under the same smallness assumption.}
Therefore, {on the same event as above}, we obtain:
\begin{equation}
     |\supp{\wh{\mathbf B}}| \lesssim {M^2} s_u.
 \end{equation}
\end{proof}

\subsection{Proof of Theorem~\ref{theorem:rrr_ols_2}}
\label{proof:theorem:rrr_ols_2}

\begin{proof}
{
The first part of this argument follows the exact same path as the proof of Theorem~\ref{theorem:cca:rrr}.
Let
\[
\mathbf V_0:=\mathbf\Sigma_Y^{1/2}\mathbf V,
\qquad
\mathbf B^*:=\mathbf U\mathbf\Lambda\mathbf V_0^\top,
\qquad
S:=\supp(\mathbf B^*)=\supp(\mathbf U).
\]
Then $\mathbf V_0^\top\mathbf V_0=\mathbf I_r$ and
\[
\operatorname{col}(\mathbf B^*)=\operatorname{col}(\mathbf U),
\qquad
\operatorname{row}(\mathbf B^*)=\operatorname{col}(\mathbf V_0).
\]
Let
\[
\mathbf B^*=\mathbf U_B\mathbf D_B\mathbf V_B^\top
\]
be the compact singular value decomposition of $\mathbf B^*$, where
$\mathbf U_B$ and $\mathbf V_B$ have orthonormal columns. Since
$\mathbf V_0$ is orthonormal and spans the same subspace as $\mathbf V_B$, there exists
$\mathbf Q_Y\in\mathcal O_r$ such that
\[
\mathbf V_B=\mathbf V_0\mathbf Q_Y.
\]
Similarly, define
\[
\mathbf U_\diamond
:=
\mathbf U_B(\mathbf U_B^\top\mathbf\Sigma_X\mathbf U_B)^{-1/2}.
\]
Then $\mathbf U_\diamond^\top\mathbf\Sigma_X\mathbf U_\diamond=\mathbf I_r$ and
$\operatorname{col}(\mathbf U_\diamond)=\operatorname{col}(\mathbf U)$. Since
$\mathbf U^\top\mathbf\Sigma_X\mathbf U=\mathbf I_r$, there exists
$\mathbf Q_X\in\mathcal O_r$ such that
\begin{equation}\label{eq:Udiamond_equals_U_current_estimator}
\mathbf U_\diamond=\mathbf U\mathbf Q_X.
\end{equation}
}

{
Let $\mathbf{\Delta}:=\widehat{\mathbf B}-\mathbf B^*$.  We first make the probability bookkeeping explicit. Fix the target constant $C'>0$ in the theorem statement and apply the concentration bounds below with an auxiliary constant $C_\star>C'$ chosen sufficiently large; after the final union bound we relabel the resulting exponent by $C'$.
}

 Let ${\mathcal{A}_Y}$ denote the event:
  \begin{equation}\label{eq:event_a}
     \begin{split}
    {
    \mathcal{A}_Y= \Big\{
\opnorm{ \widehat{\mathbf{\Sigma}}^{\frac12}_{Y}  - {\mathbf{\Sigma}}^{\frac12}_{Y}}^2 \leq \frac{C M^2q }{n},\quad
\opnorm{\widehat{\mathbf\Sigma}_Y^{-1/2}}\le 2\sqrt M,\quad
\opnorm{ \widehat{\mathbf\Sigma}_Y^{-1/2}-\mathbf\Sigma_Y^{-1/2}}
\le C M^2\sqrt{\frac qn}
\Big\}.}
     \end{split} 
 \end{equation}
 {
By Lemma~\ref{lem:cov} and the identity
\[
\widehat{\mathbf\Sigma}_Y^{-1/2}-\mathbf\Sigma_Y^{-1/2}
=
\widehat{\mathbf\Sigma}_Y^{-1/2}
(\mathbf\Sigma_Y^{1/2}-\widehat{\mathbf\Sigma}_Y^{1/2})
\mathbf\Sigma_Y^{-1/2},
\]
we have $\mathbb P(\mathcal A_Y^c)\le \exp\{-C_\star q\}$, provided $c_0$ is sufficiently small.
}

 Let $\mathcal{G}$ denote the event:
 {
 \[
 \mathcal{G} =\Big \{  \Big\|\widehat{\bS}_X^{\frac 12} \mathbf{\Delta} \Big\|_F^2 \leq   C_1 M^5 s_u \frac{q+\log p}{n},\quad
 \|\widehat \bB-\bB^*\|_F^2\le C_2M^6s_u\frac{q+\log p}{n},\quad
 |\supp(\widehat{\mathbf B})|\le C_JM^2s_u\Big\}.
 \]
By Theorem~\ref{theorem:rrr_ols}, applied with exponent $C_\star$, 
\[
\mathbb P(\mathcal G^c)
\le
\exp\{-C_\star q\}
+\exp\left\{-C_\star s_u\log\left(\frac{ep}{s_u}\right)\right\}
+\exp\{-C_\star(q+\log p)\}
+\exp\{-C_\star(r+\log p)\}.
\]
Let $s_J:=\lceil (C_JM^2+1)s_u\rceil$ and define the uniform sparse-covariance event
\[
\mathcal H_X
=
\left\{
\max_{J\subset[p]: |J|\le s_J}
\opnorm{[\widehat{\mathbf\Sigma}_X-\mathbf\Sigma_X]_{JJ}}
\le C_4M^2\sqrt{\frac{s_u\log(ep/s_u)}{n}}
\right\}.
\]
By Lemma~\ref{lem:cov} and a union bound over all subsets of size at most $s_J$,
\[
\mathbb P(\mathcal H_X^c)
\le
\exp\{-C_\star s_J\log(ep/s_J)\}
\le
\exp\left\{-C_\star' s_u\log\left(\frac{ep}{s_u}\right)\right\},
\]
where the last inequality uses that $M$ is treated as an absolute constant. Hence, for
\[
\mathcal E:=\mathcal G\cap\mathcal A_Y\cap\mathcal H_X,
\]
choosing $C_\star$ large enough and decreasing constants if necessary gives
\[
\mathbb P(\mathcal E)
\ge
1 -\exp\{-C' q\}-\exp\left\{-C' s_u\log\left(\frac{ep}{s_u}\right)\right\}-\exp\{-C' (q+\log p )\}-\exp\{-C' (r+\log p )\}.
\]
All bounds below are proved on the event $\mathcal E$.
}
\paragraph{Step 1: perturbation of the singular subspaces of $\widehat{\mathbf B}$.}
{
{By Theorem~\ref{theorem:rrr_ols}, on the event~$\mathcal E$,}
\begin{equation}\label{eq:B_error_current_estimator_t4}
\|\widehat{\mathbf B}-\mathbf B^*\|_F
\le
C_1M^{3}\sqrt{s_u\frac{q+\log p}{n}},
\qquad
|\supp(\widehat{\mathbf B})|\lesssim M^2s_u.
\end{equation}
Moreover,
\[
\sigma_r(\mathbf B^*)
=
\sigma_r(\mathbf U\mathbf\Lambda\mathbf V_0^\top)
\ge
\sigma_r(\mathbf U)\lambda_r
\ge
\frac{\lambda}{\sqrt M},
\]
{because $\mathbf V_0$ has orthonormal columns and
$\mathbf U^\top\mathbf\Sigma_X\mathbf U=\mathbf I_r$ with
$\|\mathbf\Sigma_X\|_{op}\le M$ implies
$\mathbf I_r=\mathbf U^\top\mathbf\Sigma_X\mathbf U\preceq M\mathbf U^\top\mathbf U$, hence
$\sigma_r(\mathbf U)\ge M^{-1/2}$.} For $c_0$ sufficiently small, the right-hand side of
\eqref{eq:B_error_current_estimator_t4} is smaller than a fixed fraction of
$\sigma_r(\mathbf B^*)$. Therefore Wedin's theorem (Lemma~\ref{theorem:wedin}) applied directly to
$\widehat{\mathbf B}$ and $\mathbf B^*$ gives orthogonal matrices
$\mathbf O_X,\mathbf O_Y\in\mathcal O_r$ such that
\begin{equation}\label{eq:current_estimator_direct_svd_subspace_bounds}
\|\widehat{\mathbf U}_{\widehat{B}}-\mathbf U_B\mathbf O_X\|_F
\vee
\|\widehat{\mathbf V}_{\widehat{B}}-\mathbf V_B\mathbf O_Y\|_F
\le
C_2\frac{M^{7/2}}{\lambda}
\sqrt{rs_u\frac{q+\log p}{n}}.
\end{equation}

}

\paragraph{Step 2: Error bound on $\widehat{\mathbf V}$.}
{
Since $\mathbf V_B=\mathbf V_0\mathbf Q_Y$ and
$\mathbf V_0=\mathbf\Sigma_Y^{1/2}\mathbf V$,
\[
\mathbf\Sigma_Y^{-1/2}\mathbf V_B
=
\mathbf V\mathbf Q_Y.
\]
{
Furthermore, on $\mathcal{A}_Y$,
\begin{equation}
    \opnorm{\widehat{\mathbf\Sigma}_Y^{-1/2}}\le 2\sqrt M,
    \qquad
    \opnorm{ \widehat{\mathbf\Sigma}_Y^{-1/2}-\mathbf\Sigma_Y^{-1/2}}
    \le C M^2\sqrt{\frac qn}.
\end{equation}
Thus, using \eqref{eq:current_estimator_direct_svd_subspace_bounds},
}
\begin{align*}
\|\widehat{\mathbf V}-\mathbf V\mathbf Q_Y\mathbf O_Y\|_F
&=
\|\widehat{\mathbf\Sigma}_Y^{-1/2}\widehat{\mathbf V}_{\widehat B}
 -\mathbf\Sigma_Y^{-1/2}\mathbf V_B\mathbf O_Y\|_F \\
&\le
\|\widehat{\mathbf\Sigma}_Y^{-1/2}\|_{op}
\|\widehat{\mathbf V}_{\widehat B}-\mathbf V_B\mathbf O_Y\|_F
+
\|\widehat{\mathbf\Sigma}_Y^{-1/2}-\mathbf\Sigma_Y^{-1/2}\|_{op}
\|\mathbf V_B\|_F \\
&\le
{2\sqrt M
\|\widehat{\mathbf V}_{\widehat B}-\mathbf V_B\mathbf O_Y\|_F
+
C M^2\sqrt{\frac qn}\,\|\mathbf V_B\|_F} \\
&\le
{C_3\frac{M^{11/2}}{\lambda}
\sqrt{rs_u\frac{q+\log p}{n}}.}
\end{align*}
Taking $\mathbf O=\mathbf Q_Y\mathbf O_Y\in\mathcal O_r$ proves the desired bound for
{$\widehat{\mathbf V}$ on the event $\mathcal E$.}
}

\paragraph{Step 3: Error bound on $\widehat{\mathbf U}$.}
{
Let
\[
\widehat{\mathbf A}
:=
\widehat{\mathbf U}_{\widehat B}^\top\widehat{\mathbf\Sigma}_X\widehat{\mathbf U}_{\widehat B},
\qquad
\mathbf A
:=
\mathbf U_B^\top\mathbf\Sigma_X\mathbf U_B.
\]
The matrix $\mathbf A$ is positive definite, with eigenvalues in $[1/M,M]$. Let
$J:=\supp(\widehat{\mathbf B})\cup S$. {On the event $\mathcal E$,
$|J|\le s_J$. Since the columns of $\widehat{\mathbf U}_{\widehat B}$ are supported on
$\supp(\widehat{\mathbf B})$ and the columns of $\mathbf U_B$ are supported on $S$, the event $\mathcal H_X$ gives}
\[
\|[\widehat{\mathbf\Sigma}_X-\mathbf\Sigma_X]_{JJ}\|_{op}
\le
C_4M^2\sqrt{\frac{s_u\log(ep/s_u)}{n}}.
\]
{
Together with \eqref{eq:current_estimator_direct_svd_subspace_bounds} and
$\opnorm{[\widehat{\mathbf\Sigma}_X]_{JJ}}\le C M$ on $\mathcal E$, this implies}
\begin{equation}\label{eq:Ahat_A_bound_current_estimator}
\|\widehat{\mathbf A}-\mathbf O_X^\top\mathbf A\mathbf O_X\|_{F}
\le
{C_5\frac{M^{9/2}}{\lambda}}
\sqrt{rs_u\frac{q+\log p}{n}}.
\end{equation}
{
Let
\[
\widetilde{\mathbf A}:=\mathbf O_X^\top \mathbf A\mathbf O_X .
\]
Since the eigenvalues of $\mathbf A$ lie in $[1/M,M]$, the same is true for
$\widetilde{\mathbf A}$. Moreover, by \eqref{eq:Ahat_A_bound_current_estimator}
and the smallness assumption on $c_0$, we have
\[
\lambda_{\min}(\widehat{\mathbf A})\ge \frac{1}{2M},
\qquad
\lambda_{\min}(\widetilde{\mathbf A})\ge \frac{1}{M}.
\]
In particular, for $c_0$ sufficiently small, $\widehat{\mathbf A}$ is therefore positive definite. 
We now use the standard inverse-square-root perturbation bound for positive definite
matrices: if $\mathbf H_1,\mathbf H_2\succ 0$ and
\[
\lambda_{\min}(\mathbf H_1)\wedge\lambda_{\min}(\mathbf H_2)\ge a,
\]
then
\[
\|\mathbf H_1^{-1/2}-\mathbf H_2^{-1/2}\|_F
\le
\frac{1}{2a^{3/2}}\|\mathbf H_1-\mathbf H_2\|_F.
\]
For completeness, this follows from the resolvent representation
\[
\mathbf H^{-1/2}
=
\frac{1}{\pi}\int_0^\infty t^{-1/2}(\mathbf H+t\mathbf I)^{-1}\,dt,
\]
which, using the identity $ \bA^{-1} - \bB^{-1} = 
\bA^{-1}(\bB-\bA)\bB^{-1}$ applied to $\bA = \mathbf H_1 + t\bI$ and $\bB = \mathbf H_2 +t\bI$, gives
\[
\mathbf H_1^{-1/2}-\mathbf H_2^{-1/2}
=
\frac{1}{\pi}\int_0^\infty
t^{-1/2}
(\mathbf H_1+t\mathbf I)^{-1}
(\mathbf H_2-\mathbf H_1)
(\mathbf H_2+t\mathbf I)^{-1}\,dt .
\]
Therefore,
\[
\begin{aligned}
\|\mathbf H_1^{-1/2}-\mathbf H_2^{-1/2}\|_F
&\le
\frac{1}{\pi}\|\mathbf H_1-\mathbf H_2\|_F
\int_0^\infty
\frac{t^{-1/2}}{(a+t)^2}\,dt  \\
&=
\frac{1}{2a^{3/2}}\|\mathbf H_1-\mathbf H_2\|_F .
\end{aligned}
\]
Applying this bound with
$\mathbf H_1=\widehat{\mathbf A}$,
$\mathbf H_2=\widetilde{\mathbf A}$, and $a=1/(2M)$ yields
\[
\|\widehat{\mathbf A}^{-1/2}-\widetilde{\mathbf A}^{-1/2}\|_F
\le
C M^{3/2}
\|\widehat{\mathbf A}-\widetilde{\mathbf A}\|_F .
\]
Finally, by orthogonal invariance of the functional calculus,
\[
\widetilde{\mathbf A}^{-1/2}
=
(\mathbf O_X^\top\mathbf A\mathbf O_X)^{-1/2}
=
\mathbf O_X^\top\mathbf A^{-1/2}\mathbf O_X .
\]
Hence
\[
\|\widehat{\mathbf A}^{-1/2}
   -\mathbf O_X^\top\mathbf A^{-1/2}\mathbf O_X\|_F
\le
C M^{3/2}
\|\widehat{\mathbf A}-\mathbf O_X^\top\mathbf A\mathbf O_X\|_F .
\]
Combining this with \eqref{eq:Ahat_A_bound_current_estimator} gives
\[
\|\widehat{\mathbf A}^{-1/2}
   -\mathbf O_X^\top\mathbf A^{-1/2}\mathbf O_X\|_F
\le
C_6\frac{M^{6}}{\lambda}
\sqrt{rs_u\frac{q+\log p}{n}},
\]
provided that \eqref{eq:Ahat_A_bound_current_estimator} is stated with the
corresponding $M^{9/2}/\lambda$ rate. 
}

By the
Lipschitz continuity of the inverse-square-root map on matrices whose eigenvalues are
bounded below by $1/(2M)$,
\begin{equation}\label{eq:Ahat_invsqrt_bound_current_estimator}
\|\widehat{\mathbf A}^{-1/2}
   -\mathbf O_X^\top\mathbf A^{-1/2}\mathbf O_X\|_{F}
\le
C_6\frac{M^{6}}{\lambda}
\sqrt{rs_u\frac{q+\log p}{n}}.
\end{equation}
Now
\[
\widehat{\mathbf U}=\widehat{\mathbf U}_{\widehat{B}}\widehat{\mathbf A}^{-1/2},
\qquad
\mathbf U_\diamond=\mathbf U_B\mathbf A^{-1/2}.
\]
Using \eqref{eq:current_estimator_direct_svd_subspace_bounds} and \eqref{eq:Ahat_invsqrt_bound_current_estimator},
\begin{align*}
\|\widehat{\mathbf U}-\mathbf U_\diamond\mathbf O_X\|_F
&=
\|\widehat{\mathbf U}_{\widehat{B}}\widehat{\mathbf A}^{-1/2}
   -\mathbf U_B\mathbf A^{-1/2}\mathbf O_X\|_F \\
&\le
\|\widehat{\mathbf U}_{\widehat{B}}-\mathbf U_B\mathbf O_X\|_F
\|\widehat{\mathbf A}^{-1/2}\|_{op}
+
\|\mathbf U_B\|_{op}
\|\mathbf O_X\widehat{\mathbf A}^{-1/2}-\mathbf A^{-1/2}\mathbf O_X\|_F \\
&\le
C_7\frac{M^{6}}{\lambda}
\sqrt{rs_u\frac{q+\log p}{n}}.
\end{align*}
Finally, by \eqref{eq:Udiamond_equals_U_current_estimator},
\[
\|\widehat{\mathbf U}-\mathbf U\mathbf Q_X\mathbf O_X\|_F
\le
C_7\frac{M^{6}}{\lambda}
\sqrt{rs_u\frac{q+\log p}{n}}.
\]
The last line follows from $\|U_B\|_{op}=1$.
Taking $\widetilde{\mathbf O}=\mathbf Q_X\mathbf O_X\in\mathcal O_r$ proves the bound for
$\widehat{\mathbf U}$.
{Since all estimates above hold on $\mathcal E$ and $\mathcal E$ has the probability lower bound displayed above, the theorem follows.}
This completes the proof.
}
\end{proof}

\begin{lemma}[Wedin's sin--$\Theta$ theorem, rank-$r$ signal form] \label{theorem:wedin}
Let $\mathbf A,\widehat{\mathbf A}\in\mathbb R^{p\times q}$, and suppose
$\mathbf A$ has rank $r$ with compact singular value decomposition
\[
\mathbf A=\mathbf U\mathbf D\mathbf V^\top,
\qquad
\sigma_r(\mathbf A)>0,
\]
where $\mathbf U\in\mathbb R^{p\times r}$ and
$\mathbf V\in\mathbb R^{q\times r}$ have orthonormal columns. Let
\[
\widehat{\mathbf A}
=
\widehat{\mathbf U}\widehat{\mathbf D}\widehat{\mathbf V}^{\top}
+
\widehat{\mathbf A}_{\perp}
\]
be a singular value decomposition of $\widehat{\mathbf A}$, where
$\widehat{\mathbf U}$ and $\widehat{\mathbf V}$ contain the leading $r$ left and
right singular vectors. Set
\[
\mathbf E=\widehat{\mathbf A}-\mathbf A.
\]
If
\[
\|\mathbf E\|_{op}\le \frac12\sigma_r(\mathbf A),
\]
then
\[
\|\sin\Theta(\widehat{\mathbf U},\mathbf U)\|_F
\vee
\|\sin\Theta(\widehat{\mathbf V},\mathbf V)\|_F
\le
C_W\frac{\|\mathbf E\|_F}{\sigma_r(\mathbf A)},
\]
for a universal numerical constant $C_W$.
Consequently, there exist orthogonal matrices
$\mathbf O_U,\mathbf O_V\in\mathcal O_r$ such that
\[
\|\widehat{\mathbf U}-\mathbf U\mathbf O_U\|_F
\vee
\|\widehat{\mathbf V}-\mathbf V\mathbf O_V\|_F
\le
C_W'\frac{\|\mathbf E\|_F}{\sigma_r(\mathbf A)},
\]
where $C_W'$ is another universal numerical constant.
\end{lemma}

\begin{proof}
Let
\[
\sigma:=\sigma_r(\mathbf A)>0,
\qquad
\mathbf E=\widehat{\mathbf A}-\mathbf A.
\]
Let
\[
\mathbf A=\mathbf U\mathbf D\mathbf V^\top
\]
be the compact singular value decomposition of $\mathbf A$, and extend
$\mathbf U$ and $\mathbf V$ to orthogonal matrices
$[\mathbf U,\mathbf U_\perp]$ and $[\mathbf V,\mathbf V_\perp]$.
Then
\[
\|\sin\Theta(\widehat{\mathbf U},\mathbf U)\|_F
=
\|\mathbf U_\perp^\top\widehat{\mathbf U}\|_F,
\qquad
\|\sin\Theta(\widehat{\mathbf V},\mathbf V)\|_F
=
\|\mathbf V_\perp^\top\widehat{\mathbf V}\|_F.
\]

By Weyl's inequality,
\[
\sigma_r(\widehat{\mathbf A})
\ge
\sigma_r(\mathbf A)-\|\widehat{\mathbf A}-\mathbf A\|_{op}
\ge
\frac{\sigma}{2}.
\]
Thus the leading $r$ singular values collected in
$\widehat{\mathbf D}$ are bounded below by $\sigma/2$, and
$\widehat{\mathbf D}$ is invertible with
\[
\|\widehat{\mathbf D}^{-1}\|_{op}
\le
\frac{2}{\sigma}.
\]

We first bound the left singular subspace. Since
\[
\widehat{\mathbf A}\widehat{\mathbf V}
=
\widehat{\mathbf U}\widehat{\mathbf D},
\]
we have
\[
\mathbf A\widehat{\mathbf V}
=
\widehat{\mathbf U}\widehat{\mathbf D}
-
\mathbf E\widehat{\mathbf V}.
\]
Because the columns of $\mathbf A\widehat{\mathbf V}$ lie in
$\operatorname{col}(\mathbf U)$, multiplying by $\mathbf U_\perp^\top$ gives
\[
\mathbf 0
=
\mathbf U_\perp^\top\widehat{\mathbf U}\widehat{\mathbf D}
-
\mathbf U_\perp^\top\mathbf E\widehat{\mathbf V}.
\]
Therefore,
\[
\mathbf U_\perp^\top\widehat{\mathbf U}
=
\mathbf U_\perp^\top\mathbf E\widehat{\mathbf V}
\widehat{\mathbf D}^{-1}.
\]
Hence
\[
\|\sin\Theta(\widehat{\mathbf U},\mathbf U)\|_F
=
\|\mathbf U_\perp^\top\widehat{\mathbf U}\|_F
\le
\|\mathbf U_\perp^\top\mathbf E\widehat{\mathbf V}\|_F
\|\widehat{\mathbf D}^{-1}\|_{op}
\le
\frac{2\|\mathbf E\|_F}{\sigma_r(\mathbf A)}.
\]

The right singular subspace is handled in the same way. Since
\[
\widehat{\mathbf A}^\top\widehat{\mathbf U}
=
\widehat{\mathbf V}\widehat{\mathbf D},
\]
we have
\[
\mathbf A^\top\widehat{\mathbf U}
=
\widehat{\mathbf V}\widehat{\mathbf D}
-
\mathbf E^\top\widehat{\mathbf U}.
\]
The columns of $\mathbf A^\top\widehat{\mathbf U}$ lie in
$\operatorname{col}(\mathbf V)$, so multiplying by $\mathbf V_\perp^\top$ yields
\[
\mathbf 0
=
\mathbf V_\perp^\top\widehat{\mathbf V}\widehat{\mathbf D}
-
\mathbf V_\perp^\top\mathbf E^\top\widehat{\mathbf U}.
\]
Thus
\[
\mathbf V_\perp^\top\widehat{\mathbf V}
=
\mathbf V_\perp^\top\mathbf E^\top\widehat{\mathbf U}
\widehat{\mathbf D}^{-1},
\]
and consequently
\[
\|\sin\Theta(\widehat{\mathbf V},\mathbf V)\|_F
=
\|\mathbf V_\perp^\top\widehat{\mathbf V}\|_F
\le
\frac{2\|\mathbf E\|_F}{\sigma_r(\mathbf A)}.
\]
Combining the two displays gives
\[
\|\sin\Theta(\widehat{\mathbf U},\mathbf U)\|_F
\vee
\|\sin\Theta(\widehat{\mathbf V},\mathbf V)\|_F
\le
2\frac{\|\mathbf E\|_F}{\sigma_r(\mathbf A)}.
\]
Thus the first conclusion holds with, for example, $C_W=2$.

It remains to convert the sin--$\Theta$ bound into a bound after orthogonal
alignment. For matrices with orthonormal columns, the Procrustes inequality gives
\[
\min_{\mathbf O\in\mathcal O_r}
\|\widehat{\mathbf U}-\mathbf U\mathbf O\|_F
\le
\sqrt{2}\,
\|\sin\Theta(\widehat{\mathbf U},\mathbf U)\|_F,
\]
and similarly
\[
\min_{\mathbf O\in\mathcal O_r}
\|\widehat{\mathbf V}-\mathbf V\mathbf O\|_F
\le
\sqrt{2}\,
\|\sin\Theta(\widehat{\mathbf V},\mathbf V)\|_F.
\]
Indeed, if the singular values of
$\mathbf U^\top\widehat{\mathbf U}$ are
$\cos\theta_1,\ldots,\cos\theta_r$, then
\[
\min_{\mathbf O\in\mathcal O_r}
\|\widehat{\mathbf U}-\mathbf U\mathbf O\|_F^2
=
2\sum_{j=1}^r(1-\cos\theta_j)
\le
2\sum_{j=1}^r\sin^2\theta_j.
\]
The same argument applies to $\mathbf V$ and $\widehat{\mathbf V}$.

Therefore, there exist
$\mathbf O_U,\mathbf O_V\in\mathcal O_r$ such that
\[
\|\widehat{\mathbf U}-\mathbf U\mathbf O_U\|_F
\vee
\|\widehat{\mathbf V}-\mathbf V\mathbf O_V\|_F
\le
2\sqrt{2}\,
\frac{\|\mathbf E\|_F}{\sigma_r(\mathbf A)}.
\]
Thus the second conclusion holds with, for example,
$C_W'=2\sqrt{2}$.
\end{proof}

\section{Proofs: group-sparse CCAR$^3$}\label{appendix:proof_sparse_group}

We first introduce notation for group supports. Let
\[
\mathcal G=\{g_1,\ldots,g_K\}
\]
be a non-overlapping partition of \([p]\), and write
\[
d_g=|g|,\qquad d_{\min}=\min_{g\in\mathcal G}d_g,\qquad
d_{\max}=\max_{g\in\mathcal G}d_g,\qquad
\eta=\frac{d_{\max}}{d_{\min}}.
\]
For a collection of groups \(\mathcal H\subseteq\mathcal G\), define
\[
T(\mathcal H)=\bigcup_{g\in\mathcal H}g,\qquad
d(\mathcal H)=\sum_{g\in\mathcal H}d_g.
\]
For a matrix \(\bB\in\mathbb R^{p\times q}\), let
\[
\operatorname{supp}_{\mathcal G}(\bB)
=
\{g\in\mathcal G:\|\bB_g\|_F>0\},
\]
where \(\bB_g\) denotes the submatrix of \(\bB\) formed by the rows indexed by \(g\). We use the weighted group norm
\[
\|\bB\|_{\mathcal G,21}
=
\sum_{g\in\mathcal G}\sqrt{d_g}\,\|\bB_g\|_F.
\]
Let
\[
\bB^*=\bU\bLambda \bV_0^\top,\qquad \bV_0=\bS_Y^{1/2}\bV,
\]
and define the true active group set
\[
\mathcal A
=
\operatorname{supp}_{\mathcal G}(\bB^*)
=
\{g\in\mathcal G:\|\bB_g^*\|_F>0\}.
\]
Let
\[
m_u=|\mathcal A|,\qquad
s_u=d(\mathcal A)=\sum_{g\in\mathcal A}d_g.
\]
Note that \(m_u\le s_u/d_{\min}\). Define the group rate
\[
R_{\mathcal G}
=
\frac{s_u q+(s_u/d_{\min})\log K}{n}.
\]

\begin{theorem}[Group-regularized regression error with support control]\label{thm:group-regression-support}
Consider the group-penalized estimator
\[
\widehat \bB
=
\operatorname*{argmin}_{\bB\in\mathbb R^{p\times q}}
\frac1n\|\bY_0-\bX\bB\|_F^2
+
\rho\sum_{g\in\mathcal G}\sqrt{d_g}\,\|\bB_g\|_F .
\]
Assume that \(R_{\mathcal G}\le c_0\) for a sufficiently small constant \(c_0\in(0,1)\).
Assume also that \(\eta=d_{\max}/d_{\min}\) is bounded by an absolute constant; otherwise the constants below depend on \(\eta\).
Let
\[
\mathbf\Delta=\widehat \bB-\bB^*.
\]
Choose
\[
\rho
\ge
C_uM^2
\left(
\sqrt{\frac qn}
+
\sqrt{\frac{\log K}{d_{\min}n}}
\right)
\]
for a sufficiently large constant \(C_u\). Then, for any \(C'>0\), with a probability of at least
\[
1
-
\exp(-C'q)
-
\exp\{-C'(s_u/d_{\min})\log(eK d_{\min}/s_u)\}
-
\exp\{-C'(q+\log K)\}
-
\exp\{-C'(r+\log K)\},
\]
there exist constants \(C_1,C_2,C_3>0\), depending only on \(C'\), \(c_0\), and the bounded group-size ratio, such that
\[
\left\|\widehat\bS_X^{1/2}\mathbf\Delta\right\|_F^2
\le
C_1M^5 R_{\mathcal G},
\]
and
\[
\|\mathbf{\Delta}\|_F^2
\le
C_2M^6 R_{\mathcal G}.
\]
Moreover, with the same probability, the selected group support satisfies
\[
d(\widehat{\mathcal A})
=
\sum_{g\in\widehat{\mathcal A}}d_g
\le
C_3M^2s_u,
\qquad
|\widehat{\mathcal A}|
\le
C_3M^2\frac{s_u}{d_{\min}},
\]
where
\[
\widehat{\mathcal A}
=
\operatorname{supp}_{\mathcal G}(\widehat \bB).
\]
\end{theorem}

\begin{proof}
The proof follows the same regression argument as in the sparse case, but the cone and peeling steps are performed at the group level.
Let
\[
\mathbf Z
=
\widehat\bS_{XY}\widehat\bS_Y^{-1/2}
-
\widehat\bS_X \bB^*.
\]
We work on the group score event
\[
\mathcal E_{\mathrm{score}}
=
\left\{
\max_{g\in\mathcal G}
\frac{\|\mathbf{Z}_g\|_F}{\sqrt{d_g}}
\le
\frac{\rho}{4}
\right\}.
\]
We first verify the group score event. Recall that
\[
\mathbf{Z}
=
\widehat\bS_{XY}\widehat\bS_Y^{-1/2}
-
\widehat\bS_X \bB^*,
\qquad
\bB^*=\bU\bLambda \bV_0^\top,
\qquad
\bV_0=\bS_Y^{1/2}\bV.
\]
Since
\[
\bS_X\bB^*
=
\bS_X\bU\bLambda \bV^\top\bS_Y^{1/2}
=
\bS_{XY}\bS_Y^{-1/2},
\]
we may decompose, for every group \(g\in\mathcal G\),
\[
\mathbf{Z}_g
=
\underbrace{
\big[
(\widehat\bS_{XY}-\bS_{XY})
\widehat\bS_Y^{-1/2}
\big]_g
}_{\bA_g}
+
\underbrace{
\big[
\bS_{XY}
(\widehat\bS_Y^{-1/2}-\bS_Y^{-1/2})
\big]_g
}_{\bB_g}
-
\underbrace{
\big[
(\widehat\bS_X-\bS_X)\bB^*
\big]_g
}_{\mathbf C_g}.
\tag{E.0}
\]
We work on the covariance event
\[
\mathcal E_Y
=
\left\{
\|\widehat\bS_Y^{1/2}-\bS_Y^{1/2}\|_{\text{op}}
\le
CM\sqrt{\frac qn},
\quad
\|\widehat\bS_Y^{-1/2}\|_{\text{op}}
\le
2\sqrt M
\right\}.
\]
By Lemma~\ref{lem:cov},
\[
\mathbb P(\mathcal E_Y^c)
\le
\exp(-C'q),
\]
provided \(q/n\le c_0\) and \(c_0\) is sufficiently small. On \(\mathcal E_Y\),
\[
\|\bS_Y^{1/2}\widehat\bS_Y^{-1/2}\|_{\text{op}}
\le
CM,
\]
and
\[
\|\bS_Y^{1/2}\widehat\bS_Y^{-1/2}-\bI_q\|_{\text{op}}
=
\|(\bS_Y^{1/2}-\widehat\bS_Y^{1/2})
\widehat\bS_Y^{-1/2}\|_{\text{op}}
\le
CM^{3/2}\sqrt{\frac qn}.
\tag{E.0a}
\]

We now bound the three terms in \((E.0)\).

First, set
\[
\widetilde \bY_i=\bS_Y^{-1/2}\bY_i.
\]
For a fixed group \(g\), and for any matrix
\(\mathbf H\in\mathbb R^{d_g\times q}\) with \(\|\mathbf H\|_F=1\),
the random variables
\[
\xi_i(\mathbf H)
=
\left\langle
\bX_{i,g}\widetilde \bY_i^\top
-
\mathbb E(\bX_{i,g}\widetilde \bY_i^\top),
\mathbf H
\right\rangle
\]
are centered sub-exponential with
\[
\|\xi_i(\mathbf{H})\|_{\psi_1}\le CM.
\]
Indeed, \((\bX_{i,g},\widetilde \bY_i)\) is jointly Gaussian and its covariance operator norm is bounded by \(CM\). Bernstein's inequality, followed by a \(1/4\)-net argument over the Frobenius unit sphere in
\(\mathbb R^{d_g\times q}\), gives, for every \(t>0\),
\[
\mathbb P
\left[
\left\|
\big[
(\widehat\bS_{XY}-\bS_{XY})
\bS_Y^{-1/2}
\big]_g
\right\|_F
>
CM
\sqrt{\frac{d_gq+t}{n}}
\right]
\le
\exp\{-c(d_gq+t)\},
\tag{E.0b}
\]
where the linear Bernstein term is absorbed under the small-rate condition. Taking
\(t=\log K\) and union-bounding over \(g\in\mathcal G\) gives
\[
\max_{g\in\mathcal G}
\frac{
\left\|
\big[
(\widehat\bS_{XY}-\bS_{XY})
\bS_Y^{-1/2}
\big]_g
\right\|_F
}{\sqrt{d_g}}
\le
CM
\left(
\sqrt{\frac qn}
+
\sqrt{\frac{\log K}{d_{\min}n}}
\right)
\tag{E.0c}
\]
with probability at least
\[
1-\exp\{-C'(q+\log K)\}.
\]
Combining \((E.0c)\) with
\(\|\bS_Y^{1/2}\widehat\bS_Y^{-1/2}\|_{\text{op}}\le CM\),
we obtain
\[
\max_{g\in\mathcal G}
\frac{\|\bA_g\|_F}{\sqrt{d_g}}
\le
CM^2
\left(
\sqrt{\frac qn}
+
\sqrt{\frac{\log K}{d_{\min}n}}
\right).
\tag{E.0d}
\]

Second, for the term \(\bB_g\), note that
\[
\bB_g
=
\big[
\bS_{XY}\bS_Y^{-1/2}
\big]_g
\left(
\bS_Y^{1/2}\widehat\bS_Y^{-1/2}-\bI_q
\right).
\]
Moreover,
\[
\frac{
\|
[
\bS_{XY}\bS_Y^{-1/2}
]_g
\|_F
}{\sqrt{d_g}}
\le
\|\bS_{XY}\bS_Y^{-1/2}\|_{\text{op}}
=
\|\bS_X\bU\bLambda \bV_0^\top\|_{\text{op}}
\le
\sqrt M.
\]
Therefore, using \((E.0a)\),
\[
\max_{g\in\mathcal G}
\frac{\|\bB_g\|_F}{\sqrt{d_g}}
\le
CM^2\sqrt{\frac qn}.
\tag{E.0e}
\]

Third, for the term \(\mathbf C_g\), since
\[
\bB^*=\bU\bLambda \bV_0^\top
\qquad\text{and}\qquad
\opnorm{\bLambda \bV_0^\top}\le 1,
\]
we have
\[
\|\mathbf C_g\|_F
\le
\left\|
\big[
(\widehat\bS_X-\bS_X)\bU
\big]_g
\right\|_F.
\]
For a fixed group \(g\), and for any
\(\mathbf{H}\in\mathbb R^{d_g\times r}\) with \(\|\mathbf{H}\|_F=1\),
the variables
\[
\zeta_i(\mathbf{H})
=
\left\langle
\bX_{i,g}(\bU^\top \bX_i)^\top
-
\mathbb E\{\bX_{i,g}(\bU^\top \bX_i)^\top\},
\mathbf{H}
\right\rangle
\]
are centered sub-exponential with
\[
\|\zeta_i(\mathbf{H})\|_{\psi_1}\le CM,
\]
because \((\bX_{i,g},\bU^\top \bX_i)\) is jointly Gaussian, the covariance of
\(\bX_{i,g}\) has operator norm at most \(M\), and
\[
\operatorname{Cov}(\bU^\top \bX_i)=\bU^\top\bS_X\bU=I_r.
\]
Another Bernstein plus net argument gives, for every \(t>0\),
\[
\mathbb P
\left[
\left\|
\big[
(\widehat\bS_X-\bS_X)\bU
\big]_g
\right\|_F
>
CM
\sqrt{\frac{d_gr+t}{n}}
\right]
\le
\exp\{-c(d_gr+t)\}.
\tag{E.0f}
\]
Taking \(t=\log K\), union-bounding over the \(K\) groups, and using \(r\le q\), we get
\[
\max_{g\in\mathcal G}
\frac{\|\mathbf C_g\|_F}{\sqrt{d_g}}
\le
CM
\left(
\sqrt{\frac rn}
+
\sqrt{\frac{\log K}{d_{\min}n}}
\right)
\le
CM^2
\left(
\sqrt{\frac qn}
+
\sqrt{\frac{\log K}{d_{\min}n}}
\right)
\tag{E.0g}
\]
with probability at least
\[
1-\exp\{-C'(r+\log K)\}.
\]
Combining \((E.0d)\), \((E.0e)\), and \((E.0g)\), we conclude that
\[
\max_{g\in\mathcal G}
\frac{\|\mathbf{Z}_g\|_F}{\sqrt{d_g}}
\le
CM^2
\left(
\sqrt{\frac qn}
+
\sqrt{\frac{\log K}{d_{\min}n}}
\right)
\tag{E.0h}
\]
with probability at least
\[
1
-
\exp(-C'q)
-
\exp\{-C'(q+\log K)\}
-
\exp\{-C'(r+\log K)\}.
\]
Thus, choosing
\[
\rho
\ge
C_\rho M^2
\left(
\sqrt{\frac qn}
+
\sqrt{\frac{\log K}{d_{\min}n}}
\right)
\]
with \(C_\rho\) sufficiently large ensures the group score event
\[
\mathcal E_{\mathrm{score}}
=
\left\{
\max_{g\in\mathcal G}
\frac{\|\mathbf{Z}_g\|_F}{\sqrt{d_g}}
\le
\frac{\rho}{4}
\right\}.
\]
Thus, the stated choice of \(\rho\) ensures \(\mathcal E_{\mathrm{score}}\) with high probability.

We also work on a uniform group-sparse covariance event. Specifically, for all collections
\(\mathcal \mathbf{H}\subseteq\mathcal G\),
\[
\left\|
[\widehat\bS_X-\bS_X]_{T(\mathcal H),T(\mathcal H)}
\right\|_{\mathrm{op}}
\le
CM
\sqrt{
\frac{d(\mathcal H)+|\mathcal H|\log(eK/|\mathcal H|)}{n}
}.
\]
This follows by applying the usual sub-Gaussian sample-covariance bound to each fixed group union and union-bounding over group subsets. On the group-sparse sizes used below, the right-hand side is small because \(R_{\mathcal G}\le c_0\). Hence, for all such group unions,
\[
\lambda_{\min}
\left(
[\widehat\bS_X]_{T(\mathcal H),T(\mathcal H)}
\right)
\ge
\frac{1}{2M},
\qquad
\lambda_{\max}
\left(
[\widehat\bS_X]_{T(\mathcal H),T(\mathcal H)}
\right)
\le
2M,
\]
provided \(c_0\) is small enough.

We now prove the regression error bound.
By optimality of \(\widehat \bB\),
\[
\frac1n\|\bY_0-\bX\widehat \bB\|_F^2
+
\rho\|\widehat \bB\|_{\mathcal G,21}
\le
\frac1n\|\bY_0-\bX\bB^*\|_F^2
+
\rho\|\bB^*\|_{\mathcal G,21}.
\]
Writing \(\mathbf{\Delta}=\widehat \bB-\bB^*\), this gives
\[
\left\|\widehat\bS_X^{1/2}\mathbf{\Delta}\right\|_F^2
\le
2\langle \mathbf{Z},\mathbf{\Delta}\rangle
+
\rho
\left(
\|\bB^*\|_{\mathcal G,21}
-
\|\bB^*+\mathbf{\Delta}\|_{\mathcal G,21}
\right).
\]
On \(\mathcal E_{\mathrm{score}}\),
\[
2\langle \mathbf{Z},\mathbf{\Delta}\rangle
\le
2
\max_{g\in\mathcal G}
\frac{\|\mathbf{Z}_g\|_F}{\sqrt{d_g}}
\sum_{g\in\mathcal G}\sqrt{d_g}\,\|\mathbf{\Delta}_g\|_F
\le
\frac{\rho}{2}
\sum_{g\in\mathcal G}\sqrt{d_g}\,\|\mathbf{\Delta}_g\|_F.
\]
By decomposability of the group norm,
\[
\|\bB^*\|_{\mathcal G,21}
-
\|\bB^*+\mathbf{\Delta}\|_{\mathcal G,21}
\le
\sum_{g\in\mathcal A}\sqrt{d_g}\,\|\mathbf{\Delta}_g\|_F
-
\sum_{g\notin\mathcal A}\sqrt{d_g}\,\|\mathbf{\Delta}_g\|_F.
\]
Therefore,
\[
\left\|\widehat\bS_X^{1/2}\mathbf{\Delta}\right\|_F^2
\le
\frac{3\rho}{2}
\sum_{g\in\mathcal A}\sqrt{d_g}\,\|\mathbf{\Delta}_g\|_F
-
\frac{\rho}{2}
\sum_{g\notin\mathcal A}\sqrt{d_g}\,\|\mathbf{\Delta}_g\|_F.
\tag{E.1}
\]
In particular, since the left-hand side is nonnegative, we obtain the group cone condition
\[
\sum_{g\notin\mathcal A}\sqrt{d_g}\,\|\mathbf{\Delta}_g\|_F
\le
3
\sum_{g\in\mathcal A}\sqrt{d_g}\,\|\mathbf{\Delta}_g\|_F.
\tag{E.2}
\]
Moreover,
\[
\sum_{g\in\mathcal A}\sqrt{d_g}\,\|\mathbf{\Delta}_g\|_F
\le
\left(\sum_{g\in\mathcal A}d_g\right)^{1/2}
\left(\sum_{g\in\mathcal A}\|\mathbf{\Delta}_g\|_F^2\right)^{1/2}
=
\sqrt{s_u}\,\|\mathbf{\Delta}_{\mathcal A}\|_F.
\]
Thus (E.1) implies
\[
\left\|\widehat\bS_X^{1/2}\mathbf{\Delta}\right\|_F^2
\le
\frac{3\rho\sqrt{s_u}}{2}\|\mathbf{\Delta}_{\mathcal A}\|_F.
\tag{E.3}
\]

We now use a group peeling argument to lower-bound
\(\|\widehat\bS_X^{1/2}\mathbf{\Delta}\|_F\) by \(\|\mathbf{\Delta}\|_F\).

For each inactive group \(g\notin\mathcal A\), define the normalized group size
\[
a_g=\frac{\|\mathbf{\Delta}_g\|_F}{\sqrt{d_g}}.
\]
Order the inactive groups in decreasing order of \(a_g\). Let \(m_*\) be an integer of the form
\[
m_*=\left\lceil c_*\,M^2\eta\,\frac{s_u}{d_{\min}}\right\rceil,
\]
where \(c_*>0\) is a sufficiently large universal constant. Let
\[
\mathcal J_1,\mathcal J_2,\ldots
\]
be consecutive blocks of \(m_*\) inactive groups in this ordering, with the final block possibly smaller. Define the enlarged active group set
\[
\mathcal S_e=\mathcal A\cup\mathcal J_1.
\]
For \(k\ge2\), by construction of the ordering,
\[
\max_{g\in\mathcal J_k} a_g
\le
\frac{1}{m_*d_{\min}}
\sum_{h\in\mathcal J_{k-1}}d_h a_h
=
\frac{1}{m_*d_{\min}}
\sum_{h\in\mathcal J_{k-1}}\sqrt{d_h}\,\|\mathbf{\Delta}_h\|_F.
\]
Therefore,
\[
\|\mathbf{\Delta}_{\mathcal J_k}\|_F
=
\left(\sum_{g\in\mathcal J_k}d_ga_g^2\right)^{1/2}
\le
\sqrt{m_*d_{\max}}
\max_{g\in\mathcal J_k}a_g
\le
\sqrt{\frac{\eta}{m_*d_{\min}}}
\sum_{h\in\mathcal J_{k-1}}\sqrt{d_h}\,\|\mathbf{\Delta}_h\|_F.
\]
Summing over \(k\ge2\) and using the cone condition (E.2),
\[
\sum_{k\ge2}\|\mathbf{\Delta}_{\mathcal J_k}\|_F
\le
\sqrt{\frac{\eta}{m_*d_{\min}}}
\sum_{g\notin\mathcal A}\sqrt{d_g}\,\|\mathbf{\Delta}_g\|_F
\le
3\sqrt{\frac{\eta}{m_*d_{\min}}}
\sum_{g\in\mathcal A}\sqrt{d_g}\,\|\mathbf{\Delta}_g\|_F.
\]
Using Cauchy--Schwarz again,
\[
\sum_{k\ge2}\|\mathbf{\Delta}_{\mathcal J_k}\|_F
\le
3\sqrt{\frac{\eta s_u}{m_*d_{\min}}}\,
\|\mathbf{\Delta}_{\mathcal A}\|_F.
\tag{E.4}
\]
By the definition of \(m_*\), choosing \(c_*\) large enough gives
\[
3\sqrt{\frac{\eta s_u}{m_*d_{\min}}}
\le
\frac{1}{4M}.
\tag{E.5}
\]
Now write
\[
\mathbf{\Delta}
=
\mathbf{\Delta}_{\mathcal S_e}
+
\sum_{k\ge2}\mathbf{\Delta}_{\mathcal J_k}.
\]
By the triangle inequality,
\[
\left\|\widehat\bS_X^{1/2}\mathbf{\Delta}\right\|_F
\ge
\left\|\widehat\bS_X^{1/2}\mathbf{\Delta}_{\mathcal S_e}\right\|_F
-
\sum_{k\ge2}
\left\|\widehat\bS_X^{1/2}\mathbf{\Delta}_{\mathcal J_k}\right\|_F.
\]
On the group-sparse covariance event,
\[
\left\|\widehat\bS_X^{1/2}\mathbf{\Delta}_{\mathcal S_e}\right\|_F
\ge
\frac{1}{\sqrt{2M}}\|\mathbf{\Delta}_{\mathcal S_e}\|_F,
\]
and
\[
\left\|\widehat\bS_X^{1/2}\mathbf{\Delta}_{\mathcal J_k}\right\|_F
\le
\sqrt{2M}\,\|\mathbf{\Delta}_{\mathcal J_k}\|_F.
\]
Hence, using (E.4)--(E.5) and \(\|\mathbf{\Delta}_{\mathcal A}\|_F\le\|\mathbf{\Delta}_{\mathcal S_e}\|_F\),
\[
\left\|\widehat\bS_X^{1/2}\mathbf{\Delta}\right\|_F
\ge
\left(
\frac{1}{\sqrt{2M}}
-
\frac{\sqrt{2M}}{4M}
\right)
\|\mathbf{\Delta}_{\mathcal S_e}\|_F
\ge
\frac{c}{\sqrt M}\|\mathbf{\Delta}_{\mathcal S_e}\|_F
\tag{E.6}
\]
for a universal constant \(c>0\). In addition, (E.4)--(E.5) also imply
\[
\|\mathbf{\Delta}_{\mathcal S_e^c}\|_F
\le
\sum_{k\ge2}\|\mathbf{\Delta}_{\mathcal J_k}\|_F
\le
\frac{1}{4M}\|\mathbf{\Delta}_{\mathcal S_e}\|_F.
\]
Therefore,
\[
\|\mathbf{\Delta}\|_F
\le
\|\mathbf{\Delta}_{\mathcal S_e}\|_F+\|\mathbf{\Delta}_{\mathcal S_e^c}\|_F
\le
C\|\mathbf{\Delta}_{\mathcal S_e}\|_F
\le
C\sqrt M
\left\|\widehat\bS_X^{1/2}\mathbf{\Delta}\right\|_F.
\tag{E.7}
\]

Combining (E.3) with (E.6), we get
\[
\left\|\widehat\bS_X^{1/2}\mathbf{\Delta}\right\|_F^2
\le
\frac{3\rho\sqrt{s_u}}{2}\|\mathbf{\Delta}_{\mathcal A}\|_F
\le
\frac{3\rho\sqrt{s_u}}{2}\|\mathbf{\Delta}_{\mathcal S_e}\|_F
\le
C\sqrt M\,\rho\sqrt{s_u}
\left\|\widehat\bS_X^{1/2}\mathbf{\Delta}\right\|_F.
\]
Thus
\[
\left\|\widehat\bS_X^{1/2}\mathbf{\Delta}\right\|_F^2
\le
CM\rho^2s_u.
\tag{E.8}
\]
By (E.7),
\[
\|\mathbf{\Delta}\|_F^2
\le
CM^2\rho^2s_u.
\tag{E.9}
\]
Since
\[
\rho^2
\le
CM^4
\left(
\frac qn
+
\frac{\log K}{d_{\min}n}
\right),
\]
the bounds (E.8)--(E.9) give
\[
\left\|\widehat\bS_X^{1/2}\mathbf{\Delta}\right\|_F^2
\le
CM^5
\left(
\frac{s_uq}{n}
+
\frac{(s_u/d_{\min})\log K}{n}
\right)
=
CM^5R_{\mathcal G},
\]
and
\[
\|\mathbf{\Delta}\|_F^2
\le
CM^6
\left(
\frac{s_uq}{n}
+
\frac{(s_u/d_{\min})\log K}{n}
\right)
=
CM^6R_{\mathcal G}.
\]

It remains to prove the selected-support bound.
Let
\[
\widehat{\mathcal A}
=
\operatorname{supp}_{\mathcal G}(\widehat \bB),
\qquad
\widehat T=T(\widehat{\mathcal A}),
\qquad
\widehat s=d(\widehat{\mathcal A}).
\]
The KKT conditions for the group-penalized estimator give, for every group \(g\),
\[
\frac{2}{n}\bX_g^\top(\bX\widehat \bB-\bY_0)
\in
-\rho\sqrt{d_g}\,\partial\|\widehat \bB_g\|_F.
\]
Therefore, if \(g\in\widehat{\mathcal A}\), then
\[
\frac1n
\left\|
\bX_g^\top(\bX\widehat \bB-\bY_0)
\right\|_F
=
\frac{\rho\sqrt{d_g}}{2}.
\]
On the score event,
\[
\frac1n
\left\|
\bX_g^\top(\bY_0-\bX\bB^*)
\right\|_F
=
\|\mathbf{Z}_g\|_F
\le
\frac{\rho\sqrt{d_g}}{4}.
\]
Hence, for every \(g\in\widehat{\mathcal A}\),
\[
\frac1n
\left\|
\bX_g^\top \bX\mathbf{\Delta}
\right\|_F
\ge
\frac{\rho\sqrt{d_g}}{4}.
\]
Squaring and summing over \(g\in\widehat{\mathcal A}\) yields
\[
\frac{\rho^2}{16}
\sum_{g\in\widehat{\mathcal A}}d_g
\le
\frac1{n^2}
\sum_{g\in\widehat{\mathcal A}}
\|\bX_g^\top \bX\mathbf{\Delta}\|_F^2.
\]
Because the groups are disjoint,
\[
\sum_{g\in\widehat{\mathcal A}}
\|\bX_g^\top \bX\mathbf{\Delta}\|_F^2
=
\|\bX_{\widehat T}^\top \bX\mathbf{\Delta}\|_F^2.
\]
Therefore,
\[
\frac{\rho^2}{16}\widehat s
\le
\frac1{n^2}
\|\bX_{\widehat T}^\top \bX\mathbf{\Delta}\|_F^2
\le
\left\|
[\widehat\bS_X]_{\widehat T,\widehat T}
\right\|_{\mathrm{op}}
\left\|
\widehat\bS_X^{1/2}\mathbf{\Delta}
\right\|_F^2.
\]
Using (E.8),
\[
\widehat s
\le
CMs_u
\left\|
[\widehat\bS_X]_{\widehat T,\widehat T}
\right\|_{\mathrm{op}}.
\tag{E.10}
\]
On the uniform group-sparse covariance event,
\[
\left\|
[\widehat\bS_X]_{\widehat T,\widehat T}
\right\|_{\mathrm{op}}
\le
M
+
CM
\sqrt{
\frac{
\widehat s+(\widehat s/d_{\min})\log(eK)
}{n}
}.
\]
Substituting this into (E.10),
\[
\widehat s
\le
CM^2s_u
+
CM^2s_u
\sqrt{
\frac{
\widehat s+(\widehat s/d_{\min})\log(eK)
}{n}
}.
\tag{E.11}
\]
We now solve this one-dimensional inequality. If
\[
\widehat s\le 2CM^2s_u,
\]
we are done. Otherwise, the second term in (E.11) must dominate, and hence
\[
\widehat s
\le
CM^4s_u^2
\frac{
1+\log(eK)/d_{\min}
}{n}.
\]
Since
\[
\frac{s_u}{n}
\left(
1+\frac{\log(eK)}{d_{\min}}
\right)
\le
C R_{\mathcal G}
\le
Cc_0,
\]
choosing \(c_0\) sufficiently small gives a contradiction to
\(\widehat s>2CM^2s_u\). Therefore,
\[
\widehat s=d(\widehat{\mathcal A})
\le
CM^2s_u.
\]
Finally, since each selected group has size at least \(d_{\min}\),
\[
|\widehat{\mathcal A}|
\le
\frac{d(\widehat{\mathcal A})}{d_{\min}}
\le
CM^2\frac{s_u}{d_{\min}}.
\]
This completes the proof.
\end{proof}

\begin{theorem}[Group-sparse CCAR3 direction error]
\label{thm:group-ccar3-direction}
Let
\[
\mathcal G=\{g_1,\ldots,g_K\}
\]
be a non-overlapping partition of \([p]\). Write
\[
d_g=|g|,\qquad d_{\min}=\min_{g\in\mathcal G}d_g,\qquad
d_{\max}=\max_{g\in\mathcal G}d_g,
\]
and assume that \(d_{\max}/d_{\min}\) is bounded by an absolute constant.

Assume the canonical pair model
\[
\bS_{XY}=\bS_X\bU\bLambda \bV^\top\bS_Y,
\qquad
\bU^\top\bS_X\bU=\bV^\top\bS_Y\bV=I_r,
\]
with
\[
\lambda_r(\bLambda)\ge \lambda,\qquad \lambda_1(\bLambda)\le 1-\frac1M,
\]
and
\[
\frac1M\le \lambda_{\min}(\bS_X)\wedge \lambda_{\min}(\bS_Y),
\qquad
\lambda_{\max}(\bS_X)\vee \lambda_{\max}(\bS_Y)\le M.
\]
Let
\[
\bV_0=\bS_Y^{1/2}\bV,\qquad \bB^*=\bU\bLambda \bV_0^\top.
\]
Define the active group set
\[
\mathcal A=\{g\in\mathcal G:\|\bU_g\|_F\neq 0\},
\qquad
m_u=|\mathcal A|,
\qquad
s_u=\sum_{g\in\mathcal A}|g|.
\]
Let
\[
R_{\mathcal G}
=
\frac{s_uq+(s_u/d_{\min})\log K}{n},
\qquad
\delta_{\mathcal G}=\sqrt{R_{\mathcal G}}.
\]
Assume \(R_{\mathcal G}\le c_0\) for a sufficiently small constant \(c_0\in(0,1)\), and choose
\[
\rho
\ge
C_\rho M^2
\left(
\sqrt{\frac qn}
+
\sqrt{\frac{\log K}{d_{\min}n}}
\right)
\]
for a sufficiently large constant \(C_\rho\).
Let
\[
\widehat \bB
=
\text{argmin}_{\bB\in\mathbb R^{p\times q}}
\frac1n\|\bY_0-\bX\bB\|_F^2
+
\rho\sum_{g\in\mathcal G}\sqrt{|g|}\,\|\bB_g\|_F,
\qquad
\bY_0=\bY\widehat\bS_Y^{-1/2}.
\]
Let
\[
\widehat \bB_r=\widehat \bU_{\hat B}\widehat \bLambda_{\hat B}\widehat \bV_{\hat B}^\top
\]
be the rank-\(r\) truncated singular value decomposition of \(\widehat \bB\), and define
\[
\widehat \bU
=
\widehat \bU_{\hat B}
\left(
\widehat \bU_{\hat B}^\top\widehat\bS_X\widehat \bU_{\hat B}
\right)^{-1/2},
\qquad
\widehat \bV
=
\widehat\bS_Y^{-1/2}\widehat \bV_{\hat B}.
\]
If
\[
CM^{7/2}\delta_{\mathcal G}\le \lambda/2,
\]
then, for any \(C'>0\), with probability at least
\[
1
-
\exp(-C'q)
-
\exp\left\{-C'\frac{s_u}{d_{\min}}
\log\left(\frac{eK d_{\min}}{s_u}\right)\right\}
-
\exp\{-C'(q+\log K)\}
-
\exp\{-C'(r+\log K)\},
\]
there exist orthogonal matrices \(\bO,\widetilde \bO\in\mathbb R^{r\times r}\) such that
\[
\|\widehat \bV-\bV \bO\|_F
\le
C\frac{M^{11/2}\sqrt r}{\lambda}
\sqrt{
\frac{s_uq+(s_u/d_{\min})\log K}{n}
},
\]
and
\[
\|\widehat \bU-\bU\widetilde \bO\|_F
\le
C\frac{M^6\sqrt r}{\lambda}
\sqrt{
\frac{s_uq+(s_u/d_{\min})\log K}{n}
}.
\]
In particular, when the active groups are balanced so that \(s_u/d_{\min}\asymp m_u\),
\[
\delta_{\mathcal G}
\asymp
\sqrt{
\frac{s_uq+m_u\log K}{n}
}.
\]
\end{theorem}

\begin{proof}
The proof is identical to the proof of Theorem~\ref{theorem:rrr_ols_2}: replace the sparse regression theorem by Theorem~\ref{thm:group-regression-support}, replace the sparse rate \(\sqrt{s_u(q+\log p)/n}\) by \(\delta_{\mathcal G}\), and use the selected-support bound \(d(\widehat{\mathcal A})\lesssim M^2s_u\) to obtain the same uniform covariance control in the normalization step.
\end{proof}

\begin{remark}
    We note that group sparsity and regular sparsity give very similar bounds. The main improvement comes from the replacement of the term $\sqrt{\frac{s_u(q+\log(p)) }{n}}$ in Theorem~\ref{theorem:rrr_ols_2} by $\sqrt{\frac{s_u\left(q+ \frac{\log(K)}{d_{\text{min}}}\right) }{n}}$. Consequently, when $p$ is large and in the presence of a small number of balanced groups, the group lasso is expected to improve upon the regular lasso term.
\end{remark}

\section{Proofs: graph-sparse CCAR$^3$}\label{appendix:proof_graph}
We consider now the graph-sparse CCA problem. For a graph $\mathcal{G}$ on $p$ nodes and $m$ edges indicating the relationships between the  $p$ covariates of $\mathbf{X}$, let $\mathcal{F}_{\mathcal{G}}(s_u,p, m, q,r,\lambda, M)$  be the collection of all covariance matrices $\mathbf{\Sigma}$ satisfying (\ref{eq:cca}) and 
\begin{equation}
\begin{aligned}
\text{(a)}&~~ \mbox{$\mathbf{U}\in\mathbb{R}^{p\times r}$ and $\mathbf{V}\in\mathbb{R}^{q\times r}$ with $|\supp{(\mathbf{\Gamma} \mathbf{U})}|\leq s_u$ ;}
\label{eq:family_f_g}\\
\text{(b)}&~~ \mbox{$\sigma_{\min}(\mathbf{\Sigma}_X)\wedge\sigma_{\min}(\mathbf{\Sigma}_Y)\geq \frac1M$ and $\sigma_{\max}(\mathbf{\Sigma}_X)\vee\sigma_{\max}(\mathbf{\Sigma}_Y)\leq M$;}\\
\text{(c)}&~~ \mbox{$\lambda_r\geq\lambda$ and $\lambda_1\leq 1-\frac 1M$.}
\end{aligned}
\end{equation}
We denote $\mathbf{\Gamma}$ the incidence matrix of the graph $\mathcal{G}$, and by $\mathbf{\Gamma}^{\dagger}$ its pseudo inverse. Denoting $\mathbf{\Pi} = \mathbf{I}_p - \mathbf{\Gamma}^{\dagger} \mathbf{\Gamma}$,  we decompose $\mathbf{U} \in \R^{p \times r}$ as a sum of terms carried by the orthogonal subspaces:  $\mathbf{U} = \mathbf{\Pi} \mathbf{U} + \mathbf{\Gamma}^{\dagger} \mathbf{\Gamma} \mathbf{U}.$

\medskip
\noindent We begin with a few observations on the properties of the matrices $\mathbf{\Gamma}$ and $\mathbf{\Pi}$.
\begin{enumerate}
    \item By the property of the incidence matrix, $${\mathbf{\Gamma}^\top  \mathbf{\Gamma} = \mathbf{D}-\mathbf{A}=\mathbf{L}} \hspace{3cm} (*)$$ is the (unnormalized) Laplacian of the graph \citep{hutter2016optimal}. Here $\mathbf{A}$ denotes the adjacency of the graph and $\mathbf{D}$ is a diagonal matrix such that $\mathbf{D}_{ii}$ is the degree of node $i$. {We denote by $\kappa_2=\lambda_{n_c+1}(\mathbf L)$ the smallest nonzero eigenvalue of $\mathbf L$.}
    \item $\mathbf{\Pi}$ is the projection operator of $\R^p$ onto the nullspace of $\mathbf{\Gamma}^{\dagger}\mathbf{\Gamma}$, thus, $\mathbf{\Pi}^2 =\mathbf{\Pi}.$
    Moreover, by ($*$), $\mathbf{\Pi}$ corresponds to the eigenvectors associated to the eigenvalue 0 of the Laplacian.
    \item Consequently, we have an explicit formulation for $\mathbf{\Pi}$ \citep{chung1996lectures}. Let $n_c$ be the number of connected components of $G$, and let $C_i$ denote the $i^{th}$ connected component of $G$.
Let $\mathbbm{1}_{C_i}\in \R^{p\times 1}$ be the indicator (column) vector of the $i^{th}$ connected component, i.e  $[\mathbbm{1}_{C_i}]_j=1$ if $j \in C_i$ and 0 otherwise.
Then
$$\mathbf{\Pi} = \sum_{ i\in [n_c]} \frac{\mathbbm{1}_{C_i} \mathbbm{1}_{C_i}^\top }{|C_i|}.$$
\item Since $\mathbf{\Pi}$ is a projection matrix then for any $\mathbf{B} \in \R^{p\times q}$ it holds $ \fnorm{\mathbf{\Pi} \mathbf{B}} \leq \fnorm{\mathbf{B}}.$
\item We also have:
$$ \| \mathbf{\Gamma} \mathbf{B}\|^2_F \leq \sigma_{\max}(\mathbf{L}) \| \mathbf{B}\|^2_F,  $$ where $\sigma_{\max}(\mathbf{L})$ denotes the maximum eigenvalue of the Laplacian of the graph $G$.  By property of the eigenvalues of the Laplacian, we thus have:
$$ \| \mathbf{\Gamma} \mathbf{B}\|^2_F \leq  2 d_{\max} \| \mathbf{B}\|^2_F,  $$
where $d_{\max}$ is the maximum degree of the graph \citep{chung1996lectures}.
\end{enumerate}

Similarly to the analysis of \cite{hutter2016optimal}, we introduce the scaling constant:
\begin{equation}\label{eq:scaling}
{\rho(\mathbf{\Gamma}) = \max_{e \in [m]} \big\|  \mathbf{\Gamma}^{\dagger}_{\cdot e}\big\|}
\end{equation}

\subsection{Useful Lemmas for the Graph Setting}

\begin{lemma}\label{lemma:gamma_dagger_w}
 For $i \in [n]$, let $\mathbf{W}_i\sim \mathcal N(0, \bar{\mathbf{\Sigma}})$ denote a set of multivariate random variables with covariance matrix $\bar{\mathbf{\Sigma}} =\mathbf{I}_q- \mathbf{V}_0 \mathbf{\Lambda}^2\mathbf{V}_0^\top $, where ${\mathbf{V}_0 = \mathbf{\Sigma}_Y^{\frac12}\mathbf{V}}$, with the notations introduced in section~\ref{sec:sparse-cca}. Let  $\mathbf{X}_i\sim \mathcal N(0, {\mathbf{\Sigma}}_X)$ denote a set of multivariate normal observations independent of $\mathbf{W}$.
       Assume \begin{equation}\label{assumption1G}
        M\frac{{q}+s_u\log\big(\frac{ep}{s_u}\big)}{n}\leq c_0\tag{$\mathcal{H}_{1, \mathcal{G}}$}
    \end{equation} 
    for $c_0 \in (0,1).$ Then, for a constant $C'>0$, there exists $C>0$ that solely depends on $C'$ such that:
     \begin{equation}
\max_{j \in {[m]}} \left\| \left[(\mathbf{\Gamma}^{\dagger})^\top  \frac{\mathbf{X}^\top  \mathbf{W}}{n}\right]_{j\cdot }\right\|  \leq C \cdot \left(1 + \frac{M}{\kappa_2}\right) \cdot \sqrt{\frac{q + \log m}{n}}
    \end{equation}  
with probability at least $1 -\exp\{-C'(q+\log m)\}.$  Here we assume that $c_0$ is small enough so that $C'c_0\leq c_3 -\log c_2-1$ with $c_3, c_2$ the universal constants of Theorem 6.5 of \cite{wainwright2019high}.
    
\end{lemma}
\begin{proof}
We note first that:
     \begin{equation}
\max_{j \in {[m]}} \left\| \left[(\mathbf{\Gamma}^{\dagger})^\top  \frac{\mathbf{X}^\top  \mathbf{W}}{n}\right]_{j\cdot }\right\|  \leq \max_{j \in {[m]}} \left\| \left[(\mathbf{\Gamma}^{\dagger})^\top  \frac{\mathbf{X}^\top  \mathbf{W}}{n} - \mathbf{\Sigma}_{X\Gamma^{\dagger}, W}\right]_{j\cdot}\right\|+ \max_{j \in {[m]}} \left\| \left[\mathbf{\Sigma}_{X\Gamma^{\dagger}, W}\right]_{j\cdot}\right\|
    \end{equation}
    where $\mathbf{\Sigma}_{X\Gamma^{\dagger}, W}$ is the cross-covariance matrix between the transformed variables $\mathbf{X}\mathbf{\Gamma}^{\dagger}$ and~$\mathbf{W}.$
We also note that, by independence of $\mathbf{W}$ and $\mathbf{X}$:
$$
  {\max_{j \in [m]}} \left\| \left[\mathbf{\Sigma}_{X\Gamma^{\dagger}, W}\right]_{j \cdot }\right\|= 0
$$


\noindent Let ${\mathbf{Z}_i \sim \mathcal{N}_m\big(0, (\mathbf{\Gamma}^{\dagger})^\top\mathbf{\Sigma}_X \mathbf{\Gamma}^{\dagger}\big)}$, $\mathbf{W}_i \sim \mathcal{N}_q(0, \bar{\mathbf{\Sigma}})$, and $\mathbbm{1}_j^\top \in \R^{1 \times m}$ the vector indicator of row $j$ (i.e. $[\mathbbm{1}_j]_i=0$ for all $i\neq j$, and  $[\mathbbm{1}_j]_j=1$). Thus, denoting $\widetilde{\mathbf{X}} = \mathbf{X} \mathbf{\Gamma}^{\dagger}$, we may write
$$\mathbbm{1}_j^\top(\widehat{\mathbf{\Sigma}}_{\widetilde{X}W}-\mathbf{\Sigma}_{\widetilde{X}W})=\frac{1}{n}\sum_{i=1}^n\big(\mathbbm{1}_j^\top \mathbf{Z}_i\mathbf{W}_i^\top-\mathbb{E}(\mathbbm{1}_j^\top \mathbf{Z}_i\mathbf{W}_i^\top)\big).$$
We also note that $$\opnorm{\mathbf{\Sigma}_{\widetilde{X}}} = \opnorm{(\mathbf{\Gamma}^{\dagger})^\top \mathbf{\Sigma}_X \mathbf{\Gamma}^{\dagger}}= \sup_{u \in \R^{m}} \frac{u^\top (\mathbf{\Gamma}^{\dagger})^\top \mathbf{\Sigma}_X \mathbf{\Gamma}^{\dagger} u}{\|u\|^2}  \leq M \frac{\| \mathbf{\Gamma}^{\dagger} u\|^2}{\|u\|^2} \leq \frac{M}{\kappa_2}.  $$
Here, as we defined above, constant $\kappa_2$ that appearing above is the smallest non-zero eigenvalue of the Laplacian $\mathbf{L} = \mathbf{\Gamma}^\top \mathbf{\Gamma}$. Moreover we note that
 $$\opnorm{\bar{\mathbf{\Sigma}}} = \opnorm{ \mathbf{I} - \mathbf{V}_0 \mathbf{\Lambda}^2 \mathbf{V}_0^\top} \leq 1.$$
Define for each ${j \in [m]}$ the matrix 
$\mathbf{H}_i^{(j)}=\begin{pmatrix}
\mathbbm{1}_j^\top \mathbf{Z}_i \\
\mathbf{W}_i
\end{pmatrix}.$
Since $\mathbbm{1}_j^\top \mathbf{Z}_i\mathbf{W}_i^\top$ is a submatrix of $\mathbf{H}_i^{(j)}\big(\mathbf{H}_i^{(j)}\big)^{\top}$
\begin{equation}
    \begin{split}
        \max_{j\in{[m]}}  \Big\|\mathbbm{1}_j^\top(\widehat{\mathbf{\Sigma}}_{\widetilde{X}W} - {\mathbf{\Sigma}}_{\widetilde{X}W} ) \Big\| &\leq \max_{j\in{[m]}}\opnorm{\frac{1}{n}\sum_{i=1}^n \big(\mathbf{H}_i^{(j)}\big(\mathbf{H}_i^{(j)}\big)^\top-\mathbb{E}\mathbf{H}_i^{(j)}\big(\mathbf{H}_i^{(j)}\big)^\top\big)}\\
    \end{split}
\end{equation}
Denoting $\mathbf{\mathcal{H}}^{(j)}=\mathbb{E}\mathbf{H}_i^{(j)}\big(\mathbf{H}_i^{(j)}\big)^\top  = \begin{pmatrix}
\mathbbm{1}_j^\top  \mathbf{\Sigma}_{\widetilde{X}} \mathbbm{1}_j & \mathbbm{1}_j^\top  \mathbf{\Sigma}_{\widetilde{X}W} \\
\mathbf{\Sigma}_{W\widetilde{X}} \mathbbm{1}_j & \bar{\mathbf{\Sigma}}
\end{pmatrix}=\begin{pmatrix}
\mathbbm{1}_j^\top  \mathbf{\Sigma}_{\widetilde{X}} \mathbbm{1}_j & 0 \\
0& \bar{\mathbf{\Sigma}}
\end{pmatrix},$ where the last equality follows from the fact that $\mathbf{W}$ and $\mathbf{X}$ are independent, by assumption. This implies:
\begin{equation}
    \begin{split}
        \opnorm{\mathbf{\mathcal{H}}^{(j)}} & = \sup_{\substack{u_1\in \R,~u_2 \in \R^{q}\\ u_1^2 + \|u_2\|^2=1}} \mathbbm{1}_j^\top  \mathbf{\Sigma}_{\widetilde{X}}\mathbbm{1}_j u_1^2 + {u_2^\top \bar{\mathbf{\Sigma}} u_2}\\
         &\leq \sup_{\substack{u_1\in \R,~u_2 \in \R^{q}\\ u_1^2 + \|u_2\|^2=1}} \frac{M}{\kappa_2} u_1^2 +  \|u_2\|^2
        \leq  \frac{M}{\kappa_2} +1 
    \end{split}
\end{equation}
By Theorem 6.5 of \cite{wainwright2019high}, there exist universal constants $c_1, c_2, c_3>0$ such that:
\begin{equation*}
\begin{split}
\P&\left[\opnorm{\frac{1}{n}\sum_{i=1}^n \big(\mathbf{H}_i^{(j)}\big(\mathbf{H}_i^{(j)}\big)^\top -\mathbb{E}\mathbf{H}_i^{(j)}\big(\mathbf{H}_i^{(j)}\big)^\top \big)} \geq  \opnorm{\mathbf{\mathcal{H}}^{(j)}} \left(2c_1\sqrt{\frac{q +1}{n}} + t\right)\right] \\
&\leq c_2\exp\big\{-c_3n\min\{{t},{t^2}\}\big\}.  
\end{split}
\end{equation*}
Thus, by a simple union bound, we conclude that
\begin{equation*}
    \begin{split}
       & \P\left[\max_{j \in {[m]}} \opnorm{\frac{1}{n}\sum_{i=1}^n \big(\mathbf{H}_i^{(j)}\big(\mathbf{H}_i^{(j)}\big)^\top -\mathbb{E}\mathbf{H}_i^{(j)}\big(\mathbf{H}_i^{(j)}\big)^\top \big)} \geq  \opnorm{\mathbf{\mathcal{H}}^{(j)}} \left(2c_1\sqrt{\frac{q +1}{n}} + t\right)\right] \\
       &
       \leq c_2m\exp\big\{-c_3n\min\{{t},{t^2}\}\big\}
    \end{split}
\end{equation*}  
Now, we write $c_2 = e^{\gamma_2 ({q + \log m})}$ for an appropriate constant $\gamma_2$ and  take $t^2 =   \frac{1 + C' + (\gamma_2)_+}{c_3}\cdot \frac{q + \log m}{n}$ for any $C'>0$ while assuming $c_0$ small enough so that $ t \leq 1$. This implies that
\begin{equation}
    \begin{split} 
    &\P\left[\max_{j \in {[m]}} \opnorm{\frac{1}{n}\sum_{i=1}^n \big(\mathbf{H}_i^{(j)}\big(\mathbf{H}_i^{(j)}\big)^\top -\mathbb{E}\mathbf{H}_i^{(j)}\big(\mathbf{H}_i^{(j)}\big)^\top \big)} \geq C'_3 \times \left(1  + \frac{M}{\kappa_2}\right)\sqrt{\frac{q + \log m}{n} }\right] \\
    &\leq \exp\big\{ \gamma_2 (q + \log m)+ \log m -\left(1+ C' + (\gamma_2)_+\right) (q +\log m) \big\}\\
    &\leq \exp\{-C' (q +\log m)\}
        \end{split}
        \end{equation}
where $C = 2c_1 + \sqrt{\frac{1 + C' + (\gamma_2)_+}{c_3}}. $
\end{proof}

\begin{lemma}\label{lemma:gamma_dagger_XtY}
Let $(\mathbf{X},\mathbf{Y})\sim \mathcal N(0, \mathbf{\Sigma})$ where $\mathbf{\Sigma}$ is a covariance matrix in $\mathcal{F}_{\mathcal{G}}(s_u,p, m, q,r,\lambda, M)$ as defined in (\ref{eq:family_f_g}). Let $\mathbf{\Gamma}$ be the incidence matrix corresponding to the graph $G$ on $p$ nodes and $m$ edges. Denote the support of $\mathbf{\Gamma} \mathbf{U}$ as $\supp(\mathbf{\Gamma} \mathbf{U})$. Assume that $\mathcal{H}_{1,\mathcal{G}}$ holds and that $|\supp(\mathbf{\Gamma} \mathbf{U})| \leq s_u$.
Then, for a constant $C'>0$, there exists $C>0$ that solely depends on $C'$ and $c_0$ such that:
     \begin{equation}
{\max_{j \in [m]} \left\| \left[(\mathbf{\Gamma}^{\dagger})^\top  \frac{\mathbf{X}^\top  \mathbf{Y}}{n} \mathbf{\Sigma}_Y^{-\frac12}\right]_{j\cdot}\right\|}  \leq  C \left(1 + {\frac{M}{\kappa_2}}\right) \cdot \sqrt{\frac{q + \log m}{n} } + \sqrt{\frac{M}{\kappa_2}}
    \end{equation}
    with probability at least $1 -\exp\{-C'(q+\log m)\}$. Here we assume that $c_0$ is small enough so that $C'c_0\leq c_3 -\log c_2 -1$ with $c_3, c_2$ the universal constants of Theorem 6.5 of \cite{wainwright2019high}.
    
\end{lemma}

\begin{proof}
Consider the term:
\begin{equation}\label{eq:1}
\begin{split}
\max_{j \in {[m]}} \left\| \left[(\mathbf{\Gamma}^{\dagger})^\top  \frac{\mathbf{X}^\top  \mathbf{Y}}{n} \mathbf{\Sigma}_Y^{-\frac12}\right]_{j\cdot}\right\|  
\leq &\max_{j \in {[m]}} \left\| \left[(\mathbf{\Gamma}^{\dagger})^\top  \frac{\mathbf{X}^\top  \mathbf{Y}}{n} \mathbf{\Sigma}_Y^{-\frac12} - \mathbf{\Sigma}_{X\Gamma^{\dagger}, Y\Sigma_Y^{-\frac12}}\right]_{j \cdot }\right\| \\+ &\max_{j \in {[m]}} \left\| \left[ \mathbf{\Sigma}_{X\Gamma^{\dagger}, Y\Sigma_Y^{-\frac12}}\right]_{j \cdot }\right\|
\end{split}
\end{equation}
Introducing $\mathbbm{1}_j^\top \in \R^{1 \times m}$ the vector indicator of row $j$ (i.e. $[\mathbbm{1}_j]_i=0$ for all $i\neq j$, and  $[\mathbbm{1}_j]_j=1$), we observe that:
\begin{equation}
    \max_{j \in {[m]}} \left\| \left[ \mathbf{\Sigma}_{X\Gamma^{\dagger}, Y\Sigma_Y^{-\frac12}}\right]_{j \cdot }\right\| = \max_{j \in {[m]}} \fnorm{  \mathbbm{1}_j^\top  \mathbf{\Sigma}_{X\Gamma^{\dagger}, Y\Sigma_Y^{-\frac12}}}\leq \opnorm{ \mathbf{\Sigma}_{X\Gamma^{\dagger}, Y\Sigma_Y^{-\frac12}}}\leq \sqrt{\frac{M}{\kappa_2}} \lambda_1.
\end{equation}

\noindent We now turn to the first term on the RHS of equation~(\ref{eq:1}). The proof is identical to that of Lemma~\ref{lemma:gamma_dagger_w}, by introducing the transformed variables ${\mathbf{Z}_i = \mathbf{X}_i\mathbf{\Gamma}^{\dagger}\sim \mathcal{N}_m}\big(0, (\mathbf{\Gamma}^{\dagger})^\top \mathbf{\Sigma}_X\mathbf{\Gamma}^{\dagger}\big)$ and $\widetilde{\mathbf{Y}}_i =\mathbf{Y}_i \mathbf{\Sigma}_Y^{-\frac12}\sim \mathcal{N}_q(0, \mathbf{I}_q)$, and  writing
$$\mathbbm{1}_j^\top\left((\mathbf{\Gamma}^{\dagger})^\top \widehat{\mathbf{\Sigma}}_{XY}\mathbf{\Sigma}_Y^{-\frac12}-(\mathbf{\Gamma}^{\dagger})^\top \mathbf{\Sigma}_{XY}\mathbf{\Sigma}_Y^{-\frac12}\right)=\frac{1}{n}\sum_{i=1}^n\big(\mathbbm{1}_j^\top \mathbf{Z}_i\widetilde{\mathbf{Y}}_i^\top-\mathbb{E}(\mathbbm{1}_j^\top \mathbf{Z}_i\widetilde{\mathbf{Y}}_i^\top)\big).$$
Define for each ${j \in [m]}$ the matrix 
$\mathbf{H}_i^{(j)}=\begin{pmatrix}
\mathbbm{1}_j^\top \mathbf{Z}_i \\
\widetilde{\mathbf{Y}}_i
\end{pmatrix}.$
We have then:
\begin{equation}
    \begin{split}
        \max_{j\in{[m]}}  \bigg\|\mathbbm{1}_j^\top\Big((\mathbf{\Gamma}^{\dagger})^\top \widehat{\mathbf{\Sigma}}_{XY}\mathbf{\Sigma}_Y^{-\frac12}-(\mathbf{\Gamma}^{\dagger})^\top \mathbf{\Sigma}_{XY}\mathbf{\Sigma}_Y^{-\frac12}\Big)
        \bigg\| &\leq \max_{j\in{[m]}}\opnorm{\frac{1}{n}\sum_{i=1}^n \big(\mathbf{H}_i^{(j)}\big(\mathbf{H}_i^{(j)}\big)^\top-\mathbb{E}\mathbf{H}_i^{(j)}\big(\mathbf{H}_i^{(j)}\big)^\top\big)}\\
    \end{split}
\end{equation}
Denoting $\mathbf{\mathcal{H}}^{(j)}=\mathbb{E}\mathbf{H}_i^{(j)}\big(\mathbf{H}_i^{(j)}\big)^\top  = \begin{pmatrix}
\mathbbm{1}_j^\top  (\mathbf{\Gamma}^{\dagger})^\top  \mathbf{\Sigma}_X(\mathbf{\Gamma}^{\dagger}) \mathbbm{1}_j & \mathbbm{1}_j^\top  (\mathbf{\Gamma}^{\dagger})^\top  \mathbf{\Sigma}_{XY}\mathbf{\Sigma}_{Y}^{-\frac12}\\
\mathbf{\Sigma}_{Y}^{-\frac12}\mathbf{\Sigma}_{YX} (\mathbf{\Gamma}^{\dagger})\mathbbm{1}_j & \mathbf{I}_q
\end{pmatrix} $, this implies:
\begin{equation}
    \begin{split}
        \opnorm{\mathbf{\mathcal{H}}^{(j)}} & = \sup_{\substack{u_1\in \R,~u_2 \in \R^{q}\\ u_1^2 + \|u_2\|^2=1}} \mathbbm{1}_j^\top  (\mathbf{\Gamma}^{\dagger})^\top  \mathbf{\Sigma}_X(\mathbf{\Gamma}^{\dagger})\mathbbm{1}_j u_1^2 + 2 u_1 \mathbbm{1}_j^\top (\mathbf{\Gamma}^{\dagger})^\top  \mathbf{\Sigma}_{XY} \mathbf{\Sigma}_Y^{-\frac12} u_2 + {u_2^\top u_2}\\
        & \leq \sup_{\substack{u_1\in \R,~u_2 \in \R^{q}\\ u_1^2 + \|u_2\|^2=1}} \frac{M}{\kappa_2} u_1^2 + 2  \lambda_1 \sqrt{\frac{M}{\kappa_2}} u_1 \| u_2\|+ \|u_2\|^2\\
        & \leq \sup_{\substack{u_1\in \R,~u_2 \in \R^{q}\\ u_1^2 + \|u_2\|^2=1}} {\frac{M}{\kappa_2}} u_1^2 +   \sqrt{\frac{M}{\kappa_2}}(u_1^2  + \| u_2\|^2)+  \|u_2\|^2\\
        &= 2 \left (1+\frac{M}{\kappa_2}\right)  \\
    \end{split}
\end{equation}

By the same arguments as for Lemma~\ref{lemma:gamma_dagger_w}, we can show that for a constant $C'>0$, there exists $C>0$ that solely depends on $C'$ such that, for $c_0$ small enough:
\begin{equation}
    \begin{split} 
    &\P\left[\max_{j \in {[m]}} \opnorm{\frac{1}{n}\sum_{i=1}^n (\mathbf{H}_i^{(j)}\big(\mathbf{H}_i^{(j)}\big)^\top -\mathbb{E}\mathbf{H}_i^{(j)}\big(\mathbf{H}_i^{(j)}\big)^\top )} \geq C \cdot 2\left(1 + {\frac{M}{\kappa_2}}\right) \cdot \sqrt{\frac{q + \log m}{n} }\right] \\
    &\leq \exp\big\{- C'(q + \log m)\big\}.
            \end{split}
        \end{equation}
        This concludes the proof.
\end{proof}

\begin{lemma}\label{lemma:frob_norm_SigmaXPiY}
    Let $\mathbf{X}, \mathbf{Y} \sim  \mathcal{F}_{\mathcal{G}}(s_u,p, m, q,r,\lambda, M)$ as defined in (\ref{eq:family_f_g}). Assume that
\begin{equation}
   M \frac{\log n_c+q}{n} \leq c_0, \tag{$\mathcal{H}_{1,\mathcal{G}, \Pi}$}
\end{equation}
with $c_0 \in (0,1).$
Then, for a constant $C'>0$, there exists $C>0$ that solely depends on $C'$ such that  
    \begin{equation}
        \fnorm{\mathbf{\Pi}^\top  \frac{\mathbf{X}^\top  \mathbf{Y}}{n} \mathbf{\Sigma}_Y^{-\frac12}} \leq \sqrt{M n_c} \left( C \sqrt{M}  \cdot \sqrt{\frac{q + \log n_c}{n}}   + 1\right)
    \end{equation}
with probability at least $1-\exp\big\{ -C'(q + \log n_c)\big\}$. Here we assume that $c_0$ is small enough so that $C'c_0\leq c_3 -\log c_2 -1$ with $c_3, c_2$ the universal constants of Theorem 6.5 of \cite{wainwright2019high}.
\end{lemma}

\begin{proof}
In this case:
\begin{equation*}
\begin{split}
      \fnorm{\mathbf{\Pi} \widehat{\mathbf{\Sigma}}_{ XY}\mathbf{\Sigma}_Y^{-\frac12}}^2 
      &= \sum_{j \in [n_c]} \fnorm{ \mathbbm{1}_{C_j} \frac{\mathbbm{1}_{C_j}^\top }{|C_j|} \widehat{\mathbf{\Sigma}}_{XY} \mathbf{\Sigma}_Y^{-\frac12}}^2\\
      &=  \sum_{j \in [n_c]} \Bigg\| \frac{\mathbbm{1}_{C_j}^\top }{\sqrt{|C_j|}} \widehat{\mathbf{\Sigma}}_{XY} \mathbf{\Sigma}_Y^{-\frac12}\Bigg\|^2 \qquad (\star)\\
          &\leq n_c  \max_{j \in [n_c]} \left\| \frac{\mathbbm{1}_{C_j}^\top }{\sqrt{|C_j|}} \widehat{\mathbf{\Sigma}}_{XY} \mathbf{\Sigma}_Y^{-\frac12}\right\|^2\\
\end{split}
\end{equation*}
where in $(\star)$, we have used the fact that
$$  \fnorm{ \mathbbm{1}_{C_j} \frac{\mathbbm{1}_{C_j}^\top }{|C_j|} \widehat{\mathbf{\Sigma}}_{XY} \mathbf{\Sigma}_Y^{-\frac12}}^2 
= \sum_{k \in C_j} \frac{1}{|C_j|} \left\| \left[ \mathbbm{1}_{C_j}\frac{\mathbbm{1}_{C_j}^\top }{\sqrt{|C_j|}} \widehat{\mathbf{\Sigma}}_{XY} \mathbf{\Sigma}_Y^{-\frac12}\right]_{k\cdot} \right\|^2=\left\| \frac{\mathbbm{1}_{C_j}^\top }{\sqrt{|C_j|}} \widehat{\mathbf{\Sigma}}_{XY} \mathbf{\Sigma}_Y^{-\frac12} \right\|^2. $$
We now proceed to bound the term:
\begin{equation}
    \begin{split}
   \max_{j \in [n_c]} \left\| \frac{\mathbbm{1}_{C_j}^\top }{\sqrt{|C_j|}} \widehat{\mathbf{\Sigma}}_{XY} \mathbf{\Sigma}_Y^{-\frac12}\right\|^2 
   &\leq    \max_{j \in [n_c]} \left\| \frac{\mathbbm{1}_{C_j}^\top }{\sqrt{|C_j|}}\Big( \widehat{\mathbf{\Sigma}}_{XY} \mathbf{\Sigma}_Y^{-\frac12} - \mathbf{\Sigma}_{X, Y\mathbf{\Sigma}_Y^{-\frac12}}\Big)\right\|^2  +\max_{j \in [n_c]} \left\| \frac{\mathbbm{1}_{C_j}^\top }{\sqrt{|C_j|}} \mathbf{\Sigma}_{X, Y\mathbf{\Sigma}_Y^{-\frac12}}\right\|^2\\
   &\leq    \max_{j \in [n_c]} \left\| \frac{\mathbbm{1}_{C_j}^\top }{\sqrt{|C_j|}}\Big( \widehat{\mathbf{\Sigma}}_{XY} \mathbf{\Sigma}_Y^{-\frac12} - \mathbf{\Sigma}_{X, Y\mathbf{\Sigma}_Y^{-\frac12}}\Big)\right\|^2  +\opnorm{\mathbf{\Sigma}_{X, Y\mathbf{\Sigma}_Y^{-\frac12}}}^2 \\
   &\leq    \max_{j \in [n_c]} \left\| \frac{\mathbbm{1}_{C_j}^\top }{\sqrt{|C_j|}}\Big( \widehat{\mathbf{\Sigma}}_{XY} \mathbf{\Sigma}_Y^{-\frac12} - \mathbf{\Sigma}_{X, Y\mathbf{\Sigma}_Y^{-\frac12}}\Big)\right\|^2  + {M}\lambda_1^2 
    \end{split}
\end{equation}
We follow once again the outline of the proof of Lemma 7 in \cite{gao2017sparse}. Defining $\mathbf{Z}_i \sim \mathcal{N}(0, \mathbf{\Sigma}_X)$ and $\widetilde{\mathbf{Y}}_i \sim \mathcal{N}(0, \mathbf{I}_q)$, we may write
$$\frac{\mathbbm{1}^\top _{C_j}}{\sqrt{|C_j|}}\Big(\widehat{\mathbf{\Sigma}}_{XY}\mathbf{\Sigma}_Y^{-\frac12}-\mathbf{\Sigma}_{XY}\mathbf{\Sigma}_Y^{-\frac12}\Big)=\frac{1}{n}\sum_{i=1}^n\bigg(\frac{\mathbbm{1}^\top _{C_j}}{\sqrt{|C_j|}}\mathbf{Z}_i\widetilde{\mathbf{Y}}_i^\top -\mathbb{E}\bigg(\frac{\mathbbm{1}^\top _{C_j}}{\sqrt{|C_j|}}\mathbf{Z}_i\widetilde{\mathbf{Y}}_i^\top \bigg)\bigg).$$
Then, define
$\mathbf{H}_i^{(j)}=\begin{pmatrix}
\frac{\mathbbm{1}^\top _{C_j}}{\sqrt{|C_j|}}\mathbf{Z}_i \\
\widetilde{\mathbf{Y}}_i
\end{pmatrix}$. Since $\frac{\mathbbm{1}^\top _{C_j}}{\sqrt{|C_j|}}\Big(\widehat{\mathbf{\Sigma}}_{XY}\mathbf{\Sigma}_Y^{-\frac12}-\mathbf{\Sigma}_{XY}\mathbf{\Sigma}_Y^{-\frac12}\Big)$ is a submatrix of $\mathbf{H}_i^{(j)}(\mathbf{H}_i^{(j)})^\top,$ we have\begin{equation}
    \begin{split}
        \max_{j\in [n_c]}  \left\|\frac{\mathbbm{1}^\top _{C_j}}{\sqrt{|C_j|}}(\widehat{\mathbf{\Sigma}}_{XY}\mathbf{\Sigma}_Y^{-\frac12}-\mathbf{\Sigma}_{XY}\mathbf{\Sigma}_Y^{-\frac12})\right\| 
        &\leq \max_{j\in [n_c]}\opnorm{\frac{1}{n}\sum_{i=1}^n \big(\mathbf{H}_i^{(j)}(\mathbf{H}_i^{(j)})^\top -\mathbb{E}\mathbf{H}_i^{(j)}(\mathbf{H}_i^{(j)})^\top \big)}\\
    \end{split}
\end{equation}
where $\mathbf{\mathcal{H}}^{(j)}=\mathbb{E}\mathbf{H}_i^{(j)}(\mathbf{H}_i^{(j)})^\top  = \begin{pmatrix}
\frac{\mathbbm{1}^\top _{C_j}}{\sqrt{|C_j|}}\mathbf{\Sigma}_X \frac{\mathbbm{1}_{C_j}}{\sqrt{|C_j|}} & \frac{\mathbbm{1}^\top _{C_j}}{\sqrt{|C_j|}}\mathbf{\Sigma}_{XY}\mathbf{\Sigma}_Y^{-\frac12}\\
\mathbf{\Sigma}_Y^{-\frac12}\mathbf{\Sigma}_{YX} \frac{\mathbbm{1}_{C_j}}{\sqrt{|C_j|}} & \mathbf{I}_q
\end{pmatrix} $.
We also note that 
\begin{equation}\label{eq:bound_op_norm_h}
    \begin{split}
        \opnorm{\mathcal{H}^{(j)}} 
        & = \sup_{\substack{u_1\in \R,~u_2 \in \R^{q}\\ u_1^2 + \|u_2\|^2=1}} \frac{\mathbbm{1}^\top _{C_j}}{\sqrt{|C_j|}}\mathbf{\Sigma}_X\frac{\mathbbm{1}_{C_j}}{\sqrt{|C_j|}}u_1^2 + 2 u_1 \frac{\mathbbm{1}^\top _{C_j}}{\sqrt{|C_j|}}\mathbf{\Sigma}_{XY} \mathbf{\Sigma}_Y^{-\frac12} u_2 + u_2^\top u_2\\
        & \leq \sup_{\substack{u_1\in \R,~u_2 \in \R^{q}\\ u_1^2 + \|u_2\|^2=1}} M u_1^2 + 2  \lambda_1 \sqrt{M} u_1 \| u_2\|+  \|u_2\|^2\\
        & \leq \sup_{\substack{u_1\in \R,~u_2 \in \R^{q}\\ u_1^2 + \|u_2\|^2=1}} M u_1^2 +   \lambda_1 \sqrt{M} \left(u_1^2  + \| u_2\|^2\right)+ M \|u_2\|^2\\
        &\leq 2M
    \end{split}
\end{equation}
since $\lambda_1\leq 1$ and $\sqrt{M} \leq M.$ By similar arguments to Lemma~\ref{lemma:gamma_dagger_w} (application of a union bound combined with Theorem 6.5 of \cite{wainwright2019high}, we conclude that for a constant $C'>0$, there exists $C>0$ such that:
\begin{equation}\label{eq:bound_sigma_pi_xy_graph}
    \begin{split} 
    &\P\left[\max_{j \in [n_c]} \opnorm{\frac{1}{n}\sum_{i=1}^n \big(\mathbf{H}_i^{(j)}(\mathbf{H}_i^{(j)})^\top -\mathbb{E}\mathbf{H}_i^{(j)}(\mathbf{H}_i^{(j)})^\top \big)} \geq C\cdot 2M\cdot\sqrt{\frac{q + \log n_c}{n} }\right] \\
    &\leq \exp\big\{- C'(q + \log n_c)\big\} 
            \end{split}
        \end{equation}

\noindent We conclude the proof by noting that this implies
 \begin{equation}\label{eq:bound_sigma_pi_xy_graph_conclusion}
    \begin{split} 
       \max_{j \in [n_c]} \left\| \frac{\mathbbm{1}_{C_j}^\top }{\sqrt{|C_j|}} \widehat{\mathbf{\Sigma}}_{XY} \mathbf{\Sigma}_Y^{-\frac12}\right\|^2 &\leq    C^2\cdot 4M^2\frac{q + \log n_c}{n}   + {M}\lambda_1^2 
    \end{split}  
        \end{equation}
with the same probability.
\end{proof}

\begin{lemma}\label{lemma:frob_norm_SigmaXPiW}
Let $(\mathbf{X},\mathbf{Y})\sim \mathcal{N}(0, \mathbf{\Sigma})$ where $\mathbf{\Sigma}$ is a covariance matrix in $\mathcal{F}_{\mathcal{G}}(s_u,p, m, q,r,\lambda, M)$ as defined in (\ref{eq:family_f_g}). 
Let $\mathbf{W} \sim \mathcal{N}(0, \mathbf{I}_q)$ and independent of $(\mathbf{X},\mathbf{Y})$.
Assume ($\mathcal{H}_{1,\mathcal{G}, \Pi}$)
for $c_0\in(0,1).$
Then, for a constant $C'>0$, there exists a constant $C>0$ that depends solely on $C'$ such that
    \begin{equation}
       \fnorm{ \frac{\mathbf{\Pi}^\top  \mathbf{X}^\top  \mathbf{W}}{n} }  \leq CM\sqrt{ n_c} \sqrt{\frac{q + \log n_c}{n}}   \end{equation}
with probability at least $1-\exp\big\{- C'(q + \log n_c)\big\}$.
Here we assume that $c_0$ is small enough so that $C'c_0\leq c_3 -\log c_2 -1$, with $c_3, c_2$ the universal constants of Theorem 6.5 of \cite{wainwright2019high}.
    
    
\end{lemma}

\begin{proof}
The proof is identical to that of Lemma~\ref{lemma:frob_norm_SigmaXPiY}, by replacing the matrix $\mathbf{\Sigma}_{XY}\mathbf{\Sigma}_{Y}^{-\frac12}$ with $\mathbf{\Sigma}_{XW},$ which is 0 since $\mathbf{X}$ and $\mathbf{W}$ are assumed to be independent, and  adjusting the corresponding $\opnorm{\mathbf{\mathcal{H}}^{(j)}}.$

\end{proof}

\begin{lemma}\label{lemma:max_norm}
    Let $\mathbf{X}_i \sim \mathcal{N}(0, \mathbf{\Sigma}_X)$ where $\mathbf{\Sigma}_X$ is a covariance matrix as defined in (\ref{eq:family_f_g}).  Let $\mathbf{W}_i \sim \mathcal{N}(0, \bar{\mathbf{\Sigma}})$ independently of $\mathbf{X}.$
Assume
\begin{equation}
  M  \frac{\log m + q}{n} \leq c_0 \tag{$\mathcal{H}_{1,\mathcal{G}}$}
\end{equation}
Then, for a constant $C'>0$, there exists a constant $C>0$ that depends solely on $C'$ such that
\begin{equation}
\begin{split}
\max_{j\in{[m]}}  \left\|\mathbbm{1}_j^\top\big((\mathbf{\Gamma}^{\dagger})^\top \widehat{\mathbf{\Sigma}}_{{X}W} - (\mathbf{\Gamma}^{\dagger})^\top {\mathbf{\Sigma}}_{{X}W} \big) \right\|  &\leq C\left(1 + \frac{M}{\kappa_2}\right)\sqrt{\frac{\log m + q}{n}}
    \end{split}
\end{equation}
with probability at least $1-\exp\big\{- C'(q + \log m)\big\}.$
Here we assume that $c_0$ is small enough so that $C'c_0\leq c_3 -\log c_2 -1$ with $c_3, c_2$ the universal constants of Theorem 6.5 of \cite{wainwright2019high}
\end{lemma}

\begin{proof}
Let $\mathbf{Z}_i \sim \mathcal{N}_m\big(0, (\mathbf{\Gamma}^{\dagger})^\top \mathbf{\Sigma}_X \mathbf{\Gamma}^{\dagger}\big)$ and $\widetilde{\mathbf{X}}_i \sim \mathcal{N}_q(0, \bar{\mathbf{\Sigma}})$, and $\mathbbm{1}_j^\top \in \R^{1 \times m}$ the vector indicator of row $j$ (i.e. $[\mathbbm{1}_j]_i=0$ for all $i\neq j$, and  $[\mathbbm{1}_j]_j=1$). Thus, denoting $\widetilde{\mathbf{X}} = \mathbf{X} \mathbf{\Gamma}^{\dagger}$, we may write
$$\mathbbm{1}_j^\top(\widehat{\mathbf{\Sigma}}_{\widetilde{X}W}-\mathbf{\Sigma}_{\widetilde{X}W})=\frac{1}{n}\sum_{i=1}^n\big(\mathbbm{1}_j^\top \mathbf{Z}_i\mathbf{W}_i^\top-\mathbb{E}(\mathbbm{1}_j^\top \mathbf{Z}_i\mathbf{W}_i^\top)\big).$$
We also note that 
$$\opnorm{\mathbf{\Sigma}_{\widetilde{X}}} = \opnorm{(\mathbf{\Gamma}^{\dagger})^\top  \mathbf{\Sigma}_X \mathbf{\Gamma}^{\dagger}}= \sup_{u \in \R^{m}} \frac{u^\top (\mathbf{\Gamma}^{\dagger})^\top  \mathbf{\Sigma}_X \mathbf{\Gamma}^{\dagger} u}{\|u\|^2}  \leq M \frac{\| \mathbf{\Gamma}^{\dagger} u\|^2}{\|u\|^2} \leq \frac{M}{\kappa_2}.  $$
Moreover, one can show that
$$\opnorm{\bar{\mathbf{\Sigma}}} = \opnorm{ \mathbf{I} - \mathbf{V}_0 \mathbf{\Lambda}^2 \mathbf{V}_0^\top } \leq 1.$$
Define for each ${j \in [m]}$ the matrix 
$\mathbf{H}_i^{(j)}=\begin{pmatrix}
\mathbbm{1}_j^\top \mathbf{Z}_i \\
\mathbf{W}_i
\end{pmatrix}.$
We have then
\begin{equation}
    \begin{split}
        \max_{j\in{[m]}}  \left\|\mathbbm{1}_j^\top(\widehat{\mathbf{\Sigma}}_{\widetilde{X}W} - {\mathbf{\Sigma}}_{\widetilde{X}W} ) \right\| &\leq \max_{j\in{[m]}}\opnorm{\frac{1}{n}\sum_{i=1}^n \big(\mathbf{H}_i^{(j)}\big(\mathbf{H}_i^{(j)}\big)^\top-\mathbb{E}\mathbf{H}_i^{(j)}\big(\mathbf{H}_i^{(j)}\big)^\top\big)}\\
    \end{split}
\end{equation}
Denote $\mathbf{\mathcal{H}}^{(j)}=\mathbb{E}\mathbf{H}_i^{(j)}\big(\mathbf{H}_i^{(j)}\big)^\top  = \begin{pmatrix}
\mathbbm{1}_j^\top  \mathbf{\Sigma}_{\widetilde{X}} \mathbbm{1}_j & 0 \\
0 & \bar{\mathbf{\Sigma}}
\end{pmatrix}$. This implies:
\begin{equation}
    \begin{split}
        \opnorm{\mathbf{\mathcal{H}}^{(j)}} & = \sup_{\substack{u_1\in \R,~u_2 \in \R^{q}\\ u_1^2 + \|u_2\|^2=1}} \mathbbm{1}_j^\top  \mathbf{\Sigma}_{\widetilde{X}}\mathbbm{1}_j u_1^2 +  {u_2^\top \bar{\mathbf{\Sigma}} u_2}\\
        & \leq \sup_{\substack{u_1\in \R,~u_2 \in \R^{q}\\ u_1^2 + \|u_2\|^2=1}} \frac{M}{\kappa_2} u_1^2 +  \|u_2\|^2\\
        & \leq \sup_{\substack{u_1\in \R,~u_2 \in \R^{q}\\ u_1^2 + \|u_2\|^2=1}} \frac{M}{\kappa_2} u_1^2 +  \sqrt{\frac{M}{\kappa_2}} (u_1^2  + \| u_2\|^2)+ \|u_2\|^2\\
        &\leq 1  + \frac{M}{\kappa_2},  \\
    \end{split}
\end{equation}
Therefore, by arguments similar to Lemma~\ref{lemma:gamma_dagger_w} (union bound and Theorem 6.5 of \cite{wainwright2019high}), we conclude that for any $C'>0$ such that $C'c_0\leq c_3 -\log c_2 -1$, with $c_3, c_2$ the universal constants of Theorem 6.5 of \cite{wainwright2019high}, there exists $C>0$ such that:
\begin{equation}
    \begin{split} 
    &\P\left[\max_{j \in {[m]}} \opnorm{\frac{1}{n}\sum_{i=1}^n \big(\mathbf{H}_i^{(j)}\big(\mathbf{H}_i^{(j)}\big)^\top -\mathbb{E}\mathbf{H}_i^{(j)}\big(\mathbf{H}_i^{(j)}\big)^\top \big)} \geq C \cdot \left(1  + \frac{M}{\kappa_2}\right)\sqrt{\frac{q + \log m}{n} }\right] \\
    &\leq \exp\{-C'(\log m + q) \}.
            \end{split}
        \end{equation}
   This concludes the proof.     
\end{proof}

\subsection{Proof of graph CCAR$^{3}$ consistency}\label{proof:theorem:graph}

\begin{theorem}\label{theorem:cca_rrr_graph_ols}
    Consider the family $\mathcal{F}_{\mathcal{G}}(s_u,p, m, q,r,\lambda, M)$ of covariance matrices satisfying assumptions.
    {
    Assume
    \begin{equation}\label{assumption1G_graph_corrected}
        M\frac{q+\log m}{n}\le c_0,\qquad
        M\frac{q+\log n_c}{n}\le c_0.\tag{$\mathcal{H}_{1,\mathcal{G}}$}
    \end{equation}
    Define
    \[
    K_{\Gamma,n}=C_\star\sqrt{n_c\frac{q+\log n_c}{q+\log m}}
    +4\sqrt{s_u\sigma_{\max}(\mathbf L)},
    \]
    where $C_\star$ is a numerical constant depending only on the constants in the score bounds.
    Assume in addition that
    \[
    n\ge 10\vee 128M^2\left(72n_c+576\rho(\mathbf\Gamma)^2\log p\,K_{\Gamma,n}^2\right).
    \]
    }
We write $\mathbf{\Delta} = \widehat{\mathbf{B}} - \mathbf{B}^*$ with  $\mathbf{B}^*  = \mathbf{U}\mathbf{\Lambda} \mathbf{V}^\top\mathbf{\Sigma}_{Y}^{\frac12} $, and choose ${\rho \geq  C M^2\left(1+\frac{1}{\kappa_2}\right)  \sqrt{\frac{q + \log m}{n}}}$ for some large constant $C$. Then the solution $\widehat{\mathbf B}$  of the penalized regression problem is such that, there exist constants $C_1$ and $C_2$ that do not depend on the parameters of the problem such that
\begin{equation*}
    \begin{split}
    \fnorm{\widehat{\mathbf{\Sigma}}_X^{1/2} \mathbf{\Delta}}^2 &\leq    C_1 M^5 \left( {n_c\frac{q+\log(n_c)}{n}} +  {\left(1 + \frac{1}{\kappa_2}\right)^2 \sigma_{\max}(\mathbf{L})\cdot {s_u\frac{q + \log m}{n} }}  \right) \\
    \qquad \fnorm{\mathbf{\Delta} } &\leq  C_2 M^3\left(\sqrt{n_c\frac{q+\log(n_c)}{n}}  +\sqrt{ \sigma_{\max}(\mathbf{L})}  \left(1 + \frac{1}{\kappa_2}\right) \sqrt{s_u\frac{q + \log m}{n} }\right) 
\end{split}
\end{equation*}
with probability at least
{
\[
1-\exp\{-C'q\}-2\exp\{-C'(q+\log n_c)\}
-2\exp\{-C'(q+\log m)\}-q\exp\{-c_2n\}.
\]
}
\end{theorem}

\begin{proof}
We begin by the following observation. Since $(\mathbf{X}, \mathbf{Y})$ are jointly multivariate normal and writing $\mathbf{V}_0= \mathbf{\Sigma}_Y^{\frac12}\mathbf{V}$, conditioned on $\mathbf{X}$, the following holds:
$$ \mathbf{Y} | \mathbf{X} \sim \mathcal{N}\left(\mathbf{X}\mathbf{B}^*\mathbf{\Sigma}_Y^{\frac12},~\mathbf{\Sigma}_Y^{\frac12} (\mathbf{I} - \mathbf{V}_0 \mathbf{\Lambda}^2 \mathbf{V}_0^\top ) \mathbf{\Sigma}_Y^{\frac12}\right) $$
Therefore

\begin{equation}
    \mathbf{Y} \mathbf{\Sigma}_Y^{-\frac12} | \mathbf{X} \sim \mathcal{N}\left(\mathbf{X}\mathbf{B}^*,~ (\mathbf{I} - \mathbf{V}_0 \mathbf{\Lambda}^2 \mathbf{V}_0^\top)\right) \tag{*}
\end{equation}  
{We work under the sample-size assumptions stated in the theorem.} 
{

We now turn to the analysis of the regression problem
\[
\widehat{\mathbf B}
\in
\operatorname*{argmin}_{\mathbf B\in\mathbb R^{p\times q}}
\left\{
\frac1n\|\mathbf Y_0-\mathbf X\mathbf B\|_F^2
+\rho\|\mathbf\Gamma\mathbf B\|_{21}
\right\},
\qquad
\mathbf Y_0=\mathbf Y\widehat{\mathbf\Sigma}_Y^{-1/2}.
\]
The convex first-order optimality condition gives, for every
$\mathbf B\in\mathbb R^{p\times q}$,
\[
\frac{2}{n}
\big\langle \mathbf X(\mathbf B-\widehat{\mathbf B}),
\mathbf Y_0-\mathbf X\widehat{\mathbf B}\big\rangle
\le
\rho\{\|\mathbf\Gamma\mathbf B\|_{21}
-\|\mathbf\Gamma\widehat{\mathbf B}\|_{21}\}.
\]
Since
\[
\mathbf Y\mathbf\Sigma_Y^{-1/2}=\mathbf X\mathbf B^*+\mathbf W,
\qquad
\mathbf Y_0
=\mathbf X\mathbf B^*+\mathbf W
+\mathbf Y(\widehat{\mathbf\Sigma}_Y^{-1/2}-\mathbf\Sigma_Y^{-1/2}),
\]
where $\mathbf W$ has independent rows with covariance
$\bar{\mathbf\Sigma}=\mathbf I_q-\mathbf V_0\mathbf\Lambda^2\mathbf V_0^\top$, we obtain
\begin{equation}
    \begin{split}\label{eq:basic_ineq_graph}
        \frac{2}{n}  \big\langle \mathbf{X}( \widehat{\mathbf B} -\mathbf{B}),&~\mathbf{X} (\widehat{\mathbf B} - \mathbf{B}^*) \big\rangle \\
        \leq    & \frac{2}{n}  \big\langle  \mathbf{X} (\widehat{\mathbf B} - \mathbf{B}) , \mathbf{Y} \big(\widehat{\mathbf{\Sigma}}_Y^{-\frac12}- \mathbf{\Sigma}_Y^{-\frac12}\big) + \mathbf{W}\big\rangle+  \rho \big( \| \mathbf{\Gamma} \mathbf{B}\|_{21} -  \| \mathbf{\Gamma} \widehat{\mathbf B}\|_{21}\big ).
    \end{split}
\end{equation}
}

This means that, to proceed further, we need to bound the term:
\begin{equation}
    \begin{split}\label{eq:noise}
  {\frac1n}\big\langle \mathbf{X}(\widehat{\mathbf B}- \mathbf{B}),\mathbf{Y} 
  \big({\widehat{\mathbf{\Sigma}}_Y^{-\frac12}- \mathbf{\Sigma}_Y^{-\frac12}}\big) + \mathbf{W}\big\rangle 
   &={\frac1n}\big\langle \mathbf{\Pi}(\widehat{\mathbf B}- \mathbf{B}),\mathbf{\Pi}^\top \mathbf{X}^\top \big( \mathbf{Y} \big({\widehat{\mathbf{\Sigma}}_Y^{-\frac12}- \mathbf{\Sigma}_Y^{-\frac12}}\big) + \mathbf{W}\big)\big\rangle \\
  &+  {\frac1n}\big\langle \mathbf{\Gamma} (\widehat{\mathbf B}- \mathbf{B}), (\mathbf{\Gamma}^{\dagger})^\top \mathbf{X}^\top \big( \mathbf{Y} \big({\widehat{\mathbf{\Sigma}}_Y^{-\frac12}- \mathbf{\Sigma}_Y^{-\frac12}}\big) + \mathbf{W}\big)\big\rangle \\
  &\leq   \underbrace{ \frac{1}{n} \fnorm{\mathbf{\Pi} (\widehat{\mathbf B}- \mathbf{B})}~\fnorm{\mathbf{\Pi}^\top  \mathbf{X}^\top \big( \mathbf{Y} \big({\widehat{\mathbf{\Sigma}}_Y^{-\frac12}- \mathbf{\Sigma}_Y^{-\frac12}}\big) + \mathbf{W}\big)}}_{(A)} \\
  &+  \underbrace{\big\|\mathbf{\Gamma} (\widehat{\mathbf B}- \mathbf{B})\big\|_{21}~\max_{j \in {[m]}} \left\|\left[(\mathbf{\Gamma}^{\dagger})^\top  \frac{\mathbf{X}^\top  \mathbf{Y}}{n} \big({\widehat{\mathbf{\Sigma}}_Y^{-\frac12}- \mathbf{\Sigma}_Y^{-\frac12}}\big)\right]_{j\cdot}\right\|}_{(B)} \\
  &+\underbrace{\big\|\mathbf{\Gamma} (\widehat{\mathbf B}- \mathbf{B})\big\|_{21}\max_{j \in {[m]}} \left\|\left[(\mathbf{\Gamma}^{\dagger})^\top \frac{\mathbf{X}^\top \mathbf{W}}{n}\right]_{j\cdot}\right\|}_{(C)}
    \end{split}
\end{equation}

\paragraph{Bound (A) in Equation (\ref{eq:noise}).}
We first consider the term:
\begin{equation}
    \begin{split}
       &{\fnorm{\mathbf{\Pi} (\widehat{\mathbf B}- \mathbf{B})}}~ \fnorm{\mathbf{\Pi}^\top  \frac{\mathbf{X}^\top  \mathbf{Y}}{n} \big(\widehat{\mathbf{\Sigma}}_Y^{-\frac12}- \mathbf{\Sigma}_Y^{-\frac12}\big) + \frac{\mathbf{\Pi}^\top  \mathbf{X}^\top \mathbf{W}}{n}}\\
        &\leq  \|\widehat{\mathbf B}- \mathbf{B}\|_F \cdot \bigg(  \fnorm{ \frac{1}{n} \mathbf{\Pi}^\top  \mathbf{X}^\top  \mathbf{Y} \big(\widehat{\mathbf{\Sigma}}_Y^{-\frac12}- \mathbf{\Sigma}_Y^{-\frac12}\big)} + \fnorm{ \frac{1}{n} \mathbf{\Pi}^\top  \mathbf{X}^\top  \mathbf{W}}\bigg)\\
        &\leq  \|\widehat{\mathbf B}- \mathbf{B}\|_F \cdot \Big(  \fnorm{\mathbf{\Pi}  \widehat{\mathbf{\Sigma}}_{ XY}\mathbf{\Sigma}_Y^{-\frac12}}~\opnorm{\mathbf{\Sigma}_Y^{\frac12}\widehat{\mathbf{\Sigma}}_Y^{-\frac12}- \mathbf{I}_q} +  \fnorm{  \widehat{\mathbf{\Sigma}}_{ X\Pi,W}} \Big)\\
        &\leq  \|\widehat{\mathbf B}- \mathbf{B}\|_F \cdot \Big(  \fnorm{ \widehat{\mathbf{\Sigma}}_{ X\Pi,Y}\mathbf{\Sigma}_Y^{-\frac12}}~{\opnorm{\widehat{\mathbf{\Sigma}}_Y^{\frac12}- \mathbf{\Sigma}_Y^{\frac12}}}\opnorm{\widehat{\mathbf{\Sigma}}_Y^{-\frac12}} +  \fnorm{ \widehat{\mathbf{\Sigma}}_{ X\Pi,W}} \Big)
    \end{split}
\end{equation}
Let $\mathcal{A}_Y$ be the event: $\mathcal{A}_Y = \Big\{\opnorm{\widehat{\mathbf{\Sigma}}_Y^{\frac12} - \mathbf{\Sigma}_Y^{\frac12}}^2  <  CM^2\frac{q}{n} \Big\}$, with $C$ a constant chosen such that  $\mathcal{A}_Y$ holds with probability at least $1-\exp\{ -C'q\}$ with $C'>0$ (see Lemma~\ref{lem:cov}).

\xhdr{(i) Bounding the terms in operator norm} By Weyl's inequality, letting $\delta_n = \sqrt{\frac{q}{n}}$, we know that on $\mathcal{A}_Y$
\begin{equation}\label{eq:op_norm_A}
    \opnorm{\widehat{\mathbf{\Sigma}}_Y^{-\frac12}} \leq \frac{1}{\sigma_{\min}\Big(\mathbf{\Sigma}_Y^{\frac12}\Big)- \sqrt{CM^2} \cdot \delta_n }\leq 2\sqrt{M}
\end{equation} 
as long as $\sqrt{CM^2} \cdot \sqrt{\frac{q}{n}}\leq \frac{1}{2 \sqrt{M}}$.
Therefore, on $\mathcal{A}_Y$ 
$$ {\opnorm{\widehat{\mathbf{\Sigma}}_Y^{\frac12}- \mathbf{\Sigma}_Y^{\frac12}}}\opnorm{\widehat{\mathbf{\Sigma}}_Y^{-\frac12}} \leq 2 M^{\frac32}\sqrt{C} \sqrt{\frac{q}{n}}.$$
 
\xhdr{(ii) Bounding the terms in Frobenius norm} 

We begin with the term  $\fnorm{\widehat{\mathbf{\Sigma}}_{\Pi X,Y} \mathbf{\Sigma}_{Y}^{-\frac12}}$. For $C'>0$,  we denote as $\mathcal{A}_{ X\Pi,Y}$ the event
\begin{equation}\label{eq:event_a_pi_XY_graph}
\begin{split}
    \mathcal{A}_{ X\Pi,Y}= \Big\{
\fnorm{\widehat{\mathbf{\Sigma}}_{\Pi X,Y} \mathbf{\Sigma}_{Y}^{-\frac12}}  
 &\leq \sqrt{Mn_c} \Big(1 + C\sqrt{M}\sqrt{\frac{q + \log n_c}{n} }\Big) \Big\}\\
     \end{split} 
 \end{equation}
 for some constant $C$ that depends only on $C'$.
As shown in Lemma~\ref{lemma:frob_norm_SigmaXPiY}, this event has probability at least $1-\exp\{-C'(\log n_c +q) \}.$ 

On $\mathcal{A}_Y \cap \mathcal{A}_{ X\Pi, Y}$,
 after readjusting our constant $C$ to be the maximum of the constants  in the bounds of $\mathcal{A}_Y$ and $\mathcal{A}_{ X\Pi, Y}$ we have:
\begin{equation}\label{eq:bound_noise_pi}
         \begin{split}
 \fnorm{ \widehat{\mathbf{\Sigma}}_{ X\Pi,Y}\mathbf{\Sigma}_Y^{-\frac12}}~{\opnorm{\widehat{\mathbf{\Sigma}}_Y^{\frac12}- \mathbf{\Sigma}_Y^{\frac12}}}&\opnorm{\widehat{\mathbf{\Sigma}}_Y^{-\frac12}}  \\
 &\leq \sqrt{Mn_c} \left(1 + C\sqrt{M}\sqrt{\frac{q + \log n_c}{n} }\right) \cdot 2M^{\frac32} \sqrt{C} \sqrt{\frac{q}{n}}\\
 &\leq \tilde{C} M^2 \sqrt{\frac{qn_c}{n}} \Big(1 + \sqrt{M} \cdot \sqrt{\frac{q+ \log n_c}{n} } \Big)\\
  &\leq 2\tilde{C} M^{2} \sqrt{\frac{qn_c}{n}} \qquad \text{ under assumption }\mathcal{H}_{1, \mathcal{G}}.\\
     \end{split}
    \end{equation}

We now turn to the term  $\fnorm{ \mathbf{\Pi}\widehat{\mathbf{\Sigma}}_{XW} 
 }=\fnorm{ \widehat{\mathbf{\Sigma}}_{X\Pi,W} 
 }$.  By Lemma~\ref{lemma:frob_norm_SigmaXPiW}, for $C'>0$  we know that there exists $C>0$ a constant that depends solely on $C'$ such that:
     \begin{equation}\label{eq:bound_noise_pi2}
         \begin{split}
 \fnorm{ \mathbf{\Pi} \widehat{\mathbf{\Sigma}}_{ XW} }  
 &\leq  \fnorm{ \mathbf{\Pi} \widehat{\mathbf{\Sigma}}_{ X\widetilde{W}} } \opnorm{ \bar{\mathbf{\Sigma}}^{\frac12}} \quad \text{ with } \quad \widetilde{\mathbf{W}} \sim \mathcal{N}(0, \mathbf{I}_q)\\
 &\leq CM\sqrt{n_c\cdot \frac{q + \log n_c}{n}} \quad \text{since } \quad \opnorm{ \bar{\mathbf{\Sigma}}^{\frac12}} \leq 1\\
     \end{split}
    \end{equation}
    with probability at least $1 - \exp\{-C' (q + \log n_c)\}$.
In the rest of this proof, we denote as {$\mathcal{A}_{\Pi XW}$ the event}
  \begin{equation}\label{eq:event_a_pi_XW_graph}
     \begin{split}
    \mathcal{A}_{\Pi XW}= \Big\{
 {\fnorm{ \mathbf{\Pi}^\top\mathbf X^\top\mathbf W/n} \leq}  CM\sqrt{n_c\cdot \frac{q + \log n_c}{n}}  \Big\}.\\
     \end{split} 
 \end{equation}
 
 Combining the results together and taking $C$ the maximum of all constants in the corresponding bounds we deduce that on the event $ {\mathcal{A}_{\Pi XW} \cap \mathcal{A}_Y \cap \mathcal{A}_{ X\Pi, Y}}$, \begin{equation}
    \begin{split}
       \fnorm{\mathbf{\Pi}^\top  \frac{\mathbf{X}^\top  \mathbf{Y}}{n} \Big(\widehat{\mathbf{\Sigma}}_Y^{-\frac12}- \mathbf{\Sigma}_Y^{-\frac12}\Big) + \mathbf{\Pi}^\top  \frac{\mathbf{X}^\top  \mathbf{W}}{n}}&\leq   C M^{2} \sqrt{\frac{qn_c}{n}}  +  C M\sqrt{ n_c\cdot\frac{q + \log n_c}{n}}\\
                &\leq  C M^{2} \sqrt{n_c\frac{q + \log n_c}{n}}\\
    \end{split}
\end{equation}
To characterize the corresponding probability, we note that:
\begin{align}
\begin{split}
{\P\left[ \mathcal{A}_{\Pi XW}^c \cup \mathcal{A}_Y^c \cup \mathcal{A}_{ X\Pi, Y}^c\right]} 
&\leq \P\left[ \mathcal{A}^c_{\Pi XW}\right]  +\P\left[\mathcal{A}^c_Y \right] + \P\left[ \mathcal{A}_{ X\Pi, Y}^c\right] \\
&\leq  \exp\{-C'( \log n_c+q)\} +\exp\{-C'q\} + \exp\{-C' (q+ \log n_c)\}\\
&\leq 3\exp\{-C' (q + \log n_c)\}
\end{split}
\end{align}

\paragraph{Bound (B) in Equation (\ref{eq:noise}).} We now proceed to bound the term:
\begin{equation}
    \begin{split}
    \max_{j \in {[m]}} \left\|\left[(\mathbf{\Gamma}^{\dagger})^\top \frac{\mathbf{X}^\top  \mathbf{Y}}{n} \Big(\widehat{\mathbf{\Sigma}}_Y^{-\frac12}- \mathbf{\Sigma}_Y^{-\frac12}\Big)\right]_{j\cdot}\right\|
    &\leq    \max_{j \in {[m]}} \left\|\left[(\mathbf{\Gamma}^{\dagger})^\top \frac{\mathbf{X}^\top  \mathbf{Y}}{n}\mathbf{\Sigma}_Y^{-\frac12}\right]_{j\cdot}\right\|~\opnorm{\mathbf{\Sigma}_Y^{\frac12} \widehat{\mathbf{\Sigma}}_Y^{-\frac12}-\mathbf{I}_q}\\
    \leq    \max_{j \in {[m]}}& \left\|\left[(\mathbf{\Gamma}^{\dagger})^\top \frac{\mathbf{X}^\top  \mathbf{Y}}{n}\mathbf{\Sigma}_Y^{-\frac12}\right]_{j\cdot}\right\|~\opnorm{\mathbf{\Sigma}_Y^{\frac12}- \widehat{\mathbf{\Sigma}}_Y^{\frac12}}~\opnorm{\widehat{\mathbf{\Sigma}}_Y^{-\frac12}}
        \end{split}
\end{equation}

For $C'>0$ a constant, let us denote as ${\mathcal{A}_{(\Gamma^\dagger)^\top\Sigma_{XY}}}$ the event: 
$$ {\mathcal{A}_{(\Gamma^\dagger)^\top\Sigma_{XY}}} = \bigg\{   \max_{j\in {[m]}}\left\|\left[ (\mathbf{\Gamma}^{\dagger})^\top \widehat{\mathbf{\Sigma}}_{XY}{\mathbf{\Sigma}}_{Y}^{-\frac12}\right]_{j\cdot}\right\|  \geq \sqrt{\frac{M}{\kappa_2}}  + C \left(1+ \frac{M}{\kappa_2}\right) \cdot\sqrt{\frac{q + \log m}{n}}\bigg\}.$$
In the previous equation, we've chosen $C$ to be a constant that depends solely on $C'$ such that, {with probability at least $1-\exp\{- C'(q + \log m)\}-\exp\{-C'q\}$, the event ${\mathcal{A}_{(\Gamma^\dagger)^\top\Sigma_{XY}}^c \cap \mathcal{A}_{Y}}$ occurs} (see Lemma~\ref{lemma:gamma_dagger_XtY}). Thus
\begin{equation}\label{eq:bound_sigaxy}
    \begin{split}
   \max_{j\in {[m]}}&\left\|\left[(\mathbf{\Gamma}^{\dagger})^\top \widehat{\mathbf{\Sigma}}_{XY} \Big( \widehat{\mathbf{\Sigma}}_{Y}^{-\frac12}- {\mathbf{\Sigma}}_{Y}^{-\frac12}\Big)\right]_{j\cdot}\right\| \\
   &\leq  \max_{j\in {[m]}}\left\|\left[ (\mathbf{\Gamma}^{\dagger})^\top {\widehat{\mathbf{\Sigma}}}_{XY} {\mathbf{\Sigma}}_{Y}^{-\frac12}\right]_{j\cdot}\right\| \cdot \sqrt{CM^2 \frac{q}{n}} \cdot 2\sqrt{M} \quad \text{ by Eq.(\ref{eq:op_norm_A}) }\\
    &\leq   M^2 \bigg( \sqrt{\frac{1}{\kappa_2}}+ C \left(1+ \frac{1}{\kappa_2}\right) \sqrt{M}\cdot \sqrt{\frac{q + \log m}{n}} \bigg) \cdot  \sqrt{\frac{q}{n}} \quad \text{ on } {\mathcal{A}_{(\Gamma^\dagger)^\top\Sigma_{XY}}^c} \\
      &\leq    C' M^2\left(1 +\frac{1}{\kappa_2}\right) \sqrt{\frac{q}{n}} \qquad \text{under } H_{1. \mathcal{G}}.
    \end{split}
    \end{equation}
    where $C' = 1\vee C$.
\paragraph{Bound (C) in Equation (\ref{eq:noise}).}

For $\mathbf{W} \sim \mathcal{N}(0, \bar{\mathbf{\Sigma}})$ with covariance ${\bar{\mathbf{\Sigma}} = \mathbf{I}_q - \mathbf{V}_0 \mathbf{\Lambda}^2 \mathbf{V}_0^\top} $, we now proceed to bound the term:
\begin{equation}
    \begin{split}
\max_{j \in {[m]}} \left\|\left[(\mathbf{\Gamma}^{\dagger})^\top\frac{ \mathbf{X}^\top \mathbf{W}}{n}\right]_{j\cdot}\right\|
        \end{split}
\end{equation}
For $C'>0$, let us denote as ${\mathcal{A}_{X\Gamma^\dagger,W}}$ the event: 
$$ {\mathcal{A}_{X\Gamma^\dagger,W}} = \bigg\{ \max_{j \in {[m]}} \left\|\left[(\mathbf{\Gamma}^{\dagger})^\top\frac{ \mathbf{X}^\top \mathbf{W}}{n}\right]_{j\cdot}\right\| \geq C\cdot (1 +\frac{M}{\kappa_2})\sqrt{\frac{q + \log m}{n} }\bigg\}.$$
In the previous equation, we've chosen $C$ such that the event ${\mathcal{A}_{X\Gamma^\dagger,W}^c}$ occurs with probability at least ${1-\exp\{- C'(q + \log m)\}}$  (see Lemma~\ref{lemma:gamma_dagger_w}).

\paragraph{Bounding (A) +(B) + (C) in Equation (\ref{eq:noise}).}
{
Define
\[
R_\Pi=CM^2\sqrt{n_c\frac{q+\log n_c}{n}},
\qquad
R_\Gamma=CM^2\left(1+\frac{1}{\kappa_2}\right)
\sqrt{\frac{q+\log m}{n}}.
\]
Combining the three previous bounds, we obtain on the event
\[
\mathcal E_{\rm score}
=\mathcal A_Y\cap\mathcal A_{X\Pi,Y}\cap\mathcal A_{\Pi XW}
\cap {\mathcal A_{X\Gamma^\dagger,W}^c}
\cap {\mathcal A_{(\Gamma^\dagger)^\top\Sigma_{XY}}^c}
\]
that
\begin{equation}
    \begin{split}\label{eq:noise2}
  \frac{1}{n}\left|\left\langle \mathbf{X}(\widehat{\mathbf B}- \mathbf{B}),~\mathbf{Y} \Big(\widehat{\mathbf{\Sigma}}_Y^{-\frac12}- \mathbf{\Sigma}_Y^{-\frac12}\Big) + \mathbf{W}\right\rangle\right|
\leq R_\Pi\|\widehat{\mathbf B}-\mathbf B\|_F
+R_\Gamma\|\mathbf\Gamma(\widehat{\mathbf B}-\mathbf B)\|_{21}.
    \end{split}
\end{equation}
}

\paragraph{Bounding the regression error}
For any ${S\subseteq[m]}$:
\begin{equation}
    \begin{split}
\|\mathbf{\Gamma} \mathbf{B}\|_{21} -  \| \mathbf{\Gamma} \widehat{\mathbf B}\|_{21}
&\leq \big\|[\mathbf{\Gamma} (\mathbf{B} - \widehat{\mathbf B})]_{S \cdot}\big\|_{21}  + 
\big\|[\mathbf{\Gamma} \mathbf{B}]_{S^c}\big\|_{21} - \big\|[\mathbf{\Gamma} \widehat{ \mathbf{B}}]_{S^c}\big\|_{21} \\
&\leq \big\|[\mathbf{\Gamma} (\mathbf{B} - \widehat{\mathbf B})]_{S \cdot}\big\|_{21}  + 2\big\|[\mathbf{\Gamma} \mathbf{B}]_{S^c}\big\|_{21} -   \big\|[\mathbf{\Gamma} (\widehat{ \mathbf{B}} - \mathbf{B})]_{S^c}\big\|_{21}.
    \end{split}
\end{equation}
{Since $\rho\ge 4R_\Gamma$, Equations~\eqref{eq:basic_ineq_graph} and~\eqref{eq:noise2} imply, on $\mathcal E_{\rm score}$,}
\begin{equation}
    \begin{split}
        \frac{2}{n}  \left\langle \mathbf{X}(\widehat{\mathbf B} -\mathbf{B}), \mathbf{X} (\widehat{\mathbf B} - \mathbf{B}^*) \right\rangle
        \leq 2R_\Pi\|\widehat{\mathbf B}-\mathbf B\|_F
        +\frac{\rho}{2}\|\mathbf\Gamma(\widehat{\mathbf B}-\mathbf B)\|_{21} \\
        \qquad+  \rho \Big(\big\|[\mathbf{\Gamma} (\mathbf{B} - \widehat{\mathbf B})]_{S \cdot}\big\|_{21}  + 2\big\|[\mathbf{\Gamma} \mathbf{B}]_{S^c}\big\|_{21} -   \big\|[\mathbf{\Gamma} (\widehat{ \mathbf{B}} - \mathbf{B})]_{S^c}\big\|_{21}\Big ).
    \end{split}
\end{equation}
Therefore, for $\mathbf{B}=\mathbf{B}^*$ and choosing ${S=\supp(\mathbf\Gamma\mathbf B^*)\subseteq\supp(\mathbf\Gamma\mathbf U)}$, so that $|S|\le s_u$ and $[\mathbf\Gamma\mathbf B^*]_{S^c\cdot}=0$, we have:
\begin{equation}\label{eq:BI}
    \begin{split}
        \frac{1}{n}\| \mathbf{X} \mathbf{\Delta}\|_F^2 
        \leq C R_\Pi\|\mathbf{\Delta}\|_F
        +  \rho \bigg( \frac{3}{2} \big\| [\mathbf{\Gamma} \mathbf{\Delta}]_{S\cdot}\big\|_{21}  - \frac{1}{2} \big\| [\mathbf{\Gamma} \mathbf{\Delta} ]_{S^c\cdot}\big\|_{21}\bigg ),
    \end{split}
\end{equation}
 where we denote $\mathbf{\Delta} = \widehat{\mathbf B}-\mathbf{B}^*$.
Note that, by Cauchy-Schwarz,
\begin{equation}\label{eq:CS}
    \begin{split}
         \big\| [\mathbf{\Gamma} \mathbf{\Delta}]_{S\cdot}\big\|_{21}&= 
         \sum_{j \in S}  \big\| [\mathbf{\Gamma} \mathbf{\Delta}]_{j \cdot}  \big\| \leq  \sqrt{s_u} \sqrt{\sum_{j \in S}  \big\| [\mathbf{\Gamma} \mathbf{\Delta}]_{j \cdot}  \big\|^2} \\
         &\leq \sqrt{s_u}\cdot  \| \mathbf{\Gamma} \mathbf{\Delta}\|_F\leq \sqrt{s_u\sigma_{\max}(\mathbf{L})} \cdot \| \mathbf{\Delta}\|_F.
    \end{split}
\end{equation}
The expression in Equation~(\ref{eq:BI}) can be thus simplified to
\begin{equation}\label{eq:reg_intermediary_graph}
    \begin{split}
        \frac{1}{n}\| \mathbf{X} \mathbf{\Delta}\|_F^2
        \leq   \|\mathbf{\Delta}\|_F \left(     C R_\Pi + \frac{3\rho}{2}\sqrt{s_u\sigma_{\max}(\mathbf{L})} \right)    - \frac{\rho}{2} \big\| [\mathbf{\Gamma} \mathbf{\Delta} ]_{S^c\cdot}\big\|_{21}.
    \end{split}
\end{equation}
Note that Equation~(\ref{eq:reg_intermediary_graph}) implies the following generalized cone constraint:
\begin{equation}\label{eq:cone_graph}
    \begin{split}
 \big\| [\mathbf{\Gamma} \mathbf{\Delta} ]_{S^c\cdot}\big\|_{21}
 & \leq   \|\mathbf{\Delta}\|_F \Bigg(      \frac{2C R_\Pi}{\rho} + 3 \sqrt{s_u\sigma_{\max}(\mathbf{L})}\Bigg).
    \end{split}
\end{equation}
Equation~(\ref{eq:reg_intermediary_graph}) also implies:
\begin{equation}\label{eq:reg_bound1}
    \begin{split}
        \frac{1}{n}\| \mathbf{X} \mathbf{\Delta}\|_F^2& \leq   \|\mathbf{\Delta}\|_F \left(            C R_\Pi  + \frac{3\rho}{2}\sqrt{s_u\sigma_{\max}(\mathbf{L})}  \right).  
    \end{split}
\end{equation}
To proceed any further, we must show a Restricted Eigenvalue property on our generalised cone (Eq.~\ref{eq:cone_graph}). To this end, we leverage Lemma 2.2 in \cite{tran2022generalized}.
\begin{lemma}[Lemma 2.2 in \cite{tran2022generalized}]\label{lemma:huy}
If \(\mathbf{X} \in \mathbb{R}^{n \times p}\) has i.i.d. \(\mathcal{N}(0, {\Sigma_X})\) rows and {\(p \geq 2\)}, \(n \geq 10\), then the event
\[
\left\|\frac{\mathbf{X}v}{\sqrt{n}}\right\|^2 \geq v^\top \left(\frac{1}{64}{\mathbf{\Sigma}_X} \right)v - 72\sigma_{\text{max}}(\mathbf{\Sigma}_X)n_c\frac{\|v\|^2}{n} - 576\rho(\mathbf{\Gamma})^2\frac{\sigma_{\text{max}}({\mathbf{\Sigma}_X}) \log p}{n} \|\mathbf{\Gamma} v\|_1^2
\]
holds for all \( v \in \mathbb{R}^p \) with probability at least \( 1 - c_1 \exp\{-nc_2\} \), for some universal constants \(c_1, c_2 > 0\).
\end{lemma}

\noindent Let $\mathcal{A}^j_{\kappa}$ and $\mathcal{A}_{\kappa}$ denote the events:
$$\mathcal{A}^j_{\kappa}=\left \{ \left\|\frac{\mathbf{X}\mathbf{\Delta}_{\cdot j}}{\sqrt{n}}\right\|^2 \geq \mathbf{\Delta}_{\cdot j}^\top \left(\frac{1}{64}\mathbf{\Sigma}_X \right)\mathbf{\Delta}_{\cdot j} - 72\sigma_{\text{max}}(\mathbf{\Sigma}_X)n_c\frac{\|\mathbf{\Delta}_{\cdot j}\|^2}{n} - 576\rho(\mathbf{\Gamma})^2\frac{\sigma_{\text{max}}(\mathbf{\Sigma}_X) \log p}{n} \|\mathbf{\Gamma} \mathbf{\Delta}_{\cdot j}\|_1^2\right\}$$
$$\mathcal{A}_{\kappa} = \left\{
\frac{\|{\mathbf{X}}\mathbf{\Delta}\|_F^2}{{n}} \geq\frac{1}{64} \fnorm{\mathbf{\Sigma}^{\frac12}_X\mathbf{\Delta} }^2 - \frac{72Mn_c}{n} \|\mathbf{\Delta} \|_F^2- 576 \frac{\rho(\mathbf{\Gamma})^2M\log p}{n} \sum_{j \in [q]}\big\|[\mathbf{\Gamma} \mathbf{\Delta}]_{\cdot j}\big\|_1^2
\right\} $$
By Lemma~\ref{lemma:huy}, applying a simple union bound for $q$ vectors
$\mathcal{A}_{\kappa}^c \subset \bigcup_{j \in [q]} (\mathcal{A}_{\kappa}^j)^c$,
therefore:
\begin{equation}\label{eq:proba_kappa}
    \begin{split}
 \P[\mathcal{A}_{\kappa}^c] &\leq {\sum_{j \in [q]}} \P\big[(\mathcal{A}_{\kappa}^j)^c\big]\leq qe^{-c_2n}
    \end{split}
\end{equation}
This means in particular:
\begin{equation}\label{eq:rep_graph}
    \begin{split}
\frac{1}{64} \fnorm{\mathbf{\Sigma}^{\frac12}_X\mathbf{\Delta} }^2\leq \frac{\|{\mathbf{X}}\mathbf{\Delta}\|_F^2}{{n}} + \frac{72Mn_c}{n} \|\mathbf{\Delta} \|_F^2+ 576 \frac{\rho(\mathbf{\Gamma})^2M\log p}{n} \sum_{j \in [q]}\big\|[\mathbf{\Gamma} \mathbf{\Delta}]_{\cdot j}\big\|_1^2
    \end{split}
\end{equation}
{
We note that, for any $v\in\mathbb R^{p\times q}$,
\begin{equation}
    \begin{split}
\sum_{j \in [q]}\big\|[\mathbf{\Gamma} v]_{\cdot j}\big\|_1^2 
&= \sum_{j \in [q]} \sum_{a \in [m]}\big|[\mathbf{\Gamma} v]_{a j}\big| \sum_{b \in [m]}\big|[\mathbf{\Gamma} v]_{b j}\big| \\
&=  \sum_{a \in [m]}\sum_{b \in [m]} \sum_{j \in [q]} \big|[\mathbf{\Gamma} v]_{a j}\big| \big|[\mathbf{\Gamma} v]_{b j} \big|\\
&\leq  \sum_{a \in [m]}\sum_{b \in [m]} \big\|[\mathbf{\Gamma} v]_{a\cdot}\big\|~\big\|[\mathbf{\Gamma} v]_{b\cdot}\big\| =  
\| \mathbf{\Gamma} v\|_{21}^2.
    \end{split}
\end{equation}
This is the point where no support bound on $\widehat{\mathbf B}$ is needed: the graph restricted-eigenvalue lemma is uniform in $v$, and the cone bound controls $\|\mathbf\Gamma\Delta\|_{21}$ directly. Combining Equations~\eqref{eq:CS} and~\eqref{eq:cone_graph},
\begin{equation}\label{eq:bound_gamma_delta}
    \begin{split}
 \| \mathbf{\Gamma} \mathbf{\Delta}\|_{21}
 &\leq \big\| [\mathbf{\Gamma} \mathbf{\Delta} ]_{S^c\cdot}\big\|_{21}
 +\big\| [\mathbf{\Gamma} \mathbf{\Delta} ]_{S\cdot}\big\|_{21}\\
 & \leq   \|\mathbf{\Delta}\|_F \left(\frac{2C R_\Pi}{\rho}+4 \sqrt{s_u \sigma_{\max}(\mathbf{L})} \right).
    \end{split}
\end{equation}
Let
\[
K_\Gamma=\frac{2C R_\Pi}{\rho}+4\sqrt{s_u\sigma_{\max}(\mathbf L)}.
\]
{By the lower bound on $\rho$, $K_\Gamma\le K_{\Gamma,n}$ after increasing the numerical constant $C_\star$ in the theorem statement.}
Combined with Equations~\eqref{eq:bound_gamma_delta} and~\eqref{eq:reg_bound1}, Equation~\eqref{eq:rep_graph} becomes
\begin{equation}
    \begin{split}
 \frac{1}{64} \fnorm{\mathbf{\Sigma}^{\frac12}_X\mathbf{\Delta}}^2
&\leq \|\mathbf\Delta\|_F\left(CR_\Pi+\frac{3\rho}{2}\sqrt{s_u\sigma_{\max}(\mathbf L)}\right) \\
&\quad +\frac{M}{n}\|\mathbf\Delta\|_F^2
\left(72n_c+576\rho(\mathbf\Gamma)^2\log p\,K_\Gamma^2\right).
    \end{split}
\end{equation}
Using $\sigma_{\min}(\mathbf\Sigma_X)\ge 1/M$, if
\[
 n\ge 128M^2\left(72n_c+576\rho(\mathbf\Gamma)^2\log p\,K_\Gamma^2\right),
\]
the quadratic term on the right-hand side is absorbed into the left-hand side, and hence
\begin{equation}\label{eq:graph_frob_delta_bound}
    \begin{split}
\|\mathbf\Delta\|_F
&\le C M\left(R_\Pi+\rho\sqrt{s_u\sigma_{\max}(\mathbf L)}\right)\\
&\le C M^3\left(\sqrt{n_c\frac{q+\log n_c}{n}}
+\left(1+\frac{1}{\kappa_2}\right)\sqrt{\sigma_{\max}(\mathbf L)}
\sqrt{s_u\frac{q+\log m}{n}}\right).
    \end{split}
\end{equation}
Consequently, Equation~\eqref{eq:reg_bound1} gives
\begin{equation}\label{eq:reg_bound_graph_final}
    \begin{split}
       \frac{1}{n}\| \mathbf{X} \mathbf{\Delta}\|_F^2
       &=\|\widehat{\mathbf\Sigma}_X^{1/2}\mathbf\Delta\|_F^2\\
       &\leq   C M^5 \left( {n_c\frac{q+\log(n_c)}{n}} +  {\left(1 + \frac{1}{\kappa_2}\right)^2 \sigma_{\max}(\mathbf{L})\cdot {s_u\frac{q + \log m}{n} }}  \right).
    \end{split}
\end{equation}
}
{
We conclude the proof by characterizing the probability of the event
\[
\mathcal E_{\rm graph}=\mathcal E_{\rm score}\cap\mathcal A_\kappa.
\]
By the union bound and Equation~\eqref{eq:proba_kappa},
\begin{equation}
    \begin{split}
\mathbb P(\mathcal E_{\rm graph})
&\ge 1-\mathbb P(\mathcal A_Y^c)-\mathbb P(\mathcal A_{X\Pi,Y}^c)
      -\mathbb P(\mathcal A_{\Pi XW}^c)\\
&\qquad -\mathbb P(\mathcal A_{X\Gamma^\dagger,W})
      -\mathbb P(\mathcal A_{(\Gamma^\dagger)^\top\Sigma_{XY}})
      -\mathbb P(\mathcal A_\kappa^c)\\
&\ge 1-\exp\{-C'q\}-2\exp\{-C'(q+\log n_c)\}
       -2\exp\{-C'(q+\log m)\}-q\exp\{-c_2n\}.
    \end{split}
\end{equation}
}

\end{proof}

\subsection{Consistency of the graph-regularized CCA estimator}
\label{proof:theorem:graph_2}

{

\begin{lemma}[Graph-restricted covariance for the normalization step]\label{lemma:graph_restricted_cov_norm}
For $R\ge 1$, define
\[
\mathbb T_{\mathcal G}(R)
=
\left\{
\mathbf H\in\mathbb R^{p\times r}:
\|\mathbf H\|_F\le 1,
\|\mathbf\Gamma \mathbf H\|_{21}\le R
\right\}
\]
and
\[
 w_{\mathcal G}(R)
 =
 \sqrt{r n_c}+R\rho(\mathbf\Gamma)\sqrt{r+\log m},
 \qquad
 \varepsilon_{\mathcal G}(R)
 =
 C M\left\{\frac{w_{\mathcal G}(R)}{\sqrt n}
 +\frac{w_{\mathcal G}^2(R)}{n}\right\}.
\]
Then, for a numerical constant $c>0$, with probability at least
$1-2\exp\{-c w_{\mathcal G}^2(R)\}$,
\begin{equation}\label{eq:graph_restricted_cov_norm_event}
\sup_{\mathbf H_1,\mathbf H_2\in\mathbb T_{\mathcal G}(R)}
\left\|
\mathbf H_1^\top(\widehat{\mathbf\Sigma}_X-\mathbf\Sigma_X)\mathbf H_2
\right\|_F
\le
\varepsilon_{\mathcal G}(R).
\end{equation}
\end{lemma}

\begin{proof}
Let $\mathbf G\in\mathbb R^{p\times r}$ have i.i.d. $N(0,1)$ entries.  For any
$\mathbf H\in\mathbb T_{\mathcal G}(R)$, use
$\mathbf H=\mathbf\Pi\mathbf H+\mathbf\Gamma^\dagger\mathbf\Gamma\mathbf H$ to write
\[
\langle \mathbf G,\mathbf H\rangle
=
\langle \mathbf\Pi\mathbf G,\mathbf\Pi\mathbf H\rangle
+
\langle (\mathbf\Gamma^\dagger)^\top\mathbf G,
        \mathbf\Gamma\mathbf H\rangle .
\]
Therefore
\[
\sup_{\mathbf H\in\mathbb T_{\mathcal G}(R)}
\langle \mathbf G,\mathbf H\rangle
\le
\|\mathbf\Pi\mathbf G\|_F
+
R\|(\mathbf\Gamma^\dagger)^\top\mathbf G\|_{\infty,2}.
\]
Since $\operatorname{rank}(\mathbf\Pi)=n_c$,
$\mathbb E\|\mathbf\Pi\mathbf G\|_F\le \sqrt{r n_c}$.  Also, by the definition of
$\rho(\mathbf\Gamma)$ and a union bound over the $m$ rows,
\[
\mathbb E\|(\mathbf\Gamma^\dagger)^\top\mathbf G\|_{\infty,2}
\le
C\rho(\mathbf\Gamma)\sqrt{r+\log m}.
\]
Thus the Gaussian width of $\mathbb T_{\mathcal G}(R)$ is at most
$Cw_{\mathcal G}(R)$.  The standard Gaussian quadratic-process bound for subgaussian
quadratic forms over a star-shaped class then gives
\[
\sup_{\mathbf H\in\mathbb T_{\mathcal G}(R)}
\left|
\|\widehat{\mathbf\Sigma}_X^{1/2}\mathbf H\|_F^2
-
\|\mathbf\Sigma_X^{1/2}\mathbf H\|_F^2
\right|
\le
C M\left\{\frac{w_{\mathcal G}(R)}{\sqrt n}
+\frac{w_{\mathcal G}^2(R)}{n}\right\}
\]
with probability at least $1-2\exp\{-c w_{\mathcal G}^2(R)\}$.  Finally, the bilinear
form bound \eqref{eq:graph_restricted_cov_norm_event} follows by polarization.  Indeed,
for any matrices $\mathbf A,\mathbf B\in\mathbb R^{r\times r}$ with
$\|\mathbf A\|_F\le1$ and $\|\mathbf B\|_F\le1$, the matrices
$\mathbf H_1\mathbf A$ and $\mathbf H_2\mathbf B$ remain in
$\mathbb T_{\mathcal G}(R)$, and applying the quadratic bound to
$(\mathbf H_1\mathbf A\pm\mathbf H_2\mathbf B)/2$ yields the displayed Frobenius bound.
\end{proof}
}

{\begin{theorem}
   Consider the family $\mathcal{F}_{\mathcal{G}}(s_u,p, m, q,r,\lambda, M)$ of covariance matrices satisfying assumptions.
    We assume as well that $n$ is large enough so that:
    $$n \geq 10 \vee 128 M^2 \Big( 72 n_c+   576\rho(\mathbf{\Gamma})^2\log p \big(    96     C'\sqrt{n_c} + 4 {\sqrt{s_u\sigma_{\max}(\mathbf L)}} \big)^2 \Big),$$
    and such that:
        \begin{equation}\label{assumption1G}
        M\frac{{q}+s_u\log\big(\frac{ep}{s_u}\big)}{n}\leq c_0 \qquad  \text{ and } \qquad M \frac{\log n_c+q}{n} \leq c_0 \tag{$\mathcal{H}_{1, \mathcal{G}}$}
    \end{equation} 
    {
Define
\[
\delta_{\mathcal G}
=
\sqrt{n_c\frac{q+\log n_c}{n}}
+
\left(1+\frac{1}{\kappa_2}\right)
\sqrt{\sigma_{\max}(\mathbf L)s_u\frac{q+\log m}{n}},
\]
\[
R_{\mathcal G}
=
C_R\frac{M^{7/2}}{\lambda}
\left(\sqrt{n_c}+\sqrt{s_u\sigma_{\max}(\mathbf L)}\right),
\qquad
\varepsilon_{\mathcal G}=\varepsilon_{\mathcal G}(R_{\mathcal G}),
\]
where $\varepsilon_{\mathcal G}(\cdot)$ is defined in
Lemma~\ref{lemma:graph_restricted_cov_norm}.  We also assume the normalization radius is
small enough that
\[
C\left\{\frac{M^{9/2}\sqrt r}{\lambda}\delta_{\mathcal G}
+r\varepsilon_{\mathcal G}\right\}\le \frac{1}{2M}.
\]
}
    
Then, there exist orthogonal matrices $\mathbf{O}\in\R^{r\times r}$  and $\tilde{\mathbf{O}}\in\R^{r\times r}$ such that the canonical directions $\widehat{\mathbf U}$ and $\widehat{\mathbf V}$ estimated using  the graph algorithm have errors bounded by:

\begin{equation*}
    \begin{split}
\big\| \widehat{\mathbf{V}}  - \mathbf{V} \mathbf{O}\big\|_F &\leq  CM^{\frac{11}{2}}\frac{\sqrt{r}}{\lambda} \left( \sqrt{{n_c\frac{q +\log(n_c)}{n}}}+  \left(1 + \frac{1}{\kappa_2}\right) {\sqrt{\sigma_{\max}(\mathbf{L})}}\cdot \sqrt{{s_u\frac{q + \log m}{n} }}\right)  \\
        \| \widehat{\mathbf U} - \mathbf{U} \widetilde{\mathbf{O}}\|_F  &\leq {CM^6 \frac{\sqrt r}{\lambda}} \left(\sqrt{n_c{\frac{q + \log(n_c)}{n}}}+  \left(1 + \frac{1}{\kappa_2}\right) \sqrt{\sigma_{\max}(\mathbf{L})}\cdot \sqrt{{s_u\frac{q + \log m}{n} }}\right) {+ C M^{3/2} r\varepsilon_{\mathcal G}}. \\
    \end{split}
\end{equation*}
with probability at least $$
{1-\exp\{-C'q\}-3\exp\{-C'(q+\log n_c)\}-2\exp\{-C'(q+\log m)\}-q\exp\{-c_2n\}-2\exp\{-c w_{\mathcal G}^2(R_{\mathcal G})\}}.$$ Here $C$ is a constant that depends solely on $C'$ and $c_0$, and $c_2$ is a universal constant that does not depend on the dimensions. 
\end{theorem}
}

\begin{proof}
{
Let
\[
\delta_{\mathcal G}
=
\sqrt{n_c\frac{q+\log n_c}{n}}
+
\left(1+\frac{1}{\kappa_2}\right)
\sqrt{\sigma_{\max}(\mathbf L)s_u\frac{q+\log m}{n}} .
\]
Theorem~\ref{theorem:cca_rrr_graph_ols} gives
\begin{equation}\label{eq:graph_B_error_for_directions_red}
\|\widehat{\mathbf B}-\mathbf B^*\|_F
\le C M^3\delta_{\mathcal G}.
\end{equation}
Let
\[
\widehat{\mathbf B}
=\widehat{\mathbf U}_{\widehat B}\widehat{\mathbf D}_{\widehat B}
\widehat{\mathbf V}_{\widehat B}^{\top}
\]
be the rank-$r$ truncated SVD used in the graph Algorithm, and let
$\mathbf B^*=\mathbf U_B\mathbf D_B\mathbf V_B^\top$ be the compact SVD of
$\mathbf B^*$. As in the sparse proof, $\mathbf V_B=\mathbf V_0\mathbf Q_Y$ for some
$\mathbf Q_Y\in\mathcal O_r$, where $\mathbf V_0=\mathbf\Sigma_Y^{1/2}\mathbf V$, and
\[
\sigma_r(\mathbf B^*)
\ge \lambda/\sqrt M .
\]
For sufficiently small $c_0$, \eqref{eq:graph_B_error_for_directions_red} is smaller than
a fixed fraction of $\sigma_r(\mathbf B^*)$. Therefore Wedin's theorem gives orthogonal
matrices $\mathbf O_X,\mathbf O_Y\in\mathcal O_r$ such that
\begin{equation}\label{eq:graph_wedin_subspace_red}
\|\widehat{\mathbf U}_{\widehat B}-\mathbf U_B\mathbf O_X\|_F
\vee
\|\widehat{\mathbf V}_{\widehat B}-\mathbf V_B\mathbf O_Y\|_F
\le
C\frac{M^{7/2}}{\lambda}\delta_{\mathcal G}.
\end{equation}
The $Y$-side normalization follows exactly as in the sparse proof. On the covariance event
for $\widehat{\mathbf\Sigma}_Y$,
\[
\|\widehat{\mathbf\Sigma}_Y^{-1/2}\|_{op}\le C\sqrt M,
\qquad
\|\widehat{\mathbf\Sigma}_Y^{-1/2}-\mathbf\Sigma_Y^{-1/2}\|_{op}
\le C M^2\sqrt{q/n}.
\]
Since $\mathbf\Sigma_Y^{-1/2}\mathbf V_B=\mathbf V\mathbf Q_Y$, we obtain
\[
\|\widehat{\mathbf V}-\mathbf V\mathbf Q_Y\mathbf O_Y\|_F
\le
C\frac{M^{11/2}\sqrt r}{\lambda}\delta_{\mathcal G},
\]
after absorbing the $\sqrt{qr/n}$ covariance-normalization term into
$\sqrt r\delta_{\mathcal G}$.

It remains to justify the $X$-side normalization. This is where the sparse proof cannot be
copied verbatim: the graph penalty controls $\mathbf\Gamma\widehat{\mathbf B}$, but it does
not imply that $\widehat{\mathbf B}$ has a small row support. We therefore use
Lemma~\ref{lemma:graph_restricted_cov_norm} instead of a selected-support covariance
bound.

First, we check that the singular-vector matrices to which the covariance event is applied
belong to the stated graph class. Let $S=\supp(\mathbf\Gamma\mathbf B^*)$. Since
$\operatorname{col}(\mathbf U_B)=\operatorname{col}(\mathbf U)$, we have
$\mathbf U_B=\mathbf U\mathbf R$ for some invertible $r\times r$ matrix $\mathbf R$, and
therefore $\supp(\mathbf\Gamma\mathbf U_B)\subseteq S$. Hence
\[
\|\mathbf\Gamma\mathbf U_B\|_{21}
\le
\sqrt{s_u}\|\mathbf\Gamma\mathbf U_B\|_F
\le
\sqrt{r s_u\sigma_{\max}(\mathbf L)}.
\]
Thus $\mathbf U_B/\sqrt r\in\mathbb T_{\mathcal G}(R_{\mathcal G})$. For the estimated
left singular vectors, use
$\widehat{\mathbf U}_{\widehat B}=\widehat{\mathbf B}\widehat{\mathbf V}_{\widehat B}
\widehat{\mathbf D}_{\widehat B}^{-1}$ and Weyl's inequality, which gives
$\sigma_r(\widehat{\mathbf B})\ge \lambda/(2\sqrt M)$ on the present event. Therefore
\[
\|\mathbf\Gamma\widehat{\mathbf U}_{\widehat B}\|_{21}
\le
\frac{2\sqrt M}{\lambda}
\left(\|\mathbf\Gamma\mathbf B^*\|_{21}+\|\mathbf\Gamma\mathbf{\Delta}\|_{21}\right).
\]
Moreover,
\[
\|\mathbf\Gamma\mathbf B^*\|_{21}
\le
\sqrt{s_u}\|\mathbf\Gamma\mathbf B^*\|_F
\le
\sqrt{M r s_u\sigma_{\max}(\mathbf L)},
\]
and the cone bound in the regression proof gives
\[
\|\mathbf\Gamma\mathbf{\Delta}\|_{21}
\le
C\|\mathbf{\Delta}\|_F\left(\sqrt{n_c}+\sqrt{s_u\sigma_{\max}(\mathbf L)}\right)
\le
C M^3\delta_{\mathcal G}
\left(\sqrt{n_c}+\sqrt{s_u\sigma_{\max}(\mathbf L)}\right).
\]
Since $c_0$ is chosen small enough, the last display is dominated by the radius defining
$R_{\mathcal G}$, and hence
$\widehat{\mathbf U}_{\widehat B}/\sqrt r\in\mathbb T_{\mathcal G}(R_{\mathcal G})$.
Consequently, on the event of Lemma~\ref{lemma:graph_restricted_cov_norm},
\begin{equation}\label{eq:graph_sample_cov_on_svd_vectors_red}
\left\|
\widehat{\mathbf U}_{\widehat B}^{\top}
(\widehat{\mathbf\Sigma}_X-\mathbf\Sigma_X)
\widehat{\mathbf U}_{\widehat B}
\right\|_F
\le r\varepsilon_{\mathcal G}.
\end{equation}

Define
\[
\widehat{\mathbf A}=
\widehat{\mathbf U}_{\widehat B}^{\top}
\widehat{\mathbf\Sigma}_X
\widehat{\mathbf U}_{\widehat B},
\qquad
\mathbf A=\mathbf U_B^{\top}\mathbf\Sigma_X\mathbf U_B.
\]
The eigenvalues of $\mathbf A$ lie in $[1/M,M]$. Combining
\eqref{eq:graph_wedin_subspace_red} with
\eqref{eq:graph_sample_cov_on_svd_vectors_red} gives
\begin{equation}\label{eq:graph_Ahat_A_red}
\|\widehat{\mathbf A}-\mathbf O_X^\top\mathbf A\mathbf O_X\|_F
\le
C\frac{M^{9/2}\sqrt r}{\lambda}\delta_{\mathcal G}
+C r\varepsilon_{\mathcal G}.
\end{equation}
Indeed, writing
$\mathbf E_U=\widehat{\mathbf U}_{\widehat B}-\mathbf U_B\mathbf O_X$, the population
part is bounded by $CM\|\mathbf E_U\|_F$, and the empirical covariance part is bounded by
\eqref{eq:graph_sample_cov_on_svd_vectors_red}.  By the small-radius assumption in the
statement, $\lambda_{\min}(\widehat{\mathbf A})\ge 1/(2M)$. The inverse-square-root
perturbation bound on matrices with eigenvalues bounded below by $1/(2M)$ gives
\begin{equation}\label{eq:graph_Ahat_invsqrt_red}
\|\widehat{\mathbf A}^{-1/2}
-
\mathbf O_X^\top\mathbf A^{-1/2}\mathbf O_X\|_F
\le
C\frac{M^{6}\sqrt r}{\lambda}\delta_{\mathcal G}
+C M^{3/2}r\varepsilon_{\mathcal G}.
\end{equation}
Finally, let
$\mathbf U_\diamond=\mathbf U_B(\mathbf U_B^\top\mathbf\Sigma_X\mathbf U_B)^{-1/2}$.
As before, $\mathbf U_\diamond=\mathbf U\mathbf Q_X$ for some
$\mathbf Q_X\in\mathcal O_r$. Since
\[
\widehat{\mathbf U}
=
\widehat{\mathbf U}_{\widehat B}\widehat{\mathbf A}^{-1/2},
\]
we have, using \eqref{eq:graph_wedin_subspace_red} and
\eqref{eq:graph_Ahat_invsqrt_red},
\[
\|\widehat{\mathbf U}-\mathbf U_\diamond\mathbf O_X\|_F
\le
C\frac{M^{6}\sqrt r}{\lambda}\delta_{\mathcal G}
+C M^{3/2}r\varepsilon_{\mathcal G}.
\]
Taking $\mathbf O=\mathbf Q_Y\mathbf O_Y$ and
$\widetilde{\mathbf O}=\mathbf Q_X\mathbf O_X$ completes the proof.
}
\end{proof}

\end{document}